\documentclass[11pt]{article}
\pdfoutput=1 
\usepackage{mathtools}
\usepackage{booktabs}
\usepackage[english]{babel}
\usepackage{amsmath,amssymb,amsbsy,amstext, amsthm, simplewick, amsfonts}
\usepackage{graphicx}
\usepackage[small]{caption}
\usepackage{siunitx}
\usepackage{upgreek}
\usepackage{framed}
\usepackage{wrapfig}
\usepackage{multirow}
\usepackage{bbm}
\usepackage[numbers,sort&compress]{natbib}
\usepackage[svgnames,dvipsnames,x11names]{xcolor}
\usepackage[utf8x]{inputenc}
\usepackage{selinput}
\usepackage{bm}
\usepackage{float}
\usepackage{geometry}
\usepackage{yfonts}
\usepackage{caption}
\usepackage{subcaption}
\usepackage{sidecap}
\usepackage{longtable}
\usepackage{anyfontsize}

\usepackage{genyoungtabtikz}
\Yboxdim{11pt} 
\Ylinethick{.75pt}
\usepackage{epstopdf}
\usepackage{cancel}
\usepackage{tcolorbox}

\usepackage{tocloft}

\usepackage{breqn}

\newcommand{\ba}{\begin{eqnarray}}
\newcommand{\ea}{\end{eqnarray}}
\def\xyma{\xymatrix@M.7em}
\def\xymas{\xymatrix@M.1em}

\newcommand{\Comment}[1]{{}}
\definecolor{darkblue}{rgb}{0.15,0.35,0.55}
\definecolor{reddish}{rgb}{0.65, 0.2, 0.2}
\definecolor{darkgreen}{RGB}{50,150,0}
\definecolor{greyish2}{rgb}{.96,.96,.96}
\usepackage[linktocpage=true]{hyperref}
\hypersetup{
colorlinks=true,
citecolor=darkblue,
linkcolor=reddish,
urlcolor=darkblue,
pdfauthor={},
pdftitle={},
pdfsubject={}
}

\DeclareFontFamily{OT1}{rsfs10}{}
\DeclareFontShape{OT1}{rsfs10}{m}{n}{ <-> rsfs10 }{}
\DeclareMathAlphabet{\mathscript}{OT1}{rsfs10}{m}{n}

\def\gsim{ \lower .75ex \hbox{$\sim$} \llap{\raise .27ex \hbox{$>$}} }
\def\lsim{ \lower .75ex \hbox{$\sim$} \llap{\raise .27ex \hbox{$<$}} }
\def\be{\begin{equation}}
\def\ee{\end{equation}}
\def\bea{\begin{eqnarray}}
\def\eea{\end{eqnarray}}

\newcommand{\EE}[1]{{}^{#1}\hspace{-0.5pt}{\cal I}}
\newcommand{\II}[1]{{}^{#1}\hspace{-0.5pt}{I}}

\DeclareMathOperator{\e}{e}

\newcommand{\D}{{\rm d}}

\newcommand{\Mpl}{M_{\rm Pl}}

\newcommand{\hypergeom}[2]{% #1 presubscript, #2 postsubscript
  \mathbin{_{#1}{F}_{#2}} }

\newcommand{\gbRW}{g^{\rm Sch}}

\newcommand{\Hf}{{}^{(1)}\!H}
\newcommand{\Kf}{{}^{(1)}\!K}
\newcommand{\psif}{{}^{(1)}\!\psi}
\newcommand{\hf}{{}^{(1)}\!h}
\newcommand{\Hs}{{}^{(2)}\!H}
\newcommand{\Ks}{{}^{(2)}\!K}
\newcommand{\psis}{{}^{(2)}\!\psi}
\newcommand{\hs}{{}^{(2)}\!h}
\newcommand{\Apz}{a_{2}}
\newcommand{\Apu}{a_{1}}
\newcommand{\Apd}{a_{0}}
\newcommand{\Bpd}{b_0}
\newcommand{\Amz}{c_4}
\newcommand{\Amu}{c_3}
\newcommand{\Amd}{c_2}
\newcommand{\Amt}{c_1}
\newcommand{\Amq}{c_0}
\newcommand{\Bmz}{d_2}
\newcommand{\Bmu}{d_1}
\newcommand{\Bmd}{d_0}
\newcommand{\Ez}{e_2}
\newcommand{\Eu}{e_1}
\newcommand{\Ed}{e_0}
\newcommand{\Cmz}{f_3}
\newcommand{\Cmu}{f_2}
\newcommand{\Cmd}{f_1}
\newcommand{\Dmz}{g_2}
\newcommand{\Dmu}{g_1}
\newcommand{\Cz}{\tilde a_2}
\newcommand{\Cu}{\tilde a_1}
\newcommand{\Cd}{\tilde a_0}
\newcommand{\Dz}{\tilde b_1}
\newcommand{\Du}{\tilde b_0}

\def\ddl{\delta\!\!\!{}^-}
\newcommand{\ord}[1]{^{(#1)}}

\newcommand{\cY}{\mathcal{Y}}

\usepackage{tikz}
\usepackage[compat=1.1.0]{tikz-feynman}

\newcommand{\gravh}{
\begin{tikzpicture}[baseline]
\begin{feynman}
\vertex (a);
\vertex [right=1cm of a] (b);
\diagram* {
(a) -- [photon] (b)
}; 
\end{feynman}
\end{tikzpicture}
}

\newcommand{\gravH}{
\begin{tikzpicture}[baseline]
\begin{feynman}
\vertex (a);
\vertex [right=1cm of a] (b);
\diagram* {
(a) -- [plain, insertion={0}] (b)
}; 
\end{feynman}
\end{tikzpicture}
}

\newcommand{\Csoure}{
\begin{tikzpicture}[baseline]
\begin{feynman}
\vertex [dot, minimum size=0.25cm] (a) {};
\vertex [right=0.82cm of a] (d);
\vertex [right=0.2cm of d] (e);
\diagram* {
(a) -- [photon] (d),
(d) -- [plain, -stealth] (e)
}; 
\end{feynman}
\end{tikzpicture}
}

\newcommand{\Tsoure}{
\begin{tikzpicture}[baseline]
\begin{feynman}
\vertex [square dot, minimum size=0.25cm] (a) {};
\vertex [right=0.82cm of a] (d);
\vertex [right=0.2cm of d] (e);
\vertex [below=0.5cm of a] (c);
\vertex [left=0.3cm of c] (c1);
\vertex [right=0.3cm of c] (c2);
\vertex [right=0.05cm of c1] (c11);
\vertex [left=0.05cm of c2] (c22);
\diagram* {
(a) -- [photon] (d),
(d) -- [plain, -stealth] (e), 
(c1) -- [plain, insertion={0}] (a),
(c2) -- [plain, insertion={0}] (a),
(c11) -- [dotted] (c22)
}; 
\end{feynman}
\end{tikzpicture}
}

\newcommand{\proph}{
\begin{tikzpicture}[baseline]
\begin{feynman}
\vertex [label=90:$\scriptstyle{\mu\nu}$](a);
\vertex [right=1.5cm of a, label=90:$\scriptstyle{\rho\sigma}$] (b);
\diagram* {
(a) -- [photon, momentum={[arrow shorten=0.3pt]$k$}] (b)
}; 
\end{feynman}
\end{tikzpicture}
}

\newcommand{\cubicback}{
\begin{tikzpicture}[baseline]
\begin{feynman}
\vertex (a);
\vertex [right=1cm of a] (b);
\vertex [right=1cm of b] (c);
\vertex [above=0.5cm of c, label=90:$\scriptstyle{\alpha_2\beta_2}$] (du);
\vertex [below=0.5cm of c, label=270:$\scriptstyle{\alpha_3\beta_3}$] (dd);
\diagram* {
(a) -- [plain, insertion={0}] (b),
(du) -- [photon, momentum'={[arrow shorten=0.3pt]$k_2$}] (b),
(dd) -- [photon, momentum={[arrow shorten=0.3pt]$k_3$}] (b)
}; 
\end{feynman}
\end{tikzpicture}
}

\newcommand{\quarticback}{
\begin{tikzpicture}[baseline]
\begin{feynman}
\vertex (a);
\vertex [above=0.5cm of a] (Hu);
\vertex [below=0.5cm of a] (Hd);
\vertex [right=1cm of a] (b);
\vertex [right=1cm of b] (c);
\vertex [above=0.5cm of c, label=90:$\scriptstyle{\alpha_3\beta_3}$] (du);
\vertex [below=0.5cm of c, label=270:$\scriptstyle{\alpha_4\beta_4}$] (dd);
\diagram* {
(Hu) -- [plain, insertion={0}] (b),
(Hd) -- [plain, insertion={0}] (b),
(du) -- [photon, momentum'={[arrow shorten=0.3pt]$k_2$}] (b),
(dd) -- [photon, momentum={[arrow shorten=0.3pt]$k_3$}] (b)
}; 
\end{feynman}
\end{tikzpicture}
}

\newcommand{\cubic}{
\begin{tikzpicture}[baseline]
\begin{feynman}
\vertex [label=180:$\scriptstyle{\alpha_1\beta_2}$](a);
\vertex [right=1cm of a] (b);
\vertex [right=1cm of b] (c);
\vertex [above=0.5cm of c, label=90:$\scriptstyle{\alpha_2\beta_2}$] (du);
\vertex [below=0.5cm of c, label=270:$\scriptstyle{\alpha_3\beta_3}$] (dd);
\diagram* {
(a) -- [photon, momentum={[arrow shorten=0.3pt]$k_1$}] (b),
(du) -- [photon, momentum'={[arrow shorten=0.3pt]$k_2$}] (b),
(dd) -- [photon, momentum={[arrow shorten=0.3pt]$k_3$}] (b)
}; 
\end{feynman}
\end{tikzpicture}
}

\newcommand{\Source}{
\begin{tikzpicture}[baseline]
\begin{feynman}
\vertex [dot, minimum size=0.25cm, label=180:$M$] (a) {};
\vertex [right=1cm of a] (d);
\vertex [right=0.2cm of d, label=90:$\scriptstyle{\mu\nu}$] (f);
\diagram* {
(a) -- [photon, momentum={[arrow shorten=0.3pt]$k$}] (d),
(d) -- [plain, -stealth] (f)
}; 
\end{feynman}
\end{tikzpicture}
}

\newcommand{\genhUN}{
\begin{tikzpicture}[baseline]
\begin{feynman}
\vertex [dot, minimum size=0.25cm, label=180:$M$] (a) {};
\vertex [empty dot, minimum size=0.8cm, right=1.3cm of a] (c) {$\mathcal{A}$};
\vertex [below=1cm of c] (b);
\vertex [left=0.7cm of b, label=180:$\scriptstyle{\mathcal{E}} \, $] (H1);
\vertex [right=0.4cm of H1] (Haux);
\vertex [below left=0.2cm of H1] (Hauxb1);
\vertex [right=0.7cm of b, label=0:$\; \scriptstyle{\mathcal{E}}$] (H3);
\vertex [left=0.4cm of H3] (Haux2);
\vertex [below right=0.2cm of H3] (Hauxb2);
\vertex [right=1.2cm of c, label=270:$\scriptstyle{\mu\nu}$, label=90:$x$] (d);
\diagram* {
(a) -- [photon] (c),
(c) -- [photon] (d),
(c) -- [plain, insertion={1}] (H1),
(c) -- [plain, insertion={1}] (H3),
(Haux) -- [dotted] (Haux2)
}; 
\draw [decoration={brace, mirror}, decorate] (Hauxb1.south east) -- (Hauxb2.south west) node [pos=0.5, below] {\(n \)};
\end{feynman}
\end{tikzpicture}
}

\newcommand{\genhT}{
\begin{tikzpicture}[baseline]
\begin{feynman}
\vertex [square dot, minimum size=0.25cm, label=180:$\lambda$] (a) {};
\vertex [below=1cm of a] (b);
\vertex [left=0.7cm of b, label=180:$\scriptstyle{\mathcal{E}} \, $] (H1);
\vertex [right=0.4cm of H1] (Haux);
\vertex [below left=0.2cm of H1] (Hauxb1);
\vertex [right=0.7cm of b, label=0:$\; \scriptstyle{\mathcal{E}}$] (H3);
\vertex [left=0.4cm of H3] (Haux2);
\vertex [below right=0.2cm of H3] (Hauxb2);
\vertex [right=1cm of a, label=270:$\scriptstyle{\mu\nu}$, label=90:$x$] (d);
\diagram* {
(a) -- [photon] (d),
(a) -- [plain, insertion={1}] (H1),
(a) -- [plain, insertion={1}] (H3),
(Haux) -- [dotted] (Haux2)
}; 
\draw [decoration={brace, mirror}, decorate] (Hauxb1.south east) -- (Hauxb2.south west) node [pos=0.5, below] {\(n \)};
\end{feynman}
\end{tikzpicture}
}

\usepackage{latexsym,amsmath,amssymb,epsfig}

\definecolor{greyish}{rgb}{.90,.90,.90}
\definecolor{greyish2}{rgb}{.96,.96,.96}
\usepackage{xcolor,colortbl}
\usepackage{tcolorbox}

\usepackage[all]{xy}

\usepackage{ytableau}

\title{}
\author{}

\numberwithin{equation}{section}

\begin{document}

\begin{flushright}
	  DESY\,24-141\\\phantom{~}
\end{flushright}

\vspace{-0.5cm}

\renewcommand{\thefootnote}{\fnsymbol{footnote}}

%\clearpage
%\thispagestyle{empty}
%
~
\begin{center}
{\fontsize{21.5}{18} \bf{Vanishing of Quadratic Love Numbers \\ 
  \vspace{.4cm}
  of Schwarzschild Black Holes }}
\end{center} 

\vspace{.15truecm}

\begin{center}
{\fontsize{13.5}{18}\selectfont
Simon Iteanu,${}^{\rm a,b}$
Massimiliano Maria Riva,${}^{\rm c,d}$ Luca Santoni,${}^{\rm b}$ \\
  \vspace{.3cm}
  Nikola Savi\'c,${}^{\rm e}$ and Filippo Vernizzi${}^{\rm e}$}
\end{center}
\vspace{.4truecm}

\centerline{{\it ${}^{\rm a}$\'Ecole Normale Sup\'erieure de Lyon,  Laboratoire de Physique, Lyon F-69342, France}}
 
  \vspace{.3cm}

\centerline{{\it ${}^{\rm b}$Universit\'e Paris Cit\'e, CNRS, Astroparticule et Cosmologie, }}
 \centerline{{\it  10 Rue Alice Domon et L\'eonie Duquet, F-75013 Paris, France}}
 
  \vspace{.3cm}

 \centerline{{\it ${}^{\rm c}$Deutsches Elektronen-Synchrotron DESY, Notkestr.~85, 22607 Hamburg, Germany}}
% \centerline{{\it Columbia University, New York, NY 10027}} 
 
  \vspace{.3cm}

 \centerline{{\it ${}^{\rm d}$Department of Physics, Center for Theoretical Physics, Columbia University,}}
 \centerline{{\it New York, 538 West 120th Street, NY 10027, U.S.A.}} 
 
  \vspace{.3cm}
  
 \centerline{{\it ${}^{\rm e}$Universit\'e Paris-Saclay, CNRS, CEA, }}
  \centerline{{\it Institut de Physique Th\'eorique, 91191 Gif-sur-Yvette, France}}
  \vspace{.3cm}

  \vspace{.5cm}

\begin{abstract}
\noindent
The induced conservative tidal response of self-gravitating objects in general relativity is parametrized in terms of a set of coefficients, which are commonly referred to as Love numbers. 
For asymptotically-flat black holes in four spacetime dimensions, the Love numbers are famously zero in the static regime. In this work, we show that this result continues to hold upon inclusion of nonlinearities in the theory for Schwarzschild black holes. We first solve the quadratic Einstein equations in the static limit to all orders in the multipolar expansion, including both even and odd perturbations. We show that the second-order solutions take simple analytic expressions, generically expressible in the form of finite polynomials. We then define the quadratic Love numbers at the level of the point-particle effective field theory. By performing the matching with the full solution in general relativity, we show that  quadratic Love number coefficients are zero to all orders in the derivative expansion, like the linear ones. 
\end{abstract}

\newpage

\setcounter{tocdepth}{3}
\tableofcontents
\newpage
\renewcommand*{\thefootnote}{\arabic{footnote}}
\setcounter{footnote}{0}

\section{Introduction}

Gravitational waves from coalescing binary systems represent a unique channel to study the fundamental properties of gravity and compact objects in the strong-field regime. As the number of merger events observed by the LIGO-Virgo-KAGRA network \cite{KAGRA:2021vkt} increases  and, thanks also to future facilities and envisioned new detectors, we move toward an era of precision physics with gravitational waves, ever more accurate waveform templates will be necessary~\cite{Barack:2018yly,Sathyaprakash:2019yqt,Maggiore:2019uih,Barausse:2020rsu,Kalogera:2021bya,Berti:2022wzk}. This requires in particular to have a precise understanding of  the conservative and dissipative dynamics of two-body systems (see e.g., \cite{Blanchet:2013haa,Buonanno:2022pgc,Goldberger:2022ebt}), which includes tidal effects~\cite{Flanagan:2007ix,Kalin:2020lmz,Henry:2020ski}. 

The tidal deformability of  a  compact object is described in terms of a set of coefficients, which capture the conservative  and dissipative induced response when the object is acted upon by an external long-wavelength tidal field~\cite{Fang:2005qq,Damour:2009vw,Binnington:2009bb}. The coefficients associated with the conservative response are often called Love numbers and, together with the dissipative numbers, they carry relevant information about  the object's  structure and interior dynamics.  
This makes the tidal response coefficients an important target for current and future gravitational-wave observations from merging binary systems, which can offer valuable insights  into the equation of state of neutron stars \cite{Flanagan:2007ix,Vines:2011ud,Bini:2012gu,Baiotti:2016qnr},  the physics at the horizon of black holes~\cite{Fang:2005qq,Damour:2009vw,Binnington:2009bb,Kol:2011vg,Gurlebeck:2015xpa,Hui:2020xxx,Hui:2021vcv,Rai:2024lho,LeTiec:2020spy,LeTiec:2020bos,Charalambous:2021mea,Rodriguez:2023xjd}, as well as the existence of more exotic compact objects~\cite{Franzin:2017mtq,Cardoso:2017cfl,Pani:2019cyc,Cardoso:2019rvt,Chia:2023tle,Chia:2024bwc}.

In the Post-Newtonian (PN) regime of the inspiral of two compact objects, the leading-order tidal effects are associated with tidal heating, and start affecting the gravitational waveform at 2.5PN order for spinning objects and 4PN order for non-rotating ones \cite{Poisson:1994yf,Tagoshi:1997jy,Blanchet:2013haa}.  Conservative tidal effects   
are instead estimated to become relevant from 5PN order, in the limit of static tides (see, e.g., \cite{1983LNP...124...59D,Porto:2016pyg}). Although these are the most prominent tidal effects that are expected at the level of the  gravitational waveform, there are in general very good reasons to study {\em subleading} corrections to the physics of compact sources beyond the point-particle approximation. First of all, the calculation of  subleading effects, in addition to putting leading-order results on solid grounds,  may be necessary to  probe  and get insight into interesting properties of the system that do not enter at leading order. {Moreover, subleading effects might be useful to recalibrate effective models, improve waveform and break degeneracy among observables.} They are especially relevant whenever leading-order effects happen to be vanishing, making the next-to-leading-order corrections the actual dominant contributions. One notable example is given by black holes in General Relativity (GR), for which the static Love numbers---associated with the leading-order, conservative, finite-size terms in an expansion in the number of time derivatives---are absent for all multipoles~\cite{Fang:2005qq,Damour:2009vw,Binnington:2009bb,Kol:2011vg,Gurlebeck:2015xpa,Hui:2020xxx,Rai:2024lho,LeTiec:2020spy,LeTiec:2020bos,Charalambous:2021mea,Rodriguez:2023xjd}. This fact makes the study of subleading corrections to the conservative response of black holes particularly interesting at both theoretical and observational level.

One main motivation, at the theoretical level, to compute subleading tidal effects of black holes has to do with symmetries. The vanishing of the static  Love numbers in GR has for long time been recognized as an outstanding naturalness puzzle in gravity from an effective-field-theory  perspective~\cite{Rothstein:2014sra,Porto:2016zng}. Recently, symmetry explanations  have been proposed as possible  resolutions to the puzzle~\cite{Hui:2021vcv,Hui:2022vbh,Rai:2024lho,BenAchour:2022uqo,Berens:2022ebl,Charalambous:2021kcz,Charalambous:2022rre}.   It is therefore particularly interesting to understand what is the fate of these symmetries, whether they are an `accident' of the static regime of the linearized Einstein equations for the perturbations, or whether they instead persist beyond leading order. 

At subleading order, corrections to the tidal Love numbers of black holes are of two types: they can either result from finite-frequency effects, or from gravitational nonlinearities.\footnote{In the language of effective field theories, the former correspond to higher-order time-derivative operators, while the latter correspond to operators with higher number of fields. See sec.~\ref{sec:EFT} for details.} The former, which are often referred to as \textit{dynamical} Love numbers, capture the conservative response induced by the presence of a time-dependent tidal field; in the case of black holes, they have been studied in e.g.~\cite{Chakrabarti:2013lua,Poisson:2020vap,Charalambous:2021mea,Saketh:2023bul,Perry:2023wmm}. The latter, which will be the focus of the present work, stem instead from the field nonlinearities in the Einstein equations. Similarly to electromagnetism, where nonlinear polarization theory can be used to describe nonlinear optical effects, such as the nonlinear polarization of an optical medium \cite{10.1093/acprof:oso/9780198702764.001.0001}, one can formulate an analogous question in gravity and  ask what is the \textit{nonlinear} tidal deformability of a compact object.  In this work we make progress in this direction  by explicitly computing the nonlinear Love numbers of Schwarzschild black holes at second-order in perturbation theory.

Nonlinear corrections to  the Love numbers of black holes have been previously studied in \cite{Gurlebeck:2015xpa,Poisson:2020vap,Poisson:2021yau,Riva:2023rcm,Poisson:2009qj}.\footnote{See also \cite{DeLuca:2023mio} for a  scalar-field analysis.}
In particular,   refs.~\cite{Gurlebeck:2015xpa,Poisson:2021yau} provide evidence that the nonlinear Love numbers of Schwarzschild black holes vanish to all orders in perturbation theory, but their analysis is restricted to axisymmetric perturbations only. 
On the other hand, ref.~\cite{Poisson:2020vap} examines the quadratic tidal deformation of a non-rotating compact body using a post-Newtonian definition for the tidally induced multipole moments, focusing on the parity-even sector of perturbations. Moreover, in ref.~\cite{Riva:2023rcm} some of us showed that the electric-type quadrupolar Love numbers vanish at quadratic order by explicitly performing the matching with the point-particle Effective Field Theory (EFT), focusing on the parity-even sector of perturbations.

In this work, following up  on and extending the methods proposed in~\cite{Riva:2023rcm}, we present the first comprehensive study of quadratic corrections to the tidal deformability of Schwarzschild black holes in GR, integrating and expanding upon the results in the literature in multiple directions. First, as in \cite{Riva:2023rcm}, 
 we adopt a definition for the quadratic Love numbers based on the worldline EFT \cite{Goldberger:2004jt,Goldberger:2005cd} (see also \cite{Goldberger:2006bd,Rothstein:2014sra,Porto:2016pyg,Levi:2018nxp,Goldberger:2022ebt,Goldberger:2022rqf} for some reviews).  The EFT provides a robust definition of the (nonlinear) Love numbers as coupling constants of higher-dimensional operators, which allows to systematically address potential ambiguities resulting from  gauge invariance in the theory and nonlinearities. In addition, our analysis now includes both even and odd perturbations and is not restricted to axisymmetric configurations. Finally, we complete the program initiated in~\cite{Riva:2023rcm} by computing the quadratic Love number couplings for all multipoles, beyond the leading quadrupole.  We will show explicitly that quadratic Love numbers of asymptotically-flat Schwarzschild  black holes in four-dimensional GR vanish exactly for generic multipoles.

In particular, in sec.~\ref{sec:StaticQuadraticPerturbations}, we derive the perturbation equations for the Schwarzschild metric up to second order in the Regge--Wheeler gauge, in the static limit, for both even and odd components. The quadratic terms in these equations describe how the coupling of linear perturbations sources the metric, respecting  selection rules dictated by the symmetries of the setup, which we derive and discuss in detail. These equations are then used  to derive the second-order metric in sec.~\ref{sec:sol}, where we focus on the monopole, quadrupole, and hexadecapole induced by the quadratic coupling of two quadrupoles.
Sec.~\ref{sec:EFT} reviews the definition of quadratic Love numbers within the worldline EFT formalism and the background field method. In sec.~\ref{sec:match}, we use this to study the quadratic response of the Schwarzschild metric induced by an external field, which is then compared to the second-order solutions obtained in full GR.
{Sec.~\ref{sec:HM} is instead devoted to the analysis of second-order static perturbations of arbitrary multipole: first, we check that, thanks to the special structure of the source, the quadratic solutions take a simple polynomial form; we then perform the matching to the worldline EFT to all orders in the spatial derivatives, and conclude that all quadratic Love number couplings of Schwarzschild black holes vanish in GR. 

More technical aspects of the calculations are finally collected in the appendices~\ref{App: harmonics}-\ref{app:SchwM}.
In particular, in app.~\ref{App: harmonics} we review the integrals between  three  tensor harmonics used in the main text, app.~\ref{app:source} collects the explicit numerical values of some of the coefficients  in the main text, app.~\ref{App:SphToCart} explains how to express the tidal field in Cartesian coordinates, app.~\ref{App:STFrel} derives some useful relations between trace-free tensors and, finally, app.~\ref{app:SchwM} explains how to compute the Schwarzschild metric perturbatively within our approach.

\paragraph{Notations and conventions.}   We will work  in units such that $c=\hbar=1$ and use the mostly plus signature for the metric, $(-,+,+,+)$. The reduced Planck mass is defined by $\Mpl^{-2} \equiv 8\pi G$. For convenience, we will sometimes introduce the parameter $\kappa \equiv \sqrt{32\pi G}=2\Mpl^{-1}$. For a black hole of mass $M$, we denote the Schwarzschild radius by $r_s = 2 G M$. Our convention for the time Fourier transform  is
$\Psi(t, r, \theta, \phi)=  \int \frac{\D \omega}{2\pi} \e^{-i\omega t } \tilde \Psi (\omega, r, \theta, \phi) $. Moreover, we will
decompose quantities in spherical harmonics using the convention
$\tilde \Psi(\omega, r, \theta,\phi)= \sum_{\ell,m}  \tilde \Psi^{\ell m}  (\omega, r) Y_{\ell}^m( \theta,\phi)$, where   $Y_{\ell}^m(\theta,\phi)$ are normalized as $\int \D\Omega \, Y_{\ell}^m{}^*(\theta,\phi) Y_{\ell'}^{m'}(\theta,\phi) =\delta_{\ell\ell'}\delta^{mm'}$.  To simplify the notation,  we will often omit arguments and  the tilde to denote the Fourier transform, relying on the context to discriminate between the different meanings. Greek indices run over spacetime coordinates, latin indices run over spatial coordinates and capital latin indices run over the angular variables $\theta, \phi$. 
{We define the shorthand $\int_k\equiv \int \frac{\D^4k}{(2\pi)^4}$, and the notation $\ddl{}\ord{d}(x)\equiv (2\pi)^d \delta^{(d)}(x)$. The symbol $(\cdots)$ denotes symmetrization over the enclosed indices e.g., $A_{(\mu}B_{\nu)}=\frac{1}{2}(A_{\mu}B_{\nu}+ A_{\nu}B_{\mu})$, while we use $\langle \cdots \rangle$ to denote the traceless symmetrization.}
Depending on the context, we will use different metrics to raise and lower spacetime indices. The full metric $g_{\mu \nu}$ is used for the complete GR calculations in sec.~\ref{sec:StaticQuadraticPerturbations}, and when introducing the EFT worldline approach in sec.~\ref{ppaction}. In the EFT calculation, from sec.~\ref{sec:BGF} to sec.~\ref{sec:gaugetrans}, we will use the flat Minkowski metric $\eta_{\mu \nu}$ for this purpose. Additionally, we distinguish between the 2d Levi-Civita {\em tensor} $\epsilon $, whose indices are raised and lowered with the metric on the unit sphere, and the spacetime Levi-Civita {\em symbol} $\varepsilon $, whose indices are raised and lowered with $\eta_{\mu \nu}$.

\section{Static quadratic perturbations}
\label{sec:StaticQuadraticPerturbations}

Black hole perturbation theory has a long history, dating back to the original work by Regge and Wheeler on the study of linearized  perturbations of Schwarzschild spacetime in the late 1950s~\cite{Regge:1957td}. More recently, the study of black hole perturbations has been extended to second order, see e.g.~\cite{Nicasio:1998aj,Gleiser:1998rw,Brizuela:2006ne,Brizuela:2007zza,Nakano:2007cj,Brizuela:2009qd,Lagos:2022otp,Bourg:2024jme,Bucciotti:2024zyp,Bucciotti:2024jrv}.

In this section we study the deviations from the Schwarzschild metric  up to second order in the metric perturbations in the static limit. We first  expand the Einstein--Hilbert (EH) action,
\be
S_{\rm EH} = \int  {\rm d}^4 x \sqrt{-g}\,  \frac{\Mpl^2}{2}  R \;,
\label{EHaction}
\ee
where $R$ is the Ricci scalar, up to quadratic and cubic order in the metric perturbations. We then vary the action with respect to the metric perturbations, first at linear and then at second order, and we derive the second-order equations  for  the metric degrees of freedom.

\subsection{Metric parametrization and gauge choice}
\label{sec:RWpert}

We  decompose the  metric $g_{\mu \nu}$ as
\be
g_{\mu \nu} = \gbRW_{\mu\nu} + \delta g_{\mu\nu} \;, 
\label{gpert}
\ee
where $\gbRW_{\mu\nu}$
is the static and spherically symmetric background solution. In spherical polar coordinates, $(t, r , \theta, \phi)$, this is given by  the  Schwarzschild metric,
\begin{equation}
\gbRW_{\mu\nu} =  \begin{pmatrix}
- f(r) & 0 & 0 & 0 \\
* & f^{-1}(r) & 0 & 0 \\
* & * & r^2 & 0 \\
* & * & * &  r^2 \sin^2\theta
\end{pmatrix}   \;, \qquad f(r) \equiv  1-\frac{r_s}{r} \;,
\label{gbRW}
\end{equation}
where the asterisks denote symmetric components and $r_s  \equiv 2GM$, with $M$ the mass of the black hole.
In eq.~\eqref{gpert}, $\delta g_{\mu\nu}$ describes the deviation from the  Schwarzschild  metric.

To write $\delta g_{\mu\nu}$ in four spacetime dimensions, we  follow the standard Regge--Wheeler (RW) parameterization \cite{Regge:1957td}.
It is convenient to classify the metric perturbations   according to their behavior under  parity transformation,  $(\theta, \phi)\rightarrow(\pi-\theta, \phi+\pi)$. We thus distinguish between even, or polar, perturbations, and odd, or axial, perturbations. Moreover, we  use the index $ +$ to label parity-even quantities and the index $\rm -$ for parity-odd quantities.

With this decomposition and notation, the most general parametrization of $\delta g_{\mu \nu}$ in four spacetime dimensions takes the form $\delta g_{\mu \nu} =  \delta g_{\mu\nu}^{ +} +  \delta g_{\mu\nu}^{ -}$, with
\begin{align}
\delta g_{\mu\nu}^{\rm +} &=
\begin{pmatrix}
f(r) H_0 & H_1 & \nabla_A {\mathcal H}_0 \\
* & f^{-1}(r) H_2 & \nabla_A {\mathcal H}_1 \\
* \,\,\,\, & * & \,\,\,\, r^2 { K}
                                                   \gamma_{AB} + 
                                                   r^2(\nabla_A \nabla_B - \frac{1}{2} \gamma_{AB}) G
\end{pmatrix} \;, \label{RWgeven}\\
\delta g_{\mu\nu}^{\rm -} &=
\begin{pmatrix}
0 & 0 & - \epsilon_A{}^C \nabla_C {h_0}\\
* & 0 & - \epsilon_A{}^C \nabla_C {h_1}\\
* \,\,\,\, &  * & \,\,\,\, -{1\over 2} (\epsilon_A {}^C
                                           \nabla_C \nabla_B +
                                        \epsilon_B {}^C \nabla_C
                                           \nabla_A ) {h_2}
\end{pmatrix} \;. \label{RWgodd}
\end{align}
Here $\nabla_A$ and $\epsilon_{AB}$ are the covariant derivative and Levi-Civita tensor on a sphere, respectively, as reviewed in appendix~\ref{App: harmonics}. Two-dimensional tensors labeled by capital latin letters, $A, B, C, \ldots$, are raised and lowered using the  metric on the unit sphere $\gamma_{AB}$, with line element  $ \gamma_{A B} {\rm d}x^A {\rm d}x^B  = {\rm d}\theta^2 + \sin^2\theta\, {\rm d}\phi^2$. 

The quantities $H_a$, ${\cal H}_a$, $K$, $G$ and $h_a$ above are functions of the coordinates $(t, r , \theta, \phi)$. However, since here we consider  only the static limit, they are independent of time. Moreover,
given the symmetries of the background solution $\gbRW_{\mu\nu}$, it is convenient to decompose these quantities in spherical harmonics. For instance, for $H_0$ we write
\be
H_0 (r, \theta, \phi)= \sum_{\ell = 0 }^\infty \sum_{ m = -\ell}^\ell H_0^{\ell m} ( r) Y_{\ell}^m( \theta,\phi) \;,
\label{Xdecomp}
\ee 
where $\Delta_{S^2} Y_\ell^m \equiv \nabla^A \nabla_A Y_\ell^m = - \ell (\ell + 1) Y_\ell^m$ and $i \partial_\phi Y_\ell^m = - m Y_\ell^m$. 
While the even quantities $H_a$, ${\cal H}_a$, $K$, $G$ appearing in $\delta g_{\mu \nu}^+$ do not transform under parity, $h_a$ flip sign under a parity transformation, ensuring that  $g_{\mu \nu}$ transforms  under parity as a 2-tensor. Therefore, from the behavior of $Y_{\ell}^m(\theta,\phi)$  under parity, $Y_\ell^m(\pi - \theta,\phi+\pi) = (-1)^\ell Y_\ell^m(\theta,\phi)$, 
an even harmonic component, e.g.~$H_0^{\ell m} (r)$, picks up a factor of $(-1)^{\ell}$ under parity transformation, while an odd component, e.g.~$h_0^{\ell m} (r)$, picks up a factor of $(-1)^{\ell+1}$.

For later convenience, we rewrite eqs.~\eqref{RWgeven} and \eqref{RWgodd} upon using the decomposition \eqref{Xdecomp} and expressing the covariant derivatives in terms of partial derivatives. One obtains
\begin{align}
\delta g_{\mu\nu}^{\rm +} & =
\begin{pmatrix}
f H_0^{\ell m} & H_1^{\ell m} &   \mathcal{H}_0^{\ell m} \;  \partial_\theta &     \mathcal{H}_0^{\ell m} \;  \partial_\phi\\
* & f^{-1} H_2^{\ell m} & \mathcal{H}_1^{\ell m} \;   \partial_\theta  & \mathcal{H}_1^{\ell m} \;  \partial_\phi  \\
* & * & r^2 \; \left( {K}^{\ell m} +  G^{\ell m}   \,  \mathcal{W}\right) &   r^2  \;  G^{\ell m}  \mathcal{V}     \\
* & * & * &   r^2  \;  \sin^2\theta \left( {K}^{\ell m}  + G^{\ell m}  \,  \mathcal{W} \right) 
\end{pmatrix} Y_{\ell}^m  \, , \label{hPMeven4D} \\
\delta g_{\mu\nu}^{\rm -} & =
\begin{pmatrix}
0 & 0 & - \frac{1}{\sin\theta} \;  h_0^{\ell m} \;  \partial_\phi  & \sin\theta \;  h_0^{\ell m} \;  \partial_\theta  \\
* & 0 & - \frac{1}{\sin\theta} \;  h_1^{\ell m} \;  \partial_\phi  &  \sin\theta \;  h_1^{\ell m} \;  \partial_\theta  \\
* & * & -\frac{1}{\sin\theta} \;  h_2^{\ell m} \mathcal{V}  &  \sin\theta \;  h_2^{\ell m} \; \mathcal{W} \\
* & * & * &   \sin\theta \; h_2^{\ell m} \mathcal{V} 
\end{pmatrix}  Y_{\ell}^m  \, , \label{hPModd4D}
\end{align}
where the symbols $\mathcal{V} $ and $\mathcal{W} $ stand for differential operators  defined as $\mathcal{V} \equiv  \partial_{\theta}\partial_{\phi} - \frac{\cos\theta}{\sin\theta}\partial_\phi$ and $\mathcal{W} \equiv \frac{1}{2}( \partial_\theta^2 -  \frac{\cos\theta}{\sin\theta}\partial_\theta -\frac{1}{\sin^2\theta}\partial_\phi^2 )$, respectively.

To simplify the derivation of the perturbation equations and their solutions, it is useful to fix a gauge. A convenient choice, given the symmetries of the problem, is the RW gauge \cite{Regge:1957td,Nakano:2007cj,Brizuela:2009qd}, which sets
\be
\mathcal{H}_0  = \mathcal{H}_1 = G = h_2 =0 \;.
\label{RWgauge}
\ee
This condition completely fixes the gauge freedom and can be directly applied to the action without losing any constraints \cite{Motohashi:2016prk}.\footnote{A residual (large) gauge freedom remains when considering the lowest ($\ell = 0$ and $\ell = 1$) multipoles of the metric perturbations, as discussed in more details at the end of sec.~\ref{linpert} for linear perturbations and in sec.~\ref{sec:solutionsQuadrpoleTimesQuadrupole} for quadratic perturbations.}
We then expand the action up to cubic order in the remaining metric variables and study the equations obtained from the variation of the action perturbatively.\footnote{Note that, instead of going through the action, one could have chosen to work directly at the level of the Einstein equations, expanded up to second order in the fields. In the gauge \eqref{RWgauge},  the two ways of proceeding are clearly equivalent and lead to the same result. We choose here to start from the action just for  practical convenience \cite{Franciolini:2018uyq,Hui:2021cpm,Hui:2020xxx}; having to deal with less equations, this  facilitates the analysis at second order.} Specifically, for the metric variable $H_0$, we write $H_0 = \Hf_0 + \Hs_0 + \ldots$, where $\Hf_0$ is the solution to the linear equations, obtained by varying the quadratic action.  ${}^{(2)}\! H_0$ is the solution to the second-order equations, derived by varying the cubic action and by using the linear solution in the terms that are quadratic in the metric perturbations. Note that  the expansion parameter is the amplitude of the perturbations. Unlike the EFT calculation in sec.~\ref{sec:match}, no expansion is performed in $r_s/r$. In the following sections, we first study the linear order, and then the quadratic order.

\subsection{Linear perturbations}
\label{linpert}

Let us first briefly review the derivation of the linearized equations, obtained from the variation of the quadratic action with respect to the metric variables.  
Content and results of this subsection are well known; however, it will give us the opportunity to set up the notation and introduce the ingredients that will be necessary in sec.~\ref{sec:quad} for the analysis of the second-order perturbations.

There are four metric variables in the even sector, i.e.~$H_0$, $H_1$, $ H_2 $ and $K$, and two  in the
odd sector, i.e.~$h_0$ and  $h_1$.  However, not all of them are independent degrees of freedom. Indeed, in vacuum, the gravitational field is expected to describe  two propagating degrees of freedom: one even and one odd. Since at linear order  the even and odd components do not mix \cite{Regge:1957td}, we can treat them separately. Let us first focus on the even metric quantities. 
The four equations obtained by varying the action with respect to even quantities must contain three constraints and the dynamics of a single degree of freedom, as we are going to show next.

Instead of $H_2$, it will be convenient to use the variable $\psi$ obtained  by  substituting in the action~\cite{Kobayashi:2014wsa,Franciolini:2018uyq,Franciolini:2018aad}
\begin{equation}
H_2 =  f(r) \psi  + r \partial_r K +   \frac{1}{f(r)} \left( 1- \frac{3r_s}{2 r} + \frac12 \Delta_{S^2} \right) { K}    \, ,
\end{equation}
where $\Delta_{S^2} \equiv \partial_\theta^2 +  \frac{\cos\theta}{\sin\theta}\partial_\theta  + \frac{1}{\sin^2\theta}\partial_\phi^2$.   
This field redefinition has the advantage of simplifying the  linear field equations, making them of first order in $r$ rather than second order.

The linear equations are obtained by varying  the quadratic action with respect to $H_0$, $\psi$,  $K$ and $H_1$ and expanding in spherical harmonics. We first focus on $\ell \ge 2$; we discuss the monopole and dipole cases separately,  below. Defining for convenience  the shorthand notation
\begin{equation}
 \lambda \equiv \ell(\ell+1)\;,
\label{lambdadeff}
\end{equation}
and using $\Delta_{S^2}   { Y_\ell^m}= -\ell(\ell+1) { Y_\ell^m}$ to replace the angular derivatives, we obtain 
\begin{align}
\partial_r \psif^{\ell m} + \frac{\left(\lambda+2\right) r+2 r_s}{2  f(r) r^2} \; \psif^{\ell m} - \frac{\lambda  \left[\left(\lambda-2\right) r+3 r_s\right]}{4  f(r)^3 r^2} \; \Kf^{\ell m}   &= 0 \, ,
\label{eqeveneven1_lin} \\
\partial_r \Kf^{\ell m} - 2 f(r) \partial_r \Hf_0^{\ell m}  - \frac{\lambda +1}{ f(r) r} \; \Kf^{\ell m}  -\frac{\lambda}{r_s} \Hf_0^{\ell m} + \frac{2f(r) }{r_s}  \; \psif^{\ell m}  &= 0  \, ,
\label{eqeveneven2_lin} \\
\partial_r \psif^{\ell m}  -    \frac{\lambda  \left[ ( \lambda - 2) r + 3 r_s\right]  }{2 f(r)^2 r r_s}  \; \Hf_0^{\ell m} 
+  \frac{\lambda    \left[ (\lambda - 2) r (r-2r_s)  - 
   3 r_s^2 \right] }{2f(r)^3 r^2 r_s} \; \Kf^{\ell m}    
+ \frac{ (\lambda +1 )r + r_s}{ f(r) r^2 } \; \psif^{\ell m}  &= 0  \, ,
\label{eqeveneven3_lin} \\
\lambda r^2 f(r) \; \Hf_1^{\ell m} &= 0 \;.
\label{eqeveneven4_lin}
\end{align}
 As expected from spherical symmetry, there is no $m$ dependence in these equations.

The last equation simply implies $\Hf_1=0$. 
Combining the first and third equation allows to remove $\partial_r \psif^{\ell m}$ and obtain a linear algebraic equation relating  three variables,
\be
 \psif^{\ell m} - \frac{    ( \lambda - 2) r + 3 r_s }{  f(r) r_s} \; \Hf_0^{\ell m} - \frac{
    r (\lambda - 2) (3 r_s - 2 r) + 3 r_s^2   }{
 2  f(r )^2  r r_s } \; \Kf^{\ell m}  =  0 \;.
 \label{algebraic_lin}
\ee
We choose to describe the independent even degree of freedom in terms of the quantity $ H_0 $. Thus, this equation and its first derivative with respect to $r$, together with eqs.~\eqref{eqeveneven1_lin} and \eqref{eqeveneven2_lin}, form a system of first-order differential equations which can be solved algebraically for $\psif$ and $\Kf$ in terms of $\Hf_0$ and $\partial_r \Hf_0$, i.e., 
\begin{align}
\psif^{\ell m} & = - \frac{ 3 (   r_s -2 r ) r_s^2 -  r^2 \left[ r f(r) \lambda  + r_s  )\right] (\lambda - 2)    }{2  (\lambda-2) r^3 f(r)^3} \Hf_0^{\ell m}   - \frac{
   \left[ r (2 r - 3  r_s) (\lambda- 2) -3 r_s^2  \right] }{2  (\lambda-2) r f(r)^2} \partial_r \Hf_0^{\ell m} \; , \label{psitoH0}\\
\Kf^{\ell m} & = - \frac{  r_s^2 + r r_s (\lambda-4) - r^2 (\lambda-2)      }{(\lambda-2) r^2 f(r)   } \Hf_0^{\ell m}   +  \frac{   r_s     }{\lambda-2   }   \partial_r \Hf_0^{\ell m}\;.  \label{KtoH0}
 \end{align}
Therefore, replacing $\psif$ and $\Kf$ (and their first derivatives) in eq.~\eqref{algebraic_lin} by means of these expressions, we obtain a second-order equation for $\Hf_0$,
\begin{equation}
  \partial^2_r \, \Hf_0^{ \ell m }  (r) +\frac{2 r-r_s}{r^2 f(r)} \partial_r \Hf_0^{\ell m} (r)-  \frac{ \lambda r^2 f(r)+r_s^2}{r^4 f(r)^2} \; \Hf_0^{\ell m} (r)= 0 \;.
\label{eqH0_lin}
\end{equation}
This  equation describes the parity-even sector for linear perturbations; the other even metric quantities can be derived  from the solution of this equation.

Similarly, the two equations obtained by varying the action with respect to odd quantities represent one constraint and one degree of freedom. Indeed, variation of the quadratic action with respect to $h_0$ and $h_1$ gives, respectively,  
\begin{align}
  \partial^2_r \hf^{\ell m}_0 (r) - \frac{\lambda r  - 2 r_s}{r^3 f( r )} \; \hf^{\ell m }_0 (r)&= 0  \;, \label{eqh0_lin}
\\
        \lambda(\lambda-2) r  f(r) \; \hf^{ \ell m}_1(r) &= 0 \;.
\label{eqh1_lin}
\end{align}
The second equation implies $\hf_1=0$, while the first equation describes the parity-odd  linear perturbation. 
 Note that  $h_0$ and $h_1$ appear in the metric \eqref{RWgodd} with at least one angular derivative acting on them; therefore, there is no $\ell = 0$ odd-type mode and $\lambda \neq 0$.

In conclusion, in the static limit the  parity-even and parity-odd sectors of the linear perturbations of a Schwarzschild black hole decouple and are described, respectively, by eq.~\eqref{eqH0_lin} for $\Hf_0$ and by eq.~\eqref{eqh0_lin} for $\hf_0$. The two  even variables $\Hf_2$ (or $\psif$) and $\Kf$ are given in terms of $\Hf_0$ from the constraint equations, while $\Hf_1$ and $\hf_1$ vanish.

These considerations apply for $\ell \ge 2$. The cases $\ell=0$ and $\ell=1$ (i.e., $\lambda=0$ and $\lambda=2$, respectively) are different because some of the metric components in eqs.~\eqref{hPMeven4D} and \eqref{hPModd4D} are absent, leaving a residual gauge freedom, as discussed in \cite{Regge:1957td} for the static case and in \cite{Zerilli:1970se} for the general case. Specifically, for the monopole, the components ${\cal H}_0$, ${\cal H}_1$, and $G$ are missing and there are no odd perturbations, leaving us with only four even perturbations. Moreover, eqs.~\eqref{eqeveneven1_lin} and \eqref{eqeveneven3_lin} become the same, while eq.~\eqref{eqeveneven4_lin} vanishes identically, reducing the system to only two equations instead of four.
Using time and radial coordinate transformations, two metric components can be fixed, for instance $\Kf = \Hf_1 = 0$ \cite{Zerilli:1970se}. The remaining components can be removed via a redefinition of the black hole mass, in accordance with Birkhoff's theorem.

For the dipole, the perturbations $G$ and $h_2$ are absent. Among the remaining even-type perturbations, three can be fixed by constraints, while the other three can be entirely gauged away by shifting the origin of the coordinate system. Of the two remaining odd perturbations, one can be eliminated through a coordinate transformation, and the other corresponds to a rotation of the black hole. For a non-rotating black hole, this can be set to zero. Therefore, in the following, we set all linear monopole and dipole perturbations to zero.

\subsection{Quadratic perturbations}
\label{sec:quad}

The second-order equations can be obtained by varying the cubic action with respect to the metric variables. The linear part of these equations is identical to the left-hand side of eqs.~\eqref{eqeveneven1_lin}--\eqref{eqeveneven4_lin}, \eqref{eqh0_lin} and \eqref{eqh1_lin}, upon replacement of first-order quantities  $\Hf_0$, $\psif$, etc., by second-order quantities $\Hs_0$, $\psis$, etc. Additionally, at second order there is a quadratic part, a {\em source term}, which describes how quadratic couplings of linear perturbations source the linear part of the equations. These quadratic terms include both even and odd quantities, i.e., they mix the two parity sectors. Therefore, 
at quadratic order, the even and odd components are no longer independent. 

The procedure and steps to derive the equation describing the evolution of $\Hs_0$ are analogous to the ones used to derive eq.~\eqref{eqH0_lin} described above;  the only difference is  the presence of quadratic terms in the equation, involving products of linear quantities.  Once expanded in spherical harmonics, these equations display tensor harmonics up to rank two. In particular, these are the even and odd vector harmonics, respectively defined as
\begin{align}
    Z_{\ell \ A}^{m} &\equiv  \nabla_{A} Y_{\ell}^{m} \; , \\
    X_{\ell \  A }^{m} &\equiv   \epsilon_{A}^{\  \; B} \nabla_B Y_{\ell }^{m} \; , 
\end{align}
and the even and odd 2-tensor harmonics, respectively defined as
\begin{align}
    Z_{\ell \ A B}^{m} & \equiv   \left( \nabla_{A} \nabla_{B} Y_{\ell}^{m} - \frac{1}{2} \gamma_{AB} \Delta_{S^2} Y_{\ell}^{m}\right) \; , \\
    X_{\ell \  A B }^{m} & \equiv   \epsilon_{A }^{\  \; C}  \left( \nabla_{C} \nabla_{B} Y_{\ell}^{m} - \frac{1}{2} \gamma_{CB} \Delta_{S^2} Y_{\ell}^{m}\right) \; .
\end{align}
These quantities form a basis, respectively of vectors and tensors, on the sphere.

To obtain an equation for the spherical harmonic coefficients, 
we multiply the resulting equations by $Y_\ell^m{}$ and integrate over the angles. 
Generally, the quadratic terms contain the following integrals,\footnote{In general, there are also integrals quadratic in odd tensors, such as $\int {\rm d} \Omega \; Y_{\ell}^{m}{}^*   (\theta, \phi)X_{\ell_{1} \  A }^{m_{1}}  (\theta, \phi)X_{\ell_{2}}^{m_{2} \  A } (\theta, \phi)$,  but these can be reduced to integrals at most linear in odd tensors upon contractions of Levi-Civita symbols.}  
\begin{align}
 {}^{0}\!I_{\ell \ell_1  \ell_2}^{m  m_1 m_2} &= \int {\rm d} \Omega  \; Y_\ell^m{}^* (\theta, \phi) Y^{m_1}_{\ell_1} (\theta, \phi) Y^{m_2 }_{\ell_2}   (\theta, \phi) \;, \label{intE0} \\
 {}^{1}\!I_{\ell \ell_1  \ell_2}^{m  m_1 m_2} &= \int {\rm d} \Omega \; Y_{\ell}^{m}{}^*   (\theta, \phi)Z_{\ell_{1} \  A }^{m_{1}}  (\theta, \phi)Z_{\ell_{2}}^{m_{2} \  A } (\theta, \phi) \;, \\ 
 {}^{2}\!I_{\ell \ell_1  \ell_2}^{m  m_1 m_2} &= \int {\rm d} \Omega \; Y_{\ell}^{m}{}^*   (\theta, \phi)Z_{\ell_{1} \  A B}^{m_{1}}  (\theta, \phi)Z_{\ell_{2}}^{m_{2} \  A B} (\theta, \phi) \;,   \\
 {}^{1}\!J_{\ell \ell_1  \ell_2}^{m  m_1 m_2} &= \int {\rm d} \Omega \; Y_{\ell}^{m}{}^* (\theta, \phi)  X_{\ell_{1} \  A }^{m_{1}}(\theta, \phi)  Z_{\ell_{2}}^{m_{2} \ A } (\theta, \phi) \;, \\
 {}^{2}\!J_{\ell \ell_1  \ell_2}^{m  m_1 m_2} &= \int {\rm d} \Omega \; Y_{\ell}^{m}{}^* (\theta, \phi)  X_{\ell_{1} \  A B }^{m_{1}}(\theta, \phi)  Z_{\ell_{2}}^{m_{2} \ A B} (\theta, \phi) \;, \label{intJ2}
\end{align}
where ${\rm d} \Omega = \sin\theta \, {\rm d} \theta {\rm d} \phi$.
We will explain how to solve these integrals in appendix~\ref{App: harmonics}. There we show that
\begin{align}
{}^{s}\!I_{\ell \ell_1  \ell_2}^{m  m_1 m_2} &=  \frac12 \left[ 1+ (-1)^{\ell + \ell_1 + \ell_2} \right] \EE{s}_{\ell \ell_1  \ell_2}^{m  m_1 m_2} = \frac12 (\EE{s}_{\ell \ell_1  \ell_2}^{m  m_1 m_2} + \EE{s}_{\ell \ell_2  \ell_1}^{m  m_2 m_1}) \;,  \label{Iint}\\ 
{}^{s}\!J_{\ell \ell_1  \ell_2}^{m  m_1 m_2}  &= - \frac{i}2 \left[ 1- (-1)^{\ell + \ell_1 + \ell_2} \right] \EE{s}_{\ell \ell_1  \ell_2}^{m  m_1 m_2}  = - \frac{i}{2}  (   \EE{s}_{\ell \ell_1  \ell_2}^{m  m_1 m_2} -  \EE{s}_{\ell  \ell_2   \ell_1}^{m   m_2  m_1}) \label{Jint} \;,
\end{align}
where we  defined
\be
\EE{s}_{\ell \ell_1  \ell_2}^{m  m_1 m_2} \equiv (2 - \delta_{0s}) 2^{-s} (-1)^s \sqrt{\frac{(1+2 \ell_1) (1+2 \ell_2)}{4 \pi (1+2 \ell)} \frac{(\ell_1+|s|)! (\ell_2+|s|)!}{(\ell_1-|s|)! (\ell_2-|s|)!}} \; C^{m_{1}  m_{2}   m }_{\ell_{1}  \ell_{2}  \ell} \; C^{-s   s   0}_{\ell_{1} \ell_{2}  \ell} \;,
\ee
and  where \(C^{m_{1}  m_{2}  m}_{\ell_{1}  \ell_{2} \ell}\) are the Clebsch--Gordan coefficients, often denoted  as \(\langle  \ell_1 \; m_1 \; \ell_2 \; m_2 | \ell \; m \rangle\).  
We note that transformation of spherical harmonics under parity implies that ${}^{s}\!I_{\ell \ell_1  \ell_2}^{m  m_1 m_2}$ vanishes for $\ell + \ell_1 + \ell_2$ odd, while ${}^{s}\!J_{\ell \ell_1  \ell_2}^{m  m_1 m_2}$ vanishes for $\ell + \ell_1 + \ell_2$ even. In addition, from their definitions it is manifest that ${}^{s}\!I_{\ell \ell_1  \ell_2}^{m  m_1 m_2}$ is symmetric while ${}^{s}\!J_{\ell \ell_1  \ell_2}^{m  m_1 m_2}$ is antisymmetric under the exchange of $\ell_1 m_1$ and $\ell_2 m_2$.

In the end, from the same manipulations that we used  in the linear case, with the inclusion of the quadratic terms, we find  the differential equation
\begin{equation}
\partial^2_r \, \Hs_0^{ \ell m }  (r) +\frac{2 r-r_s}{r^2 f(r)} \partial_r \Hs_0^{\ell m} (r)-  \frac{ \lambda r^2 f(r)+r_s^2}{r^4 f(r)^2} \; \Hs_0^{\ell m} (r) =  \sum_{\ell_1, m_1, \ell_2, m_2} {\cal S}_{H_0 \ \ell \ell_1  \ell_2}^{m  m_1 m_2} (r) \;.
\label{eqH0_sec}
\end{equation}
As expected,  the left-hand side is the same as in the linear equation~\eqref{eqH0_lin},  with $\Hf_0$ replaced by $\Hs_0$. The source term on the right-hand side is given by a sum over angular and magnetic numbers $\ell_1$, $ m_1$, $\ell_2$ and $m_2$, and  the selection rules due to the Clebsch--Gordan coefficients imply that  $|\ell_{1}-\ell_{2}| \leq \ell \leq \ell_{1} + \ell_{2} $, $|m_{1}|\leq \ell_{1}$,  $|m_{2}|\leq \ell_{2}$ and $m =  m_{1} + m_{2}$. 
The source term is quadratic in both the even and odd linear harmonic variables $\Hf_0^{\ell m}$, $\psif^{\ell m}$, $\Kf^{\ell m}$ and $\hf_0^{\ell m}$, and their first and second radial derivatives, multiplied by one of the  integrals in eqs.~\eqref{intE0}--\eqref{intJ2}. The explicit expression of the source  is cumbersome, but it can be  simplified  by using eqs.~\eqref{psitoH0} and \eqref{KtoH0} to express $\psif$ and $\Kf$ in terms of $\Hf_0$ and $\partial_r \Hf_0$. 
Additionally, one can use the linear equations \eqref{eqH0_lin} and \eqref{eqh0_lin} to eliminate the second-order radial derivatives and write the source term solely in terms of $\Hf_0$, $\hf_0$, $\partial_r \Hf_0$, and $\partial_r \hf_0$. By doing so, one obtains a lengthy but manageable expression,  
\begin{equation}
\begin{split}
{\cal S}_{H_0 \ \ell \ell_1 \ell_2}^{ m m_1 m_2}  (r) = \ & \frac{r_s^2}{\lambda (\lambda - 2)  f(r)^2 r^4} \bigg\{ {}^{0}\!I_{\ell \ell_1  \ell_2}^{m  m_1 m_2}  \bigg[   \bigg( \Apz \frac{r^2}{r_s^2} + \Apu  \frac{r}{r_s} + \Apd   \bigg) \; \Hf_0^{\ell_1 m_1} \;  \Hf_0^{\ell_2 m_2}   \\
& +   f(r) \frac{r^2}{r_s} \;   \Bpd \partial_r \Hf_0^{\ell_1 m_1} \;  \Hf_0^{\ell_2 m_2}        -  \frac12 \lambda (\lambda - 2)  f(r)^2 \frac{r^4}{r_s^2}  \;  \partial_r \Hf_0^{\ell_1 m_1} \; \partial_r \Hf_0^{\ell_2 m_2}  \\
& +   \frac{r_s^2 }{ f(r)^2 r^4} \bigg( \Amz \frac{r^4}{r_s^4} + \Amu  \frac{r^3}{r_s^3}   + \Amd \frac{r^2}{r_s^2}
  +\Amt \frac{r }{r_s}  + \Amq  \bigg)   \; \hf_0^{\ell_1 m_1} \;  \hf_0^{\ell_2 m_2}  \\
& +    \frac{r_s}{f(r) r^2} 
\bigg(  \Bmz \frac{r^2}{r_s^2} +  \Bmu \frac{r}{r_s} + \Bmd   \bigg)   \partial_r \hf_0^{\ell_1 m_1} \;  \hf_0^{\ell_2 m_2}  \\
& +   \left(\Ez \frac{r^2}{r_s^2} + \Eu  \frac{r}{r_s} + \Ed \right)   \;   \partial_r \hf_0^{\ell_1 m_1}  \partial_r \hf_0^{\ell_2 m_2} + (\ell_1 m_1 \leftrightarrow \ell_2 m_2 )   \bigg] \\
& +  {}^{2}\!I_{\ell \ell_1  \ell_2}^{m  m_1 m_2}   \bigg[   \frac{r_s^2}{f(r)^2 r^4} \bigg( \Cmz   \frac{r^3}{r_s^3} + \Cmu \frac{r^2}{r_s^2} + \Cmd \frac{r }{r_s} -\frac{19}{2} \bigg)  \; \hf_0^{\ell_1 m_1} \;  \hf_0^{\ell_2 m_2}  \\
& +   \frac{r_s}{f(r) r^2 } \left( \Dmz \frac{r^2}{r_s^2}  + \Dmu  \frac{r }{r_s} -11   \right) \;  \partial_r \hf_0^{\ell_1 m_1} \;  \hf_0^{\ell_2 m_2}  \\
& +   \left(4\frac{r}{r_s}  - 2  \right)   \partial_r \hf_0^{\ell_1 m_1}  \partial_r \hf_0^{\ell_2 m_2} + (\ell_1 m_1 \leftrightarrow \ell_2 m_2 )   \bigg] \bigg\} \;, \label{sourceH0}
\end{split}
\end{equation}
where the coefficients $\Apz$, $\Apu$, etc.,   are functions of $\lambda$, $\lambda_1$ and $\lambda_2$, i.e., $\Apz= \Apz(\lambda, \lambda_1, \lambda_2)$, etc.,\footnote{The denominator in the prefactor of the curly brackets vanishes for $\lambda=0$ and $\lambda=2$, corresponding to $\ell=0$ and $\ell=1$, respectively. However, as the curly brackets also vanish in these cases, the source remains finite, as we will show in sec.~\ref{sec:solutionsQuadrpoleTimesQuadrupole}. Similar apparent divergences occur in the other sources discussed in this section.} 
with $\lambda_1$ and $\lambda_2$ defined as $\lambda_i \equiv \ell_i(\ell_i+1)$ in analogy with \eqref{lambdadeff}, with an extra subscript to distinguish  the two different linear modes that combine to generate the second-order one.
Explicit expressions for all the coefficients can be found in appendix~\ref{app:source}. 

The source for $\Hs_0$ contains both even and odd linear modes; however, only
quadratic terms of the type {\em even} $\times$ {\em even} or {\em odd} $\times$ {\em odd} are present. 
As already noted, the symbols ${}^{s}\!I_{\ell \ell_1  \ell_2}^{m  m_1 m_2} $ are symmetric under the exchange $\ell_1 m_1  \leftrightarrow \ell_2 m_2 $ and vanish for $\ell+\ell_1 + \ell_2=$ odd. Therefore the source for $\Hs_0$ receives contribution only when $\ell+\ell_1 + \ell_2$ is even and has to be symmetric under the exchange $\ell_1 m_1  \leftrightarrow \ell_2 m_2 $.
The other even variables of the second-order metric, $\psis$ and $\Ks$, can be obtained from equations analogous to eqs.~\eqref{psitoH0} and \eqref{KtoH0} extended at second order, which we do not write explicitly here.   We have verified that, like $\Hs_0$, they are sourced by {\em even} $\times$ {\em even} or {\em odd} $\times$ {\em odd} quadratic terms for $\ell+\ell_1 + \ell_2=$ even.

The remaining even variable,  $H_1$, no longer vanishes at second order but it is determined by quadratic terms of the linear perturbations. Indeed, varying the cubic action with respect to $H_1$, multiplying  the resulting equation by $Y_\ell^m{}^*$   and  integrating over the angles, yields
\begin{equation}
 \Hs_1^{\ell m} = \sum_{\ell_1, m_1, \ell_2, m_2} {\cal S}_{H_1 \ \ell \ell_1 \ell_2}^{ m m_1 m_2}  (r)  \;, 
\label{eqH1withsource}
\end{equation}
where the sources on the right-hand side are given by   
\begin{equation}
\begin{split}
  {\cal S}_{H_1 \ \ell \ell_1 \ell_2}^{ m m_1 m_2}  (r)   =  \frac{ 1 }{2 \lambda  f(r)  r^2} & \;  {}^{1}\!J_{\ell \ell_1  \ell_2}^{m  m_1 m_2}   \Big[ r_s  \Hf^{\ell_{1} m_{1}}_0 \hf^{\ell_{2} m_{2}}_0
\\  
&   + r^2 f(r) \left( \partial_r \Hf^{\ell_{1} m_{1}}_0 \hf^{ \ell_{2} m_{2}}_0 -   \Hf^{\ell_{1} m_{1} }_0 \partial_r \hf^{\ell_{2} m_{2}}_0 \right) 
  - (\ell_1 m_1 \leftrightarrow \ell_2 m_2 ) \Big]\, .
\end{split}
\label{sourceH1}
\end{equation}
The source above contains only
quadratic terms of the type {\em even} $\times$ {\em odd}.
The symbols $ {}^{s}\!J_{\ell \ell_1  \ell_2}^{m  m_1 m_2}$ are antisymmetric under the exchange $\ell_1 m_1  \leftrightarrow \ell_2 m_2 $ and vanish whenever $\ell+\ell_1 + \ell_2$ even or $m_1 = m_2 =0$ or $\ell_1 m_1 = \ell_2 m_2$ (see app.~\ref{App: harmonics}) . Hence, $\Hs_1$  does not receive any contribution in such cases. The only contribution can come from modes for which $\ell+\ell_1 + \ell_2$ is odd.

We can now turn to  the  odd quantities  $h_0$ and $h_1$ at second order. The equations governing their behavior can be found by proceeding similarly to what we described above for the even sector.
For $\hs_0$ we find   
\begin{align}
  \partial^2_r \hs^{\ell m}_0 (r) - \frac{\lambda r  - 2 r_s}{r^3 f( r )} \; \hs^{\ell m }_0 (r)  = \sum_{\ell_1, m_1, \ell_2, m_2} {\cal S}_{h_0 \ \ell \ell_1 \ell_2}^{ m m_1 m_2}  (r)  \;, \label{eqh0oddodd}
\end{align}
where the left-hand side has the same form as in the linear equation~\eqref{eqh0_lin}, and the source term is given by 
\begin{equation}
\begin{split}
   {\cal S}_{h_0 \ \ell \ell_1 \ell_2}^{ m m_1 m_2}  (r)  = \frac{1}{2 \lambda  f(r)  r ^2}  
    \bigg\{     {}^{0}\!I_{\ell \ell_1  \ell_2}^{m  m_1 m_2}  &  \bigg[  \frac{r_s^2 }{  f(r) r^2}    \bigg(  \Cz   \frac{r^2}{ r_s^2} 
+ \Cu   \frac{r}{r_s}  + \Cd    \bigg) \;   \Hf^{\ell_1 m_1}_0 \hf^{ \ell_2 m_2}_0  
\\
&  + r_s   \bigg( \Dz \frac{r}{ r_s}    +  \Du    \bigg)  \;    \partial_r \Hf^{\ell_1 m_1}_0  \hf^{ \ell_2 m_2}_0 
 + (\ell_1 m_1 \leftrightarrow \ell_2 m_2)   \bigg] 
 \\
&   -  2 \;  {}^{2}\!I_{\ell \ell_1  \ell_2}^{m  m_1 m_2}  \;  \bigg[ \Hf^{\ell_1 m_1}_0 \hf^{ \ell_2 m_2}_0  + (\ell_1 m_1 \leftrightarrow \ell_2 m_2) \bigg]
      \bigg\} \;, 
\end{split}
\label{sourceh0}
\end{equation}
where the  explicit expressions of the coefficients $\Cz$, $\Cu$, $\Cd$, $\Dz$ and $\Du$ are given in appendix~\ref{app:source}.  The odd quantity $\hs^{\ell m}_0$ is sourced by quadratic terms of the type {\em even} $\times$ {\em odd}, proportional to the ${}^{s}\!I_{\ell \ell_1  \ell_2}^{m  m_1 m_2} $ symbols. Thus, by a similar logic as above, we conclude that ${\cal S}_{h_0 \ \ell \ell_1 \ell_2}^{ m m_1 m_2}$ must vanish for $\ell+\ell_1 + \ell_2=$ odd and must be symmetric under the exchange of $\ell_1 m_1$ and $\ell_2 m_2$.

The remaining second-order odd metric quantity, $\hs_1$, is determined by  {\em even} $\times$ {\em even} and {\em odd} $\times$ {\em odd} quadratic terms, which are nonzero only for $\ell + \ell_1 + \ell_2 = $ odd and antisymmetric under $\ell_1 m_1 \leftrightarrow \ell_2 m_2$. Indeed, we find
\be
        \hs^{ \ell m}_1 = \sum_{\ell_1 ,m_1 ,\ell_2, m_2} {\cal S}_{h_1 \ \ell \ell_1 \ell_2}^{ m m_1 m_2}  (r) \;,
\label{eqh1oddodd}
\ee
with   
\begin{equation}
\begin{split}
    {\cal S}_{h_1 \ \ell \ell_1 \ell_2}^{ m m_1 m_2}  (r)  =    \frac{1}{2  \lambda(\lambda-2) f(r)   r } \;  {}^{1}\!I_{\ell \ell_1  \ell_2}^{m  m_1 m_2}  &  \Big[ 
    r^3 f(r) \Hf^{\ell_{1} m_{1}}_0 \partial_r \Hf^{ \ell_{2} m_{2} }_0
   \\
&     +     
 \lambda_2 \hf^{\ell_{1} m_{1}}_0 (2 - r \partial_r ) \hf^{\ell_{2} m_{2}}_0 
  -  (\ell_1 m_1 \leftrightarrow \ell_2 m_2)  \Big] \;.
\end{split}
\label{sourceh1}
\end{equation}

\subsection{Selection rules}

\begin{table}
\setlength{\tabcolsep}{10pt}
\renewcommand{\arraystretch}{1.5}
\centering
\begin{tabular}{|c || c|  c |}
\hline
$\ell + \ell_1 + \ell_2 $ even & $\Hf_0^{\ell_2 m_2} $ & $\hf_0^{\ell_2 m_2} $  \\  [0.5ex]  
\hline \hline
$\Hf_0^{\ell_1 m_1} $ & $\Hs_0^{\ell m} $ & $\hs_0^{\ell m} $    \\
\hline
$\hf_0^{\ell_1 m_1} $ & $\hs_0^{\ell m} $ & $\Hs_0^{\ell m} $  \\
\hline
\end{tabular}
\quad 
\begin{tabular}{|c || c|  c |}
\hline
$\ell + \ell_1 + \ell_2 \text{ odd}$   & $\Hf_0^{\ell_2 m_2} $ & $\hf_0^{\ell_2 m_2} $  \\  [0.5ex]  
\hline \hline
$\Hf_0^{\ell_1 m_1} $ & $\hs_1^{\ell m} $ & $\Hs_1^{\ell m} $    \\
\hline
$\hf_0^{\ell_1 m_1} $ & $\Hs_1^{\ell m} $ & $\hs_1^{\ell m} $  \\
\hline
\end{tabular}
\caption{Selection rules for quadratic sources, for the case $\ell + \ell_1 + \ell_2  = $ even, left table, and  $\ell + \ell_1 + \ell_2  = $ odd, right table. Here by $H_0$ we generically denote the trio $H_0$, $H_2 $ and $K$, which are related by constraints. }
\label{table:SR}
\end{table}
Let us conclude this subsection with a brief summary of the selection rules encountered above. They arise from the symmetries of the system, specifically, the {\em staticity} (stationarity and time-reversal invariance) and {\em spherical symmetry} of the background, along with the invariance of the action under parity and time-reversal. As previously discussed, since $H_2$ and $K$ can be expressed in terms of $H_0$ and its derivatives using constraint equations, only four metric variables remain in the RW action, leading to 20 possible cubic combinations of metric variables (without counting derivatives) in the action.  However, because $\Hf_1 = 0 = \hf_1$, cubic interactions involving at least two of these fields do not contribute to the quadratic equations of motion, excluding 10 cubic combinations. Additionally, due to the static background and the time-independence of the perturbations, it is impossible to construct cubic interactions with an odd number of $\delta g_{i t}$ perturbations, such as $H_1 H_0^2$ or $h_0^3$. This leaves us with only five combinations, which, upon variation of the action, are the ones that yield the terms  in the sources \eqref{sourceH0}, \eqref{sourceH1}, \eqref{sourceh0} and \eqref{sourceh1}. 

The spherical symmetry of the background allows for the decomposition of the fields into spherical harmonics, with angular momentum conservation in the cubic operators enforced by the Clebsch--Gordan coefficients. In general, the selection rules imposed by the Clebsch--Gordan coefficients dictate that $| \ell_1 - \ell_2 | \le \ell \le \ell_1 + \ell_2$ and $m = m_1 + m_2$. If we define $p$ as the parity of a quadratic mode---where $p=+1$ denotes an even mode and $p=-1$ denotes an odd mode---then we have seen that  $p = (-1)^{\ell + \ell_1 + \ell_2} \times p_1 \times p_2$ \cite{Brizuela:2006ne,Brizuela:2007zza,Brizuela:2009qd,Lagos:2022otp,Bucciotti:2024jrv}, where $p_1$ and $p_2$ are the parities of the first and second linear modes generating the quadratic mode, respectively. Indeed, we can consider two cases:
\begin{itemize}
\item $\ell + \ell_1 + \ell_2 = \text{even}$: By parity and the properties of the integrals involving three tensor harmonics, this case arises from cubic couplings in the action with an even number of odd tensor harmonics, i.e.~$H_0^3$ and $H_0 h_0^2$. In this case, quadratic terms with $p_1 = p_2 = \pm 1$ can source $p = +1$, corresponding to $\Hs_0$ (or $\Hs_2$ and $\Ks$, using the constraints). Conversely, quadratic terms with $p_1 = - p_2 = \pm 1$ can source $p = -1$, corresponding to $\hs_0$. See table~\ref{table:SR}, left-hand side.
\item $\ell + \ell_1 + \ell_2 = \text{odd}$: This case arises from the  cubic couplings involving an odd number of odd tensor harmonics, i.e.~$H_0^2 h_1$, $h_0^2 h_1$, and $H_0 H_1 h_0$.
In this case, quadratic terms with $p_1 = p_2 = \pm 1$ can source $p = -1$, corresponding to $\hs_1$ (we recall that $\Hf_1=0$, so that $\hs_0$ is not sourced at second order in this case), while quadratic terms with $p_1 = - p_2 = \pm 1$ can source $p = +1$, corresponding to $\Hs_1$. See table~\ref{table:SR}, right-hand side. Additionally, as discussed above, the selection rules of the Clebsch--Gordan coefficients in this case imply that when $m_1 = m_2 = 0$ and $\ell_1 m_1 = \ell_2 m_2$, the quadratic sources vanish.
\end{itemize}

\section{Solutions}
\label{sec:sol}
In this section we solve the equations derived previously, upon imposing suitable boundary conditions at the horizon. We start by reviewing the derivation of  the linear solutions.

\subsection{Linear solutions}
\label{subsec:linearsolutions}

We solve here eqs.~\eqref{eqH0_lin} and \eqref{eqh0_lin}, respectively for $\Hf_0^{\ell m} (r)$ and $\hf_0^{\ell m} (r)$, and find the regular solutions  at the horizon $r=r_s$. 
The other even variables, $\psif$ and $\Kf$, can be extracted using eqs.~\eqref{psitoH0} and \eqref{KtoH0}, once $\Hf_0$ is known, while we remind that $\Hf_1=0 = \hf_1$.  Let us start with $\Hf_0$. Whenever the $m$-dependence is trivial, we will suppress it from our notation in this section.

 Equation~\eqref{eqH0_lin} is a second-order  ordinary differential equation with three regular singular points, i.e.~$r = 0, \, r_s, \, \infty $.  
It is a special case of the hypergeometric equation, and it can be recast in the the form of an associated Legendre equation \cite{1967ApJ...149..591T,Hinderer:2007mb}:
\begin{equation}
(1-z^2)  \partial^2_z \, \Hf_0^{ \ell m }  -2 z\partial_z \Hf_0^{\ell m} + \left[\ell(\ell+1 ) -\frac{4}{1-z^2}  \right]   \Hf_0^{\ell m} = 0\, ,
\qquad
z\equiv \frac{2r}{r_s}-1 \, .
\label{eqH0_linLeg}
\end{equation}
The two independent solutions are thus\footnote{In terms of hypergeometric functions, the two independent solutions can be  written as \cite{Riva:2023rcm}
\begin{align}
H_{0,1}^\ell (r) &= \frac{\Gamma (\ell+2) \Gamma (\ell+3)}{\Gamma (\ell)} (-1)^{\ell+1} \frac{r}{r-r_s} \hypergeom{2}{1} \left(-\ell,  \ell+1,  3; \frac{r}{r_s} \right) ,
\label{H04ell}
\\
H_{0,3}^\ell (r) &=  \left(\frac{r_s}{r}\right)^{\ell+1} \frac{r}{r-r_s} \hypergeom{2}{1} \left(\ell-1,  \ell+1, 2\ell+2; \frac{r_s}{r} \right) .
\label{H01ell}
\end{align}
}
\begin{align}
H_{0,1}^\ell (r) &= P_\ell^2 \left(2r/r_s-1\right) \;,
\label{H01Pell}
 \\
H_{0,2}^\ell (r) &= Q_\ell^2 \left(2r/r_s-1\right)\;,
\label{H02Qell}
\end{align}
where $P_\ell^2$ and $Q_\ell^2$ are the associated Legendre functions of the first and second kind respectively, of degree $\ell$ and order $2$, with near-horizon behaviors  
\begin{align}
H_{0,1}^\ell (r) &= - (\ell-1) \ell (\ell+1) (\ell+2) \frac{ r-r_s}{2 r_s} + \mathcal{O}((r-r_s)^2)   \;,
 \\
H_{0,2}^\ell (r) &= -\frac{r_s}{2 (r-r_s)} + \mathcal{O}((r-r_s)^0)  \;.
\end{align}
The second solution is divergent at the horizon, so we discard it, while the first  one is regular. In particular, since $\ell$ takes integer values, $H_{0,1}^\ell$ is a finite polynomial in $r$, while $H_{0,2}^\ell$ contains instead  $\log\left(\frac{r-r_s}{r} \right)$ terms.
Imposing regularity at the horizon leads us to discard the solution containing logarithms.\footnote{One can express the Zerilli variable  \cite{Zerilli:1970se} in terms of $H_0$ and $\partial_r H_0$, and check that the logarithmic part of the homogeneous solution gives a divergent Zerilli variable \cite{Hui:2020xxx}.} 
Including a constant  amplitude factor, we will write the physical linear solution for the even variable $\Hf_0^{\ell m}$ as
\begin{equation}
    \Hf_0^{\ell m} (r) = - \mathcal{E}^{\rm +}_{\ell m} \frac{\Gamma(\ell+1)^2}{\ell(\ell-1)\Gamma(2\ell+1)}P_{\ell}^{2} \big(2r/r_s-1\big) = \mathcal{E}^{\rm +}_{\ell m} \frac{\Gamma(\ell+1)^2}{\ell(\ell-1)\Gamma(2\ell+1)} r(r-r_s) \partial_r^2 P_{\ell}\big(2r/r_s-1\big) \;,
\label{H0linsol}
\end{equation}
where the normalization is chosen to ensure that $H^{\ell m}_{0} (r) = \mathcal{E}^{\rm +}_{\ell m} (r/r_s)^{\ell} + \mathcal{O}(r^{\ell-1})$ for large $r$, with $\mathcal{E}^{\rm +}_{\ell m}$ being the amplitude of the external even tidal field.

Let us now turn to $\hf_0$. Through the field redefinition 
\be
\hf_0^{\ell m}(r(x)) = x^{-1} (1-x) u(x) \;, \qquad x \equiv \frac{r_s}{r} \;,
\ee
eq.~\eqref{eqh0_lin} can be recast in the standard hypergeometric form

\be
x(1-x)u''(x) +\left[ c-(a+b+1) x \right] u'(x) - a b \; u(x) =0 \;,
\label{HGE}
\ee
with  parameters 
\be
a = -\ell - 1, \qquad b = \ell, \qquad c = -2 \; ,
\ee
which satisfy $a+ b - c =1$. For this combination of parameters, the two independent solutions are\footnote{See, e.g., ref.~\cite{Bateman:100233}. This corresponds to case 27 in the table in section 2.2.2 of that reference.}   
\begin{align}
    h_{0,4}^\ell (r) &= -\frac{r_s(r-r_s)}{r} \left( -\frac{r_s}{r} \right)^{\ell} \, _2F_1 \left(\ell,\ell+3;2 \ell+2;\frac{r_s}{r} \right) \; , 
\label{h04ell}
\\
    h_{0,5}^\ell (r) &= -\frac{r^2 (r-r_s)}{r_s^2} \, _2F_1 \left( 2-\ell,\ell+3;4;\frac{r}{r_s} \right) \;.
\label{h05ell}
\end{align}
Since the first argument of $_2F_1 ( 2-\ell,\ell+3;4;\frac{r}{r_s} )$ in $h_{0,5}$ is a non-positive integer, while its third argument is a  positive integer, it follows from the properties of the hypergeometric functions that   $h_{0,5}$ is a finite polynomial with only positive powers of $\frac{r}{r_s}$.\footnote{\label{ft:h0pol} To see this, it might be convenient to rewrite $h_0$ in \eqref{h05ell} (or, equivalently, \eqref{h0linsol} up to an overall constant prefactor)  as
\begin{equation}
    h_{0,5}^\ell (r) = -\frac{r^2 (r-r_s)}{r_s^2} \sum _{j=0}^{\ell-2} (-1)^j \frac{ 6(\ell-2)! }{ j! (\ell-j-2)!} \frac{ \Gamma (\ell+j+3) }{\Gamma (\ell+3) \Gamma (j+4) }  \left(\frac{r}{r_s}\right)^j \;.
\label{h05ell-2}
\end{equation}
}
  On the other hand, $_2F_1 \left(\ell,\ell+3;2 \ell+2;\frac{r_s}{r} \right)$ in $h_{0,4}$ contains an inverse power of $r-r_s$ and a $\log( {r} - {r_s})$, going as $_2F_1\left(\ell,\ell+3;2 \ell+2;\frac{r_s}{r}\right) =    \frac{r_s}{r-r_s} \frac{\Gamma{(2\ell+2)} }{\Gamma(\ell)\Gamma(\ell+3)}  + \frac{\Gamma{(2\ell+2)} }{\Gamma(\ell-1)\Gamma(\ell+2)}  \log(\frac{r}{r_s}-1) +{\cal O}((r-r_s)^0)$ for  $r\rightarrow r_s$. Note that, despite the divergent behavior of $_2F_1\left(\ell,\ell+3;2 \ell+2;\frac{r_s}{r}\right)$ near the horizon, $h_{0,4}$ remains finite because of the $r-r_s$ prefactor in \eqref{h04ell}.
Following~\cite{Hui:2020xxx}, we will demand that the Regge--Wheeler variable $\Psi_{\rm RW}$, which is related to the odd field $h_0$ via $ \Psi_{\rm RW} \propto (2 h_0 - r \partial_r h_0)$ \cite{Zerilli:1970se,Hui:2020xxx}, does not diverge at the horizon. This excludes $h_{0,4}$, selecting  $h_{0,5}$ as the correct physical odd solution, which we will rewrite for convenience as  
\begin{equation}
    \hf^{ \ell m}_{0} (r) = \mathcal{E}^{\rm -}_{\ell m} \frac{(4)_{\ell-2} \Gamma (\ell-1)}{(2-\ell)_{\ell-2} (\ell+3)_{\ell-2}}\frac{r^2 (r-r_s)}{r_s^2} \, _2F_1\left(2-\ell,\ell+3;4;\frac{r}{r_s}\right) \;,
\label{h0linsol}
\end{equation}
where $\mathcal{E}^{\rm -}_{\ell m}$ is a  dimensionless constant, corresponding to the amplitude of the linear odd field,  and $(a)_n \equiv a(a+1) \cdots (a+n-1)$ is the  Pochhammer symbol.   
The normalization factor is chosen in such a way that  $\hf^{\ell}_{0} (r) = r_s \mathcal{E}^{\rm -}_{\ell m} (r/r_s)^{\ell+1} + \mathcal{O}(r^{\ell})$ for large $r$. 

Equations~\eqref{H0linsol} and \eqref{h0linsol} are the final expressions for the linear solutions that we will use in the next section to evaluate the quadratic sources in the second-order equations, to compute the  metric perturbations at second order.

\subsection{Quadratic solutions}
\label{sec:QuadraticSolutions}

We have now all the ingredients to solve eqs.~\eqref{eqH0_sec}  and \eqref{eqh0oddodd} for the metric at second order, induced by the coupling of two linear modes. The right-hand sides of \eqref{eqH0_sec}  and \eqref{eqh0oddodd} are computed on the linear solutions \eqref{H0linsol} and \eqref{h0linsol}, and are therefore known functions of $r$. From the standard properties of linear ordinary differential equations,  the most general solutions to the inhomogeneous equations \eqref{eqH0_sec}  and \eqref{eqh0oddodd} are obtained from a linear superposition of homogeneous and particular solutions. The particular solutions can be found  using standard Green's function methods:  
\begin{align}
\Hs_0^{ \ell m }  (r) & =  \sum_{\ell_1 ,m_1, \ell_2, m_2} \int_{r_s}^\infty  {\rm d}r' \, {\cal G}_{H_0}^{\ell }(r,r')     {\cal S}_{H_0 \ \ell \ell_1  \ell_2}^{m  m_1 m_2} (r') \;,
\label{H0_Greens}
\\
  \hs^{\ell m}_0 (r)  & = \sum_{\ell_1, m_1, \ell_2, m_2}  
  \int_{r_s}^\infty  {\rm d}r' \, {\cal G}_{h_0}^{\ell }(r,r') 
   {\cal S}_{h_0 \ \ell \ell_1 \ell_2}^{ m m_1 m_2}  (r')  \;,
\label{h0_Greens}
\end{align}
where ${\cal G}_{H_0}^{\ell }(r,r')$ and ${\cal G}_{h_0}^{\ell }(r,r')$ are the Green's functions associated with \eqref{eqH0_sec}  and \eqref{eqh0oddodd}, satisfying  
\begin{align}
\partial^2_r {\cal G}_{H_0}^{\ell }(r,r') +\frac{2 r-r_s}{r^2 f(r)} \partial_r {\cal G}_{H_0}^{\ell }(r,r') -  \frac{ \lambda r^2 f(r)+r_s^2}{r^4 f(r)^2} {\cal G}_{H_0}^{\ell }(r,r') & =  \delta (r-r')\;,
\\
\partial^2_r {\cal G}_{h_0}^{\ell }(r,r')  -\frac{\lambda r  - 2 r_s}{r^3 f( r )} {\cal G}_{h_0}^{\ell }(r,r') & = \delta(r-r')  \;,
\end{align}
respectively.
Following the standard procedure, it is easy to check that the most general solutions for ${\cal G}_{H_0}^{\ell }(r,r')$ and ${\cal G}_{h_0}^{\ell }(r,r')$ that respect the correct boundary condition at the horizon, and are continuous across $r=r'$, are given by \cite{Riva:2023rcm}:\footnote{Recall that, in sec.~\ref{subsec:linearsolutions}, we argued that the static solutions associated with regular behavior at the horizon are $H_{0,1}^\ell (r)$ in \eqref{H01Pell} and $h_{0,5}^\ell (r)$ in \eqref{h05ell}.}  
\begin{align}
{\cal G}_{H_0}^{\ell }(r,r') & = \frac{r'(r'-r_s)}{r_s W^\ell_{H_0} } \left[ H_{0,2}^\ell (r) H_{0,1}^\ell (r')\Theta(r-r') + H_{0,1}^\ell (r) H_{0,2}^\ell (r')\Theta(r'-r)
\right]\, ,
\label{GH0reg}
\\
{\cal G}_{h_0}^{\ell}(r,r') & = \frac{1}{r_s W^\ell_{h_0} } \left[ h_{0,4}^\ell (r) h_{0,5}^\ell (r')\Theta(r-r') + h_{0,5}^\ell (r) h_{0,4}^\ell (r')\Theta(r'-r)
\right]\, ,
\label{Gh0reg}
\end{align}
where $\Theta$ is the Heaviside step function, while $H_{0,1}^\ell$, $H_{0,2}^\ell$, $h_{0,4}^\ell$ and $h_{0,5}^\ell$ are the homogeneous solutions \eqref{H01Pell}--\eqref{H02Qell} and \eqref{h04ell}--\eqref{h05ell}. The quantities $W^\ell_{H_0}$ and $W^\ell_{h_0}$ in the expressions above are $r$-independent (but, $\ell$-dependent) dimensionless constants associated  to the Wronskians of the differential equations, and can be expressed as
\begin{equation}
W^\ell_{H_0} = -\frac{1}{2} (\ell-1) \ell (\ell+1) (\ell+2)  \, ,
\qquad
W^\ell_{h_0} = -\frac{3  \Gamma (2 \ell+3)}{(\ell+1) \Gamma (\ell+2) \Gamma (\ell+3)}
\, .
\end{equation}

The homogeneous solutions at second order to add to the particular ones are already known: they are given by eqs.~\eqref{H01Pell}--\eqref{H02Qell} and \eqref{h04ell}--\eqref{h05ell}, since the left-hand sides of \eqref{eqH0_sec}  and \eqref{eqh0oddodd} are identical to the left-hand sides in the linear equations \eqref{eqH0_lin} and \eqref{eqh0_lin}. The corresponding arbitrary integration constants are chosen as follows \cite{Riva:2023rcm,DeLuca:2023mio}: at second order, we will set to zero the constant associated to the solution that has growing profile at large $r$, since it is degenerate with the linear tidal field solution (in other words, it can always  be reabsorbed through a redefinition of the  linear tidal field's amplitude); we will choose the remaining constant, associated to the amplitude of the decaying solution at infinity, in such a way to remove any possible divergence at the horizon in the particular solutions \eqref{H0_Greens} and \eqref{h0_Greens}. Since the Green's functions in \eqref{H0_Greens} and \eqref{h0_Greens} have already been chosen in \eqref{GH0reg} and \eqref{Gh0reg} in such a way to respect regularity, we will also set the second integration constant to zero.

All in all, the inhomogeneous solutions for $\Hs_0^{ \ell m }$ and $\hs^{\ell m}_0$ that satisfy regularity conditions at the horizon can be written in full generality as \cite{Riva:2023rcm}:  
\begin{align}
\Hs_0^{ \ell m }  (r) & =  \frac{1}{r_s W^\ell_{H_0} }  \sum_{\ell_1 ,m_1, \ell_2, m_2}  \bigg[ H_{0,2}^\ell (r) \int_{r_s}^r  {\rm d}r' \, r'(r'-r_s)   H_{0,1}^\ell (r')
 {\cal S}_{H_0 \ \ell \ell_1  \ell_2}^{m  m_1 m_2} (r') 
\nonumber \\
& \qquad\qquad\qquad\qquad\quad  -
H_{0,1}^\ell (r) \int^r  {\rm d}r' \,  r'(r'-r_s)   H_{0,2}^\ell (r')
 {\cal S}_{H_0 \ \ell \ell_1  \ell_2}^{m  m_1 m_2} (r') 
\bigg] \;,
\label{H0_Greensreg}
\\
  \hs^{\ell m}_0 (r)  & = \frac{1}{ r_s W^\ell_{h_0} }  \sum_{\ell_1 ,m_1, \ell_2, m_2}  
  \bigg[ h_{0,4}^\ell (r)
  \int_{r_s}^r  {\rm d}r' \,  h_{0,5}^\ell (r')
   {\cal S}_{h_0 \ \ell \ell_1 \ell_2}^{ m m_1 m_2}  (r') 
\nonumber \\
& \qquad\qquad\qquad\qquad\quad
- h_{0,5}^\ell (r) \int^r  {\rm d}r' \,  h_{0,4}^\ell (r')
   {\cal S}_{h_0 \ \ell \ell_1 \ell_2}^{ m m_1 m_2}  (r') 
\bigg] \;,
\label{h0_Greensreg}
\end{align}
where the integrals on the second line of each equation are indefinite integrals.\footnote{Alternatively, they can also equivalently rewritten as integrals between $r$ and some other arbitrary radius $r_0$. The latter is however completely immaterial, because it can be removed in general via a  redefinition of the integration constant of the growing homogeneous solutions $H_{0,1}^\ell (r)$ and $h_{0,5}^\ell (r)$ \cite{Riva:2023rcm}.}

\subsection{Quadratic solutions induced by {\em quadrupole} $\times$ {\em quadrupole}}
\label{sec:solutionsQuadrpoleTimesQuadrupole}

 It is instructive to provide an explicit example of the previous expressions. We will focus here on the case of two linear quadrupolar fields (i.e., $\ell_1=2$ and $\ell_2=2$), generating a second-order mode $\ell$ \cite{Riva:2023rcm}. 
The $\ell=2$  linear even and odd solutions can be read off from
 the general expressions \eqref{H0linsol} and \eqref{h0linsol}:  
\begin{align}
    \Hf^{2 m}_0(r) &= \mathcal{E}^{\rm +}_{2 m } \frac{r}{r_s} \bigg(\frac{r}{r_s} - 1\bigg) \;, \label{evenLinearQuadrupole} \\
    \hf^{2 m }_0(r) &= \mathcal{E}^{\rm -}_{ 2 m } \frac{r^2}{r_s} \bigg(\frac{r}{r_s} - 1\bigg) \;,
\label{oddLinearQuadrupole}
\end{align}
where in the amplitudes of the even and odd linear modes, respectively $\mathcal{E}^+$ and $\mathcal{E}^-$, we kept the dependence on $m$ generic.

Based on the angular momentum and parity selection rules, the induced quadratic mode can only take the values $\ell = 0$, $\ell = 2$, or $\ell = 4$.\footnote{Quadratic perturbations with $\ell = 1$ and $\ell = 3$ vanish. This occurs because in these cases $\ell + \ell_1 + \ell_2$ is odd and, according to the selection rules discussed in table~\ref{table:SR}, only $\Hs_1$ and $\hs_1$ can be generated. However, these also vanish because the radial profiles of their sources are antisymmetric under $\ell_1 m_1 \leftrightarrow \ell_2 m_2$ and do not explicitly depend on the magnetic quantum numbers.}
In the following, we will discuss each case separately, starting with the quadrupole and addressing the somewhat special case of the monopole at the end. The analysis of the second-order solutions induced by generic $\ell_1$ and $\ell_2$ modes will be considered in sec.~\ref{sec:HM}.

\subsubsection{Quadrupole} Let us start by analyzing the second-order $\ell=2$ mode induced by  two linear quadrupoles.
In the even sector, from the general formula \eqref{H0_Greensreg}, we obtain for $\Hs_0^{2 m}$:
\be
\begin{split}
\Hs_0^{2 m} =  \sum_{m_1,m_2} \II{0}_{222}^{m m_1 m_2} \frac14 \bigg[ \bigg. \mathcal{E}^{\rm -}_{2 m_1} \mathcal{E}^{\rm -}_{2 m_2} \frac{3}{2} \frac{r}{r_s}\bigg(12 \frac{r^3}{r_s^3} - 14 \frac{r^2}{r_s^2} + \frac{r}{r_s} - 1\bigg) 
-\mathcal{E}^{\rm +}_{ 2 m_1 } \mathcal{E}^{\rm +}_{2 m_2 }   \frac{r^2}{r_s^2} \bigg(2 \frac{r^2}{r_s^2} + \frac{r}{r_s}-3 \bigg)    \bigg] \;,
\end{split}
\label{eq:H02m}
\ee
where the sums run over $|m_1|\le 2$ and $|m_2|\le 2$. 
By solving the constraint equations at second order one can also easily find the other even scalars in the metric,
\begin{align}
\Hs_2^{2 m} &=  \sum_{m_1,m_2} \II{0}_{222}^{m m_1 m_2} f(r) \frac14 \bigg[ \bigg. \mathcal{E}^{\rm -}_{2 m_1} \mathcal{E}^{\rm -}_{2 m_2} \frac{3}{2} \frac{r}{r-r_s  }\bigg(56 \frac{r^3}{r_s^3} - 54 \frac{r^2}{r_s^2} + \frac{r}{r_s} - 1\bigg) 
-\mathcal{E}^{\rm +}_{ 2 m_1} \mathcal{E}^{\rm +}_{2 m_2 }   \frac{r^3}{r_s^3}\left(4 \frac{r}{r_s} + 1 \right)  \bigg]  \;, \\
\Ks^{2 m } &= \sum_{m_1,m_2} \II{0}_{222}^{m m_1 m_2}  \frac{1}{16} \bigg[  \mathcal{E}^{\rm -}_{2 m_1} \mathcal{E}^{\rm -}_{2 m_2} 3 \bigg(42 \frac{r^4}{r_s^4} - 24 \frac{r^3}{r_s^3} + 2\frac{r^2}{r_s^2} -1 \bigg)
-\mathcal{E}^{\rm +}_{2 m_1} \mathcal{E}^{\rm +}_{2 m_2} \bigg(2 \frac{r^4}{r_s^4} + 8\frac{r^2}{r_s^2}-7 \bigg)  \bigg]  \;.
\end{align}
Since $\ell + \ell_1 + \ell_2$ is even, $\Hs_1^{2 m}=0$, for all $m$.

In the odd sector, from \eqref{h0_Greensreg} and eq.~\eqref{sourceh0} for the source, we find instead  
\begin{equation}
\hs_0^{2 m}= - \sum_{m_1,m_2} \II{0}_{222}^{m m_1 m_2} \mathcal{E}^{\rm -}_{2 m_1} \mathcal{E}^{\rm +}_{2 m_2} \frac{1}{2} \frac{r^3}{r_s^2} \bigg(\frac{r^2}{r_s^2} + \frac{r}{r_s} - 2\bigg) \,. 
\label{eq:h02m}
\end{equation}
In addition, the other odd variable in the metric vanishes, $\hs_1^{2 m} =0$.

\subsubsection{Hexadecapole} For the induced $\ell=4$ mode, the derivation follows the same logic as for the quadrupole. The second-order solutions are  
\begin{align}
\Hs_0^{4 m} &= \sum_{m_1,m_2} \II{0}_{422}^{m m_1 m_2}  \frac{1}{36} \bigg[ \mathcal{E}^{\rm +}_{2 m_1} \mathcal{E}^{\rm +}_{2 m_2} \frac{r^2}{r_s^2} \bigg( 73 \frac{r^2}{r_s^2} - 107\frac{r}{r_s}+34 \bigg)  \nonumber \\ & \qquad \qquad \qquad \qquad  +   \mathcal{E}^{\rm -}_{2 m_1} \mathcal{E}^{\rm -}_{2 m_2}  \frac{r}{r_s} \bigg( -139 \frac{r^3}{r_s^3} + 119 \frac{r^2}{r_s^2} + 17\frac{r}{r_s} - 3 \bigg)   \bigg]  \;,  
\label{eq:H04m}
\\
\Hs_2^{4 m} & = \sum_{m_1,m_2}  \II{0}_{422}^{m m_1 m_2} f(r) \frac{1}{36} \bigg[ \mathcal{E}^{\rm +}_{2 m_1} \mathcal{E}^{\rm +}_{2 m_2} \frac{r^3}{r_s^3} \bigg(97 \frac{r}{r_s} - 58 \bigg)  
\nonumber \\ 
& \qquad \qquad \qquad \qquad -  \mathcal{E}^{\rm -}_{2 m_1} \mathcal{E}^{\rm -}_{2 m_2}   \frac{r^2}{r_s(r-r_s)} \bigg(7 \frac{r^3}{r_s^3} +\frac{r^2}{r_s^2} -17 \frac{r}{r_s} + 3 \bigg)  \bigg] \;, \\
\Ks^{ 4 m } & =  \sum_{m_1,m_2}  \II{0}_{422}^{m m_1 m_2}  \frac{1}{108} \bigg[ \mathcal{E}^{\rm +}_{2 m_1} \mathcal{E}^{\rm +}_{2 m_2}  \bigg( 291 \frac{r^4}{r_s^4} - 266\frac{r^3}{r_s^3}-12\frac{r^2}{r_s^2} +21  \bigg)  \nonumber \\
& \qquad \qquad \qquad \qquad -  \mathcal{E}^{\rm -}_{ 2 m_1} \mathcal{E}^{\rm -}_{2 m_2} \bigg(21 \frac{r^4}{r_s^4} + 52 \frac{r^3}{r_s^3} -30 \frac{r^2}{r_s^2} + 1\bigg)   \bigg] \;, \\
\hs_0^{ 4m} & = \sum_{m_1,m_2}  \II{0}_{422}^{m m_1 m_2} \mathcal{E}^{\rm -}_{2 m_1} \mathcal{E}^{\rm +}_{2 m_2} \frac{1}{5} \frac{r^3}{r_s^2} \bigg(8\frac{r^2}{r_s^2} -13 \frac{r}{r_s} + 5 \bigg) \;.
\label{eq:h04m}
\end{align}
By parity, $\Hs^{4 m}_1 = 0 = \hs_1^{4 m}$.

\subsubsection{Monopole} 
As discussed at the end of sec.~\ref{linpert}, the case of the monopole is special because the metric before gauge fixing contains only four (even) components: $H_0$, $H_1$, $H_2$ and $K$. At the linear level, a monopolar perturbation can be absorbed into the definition of the Schwarzschild mass, consistently with Birkhoff's theorem. However, this is no longer the case at second order because the two linear quadrupoles, $\ell_1 = 2$ and $\ell_2 = 2$, break the spherical symmetry. Therefore, the generated quadratic  monopole in RW gauge is a physical perturbation.

Despite its apparent singular behavior, the source in RW gauge in eq.~\eqref{sourceH0} remains finite for $\ell = 0$ once the linear solutions are inserted and the limit $\lambda \to 0$ is appropriately taken. Therefore, the quadratic equation \eqref{eqH0_sec} can be used to compute the monopole of $\Hs_0$. The result is
\be
\begin{split}
\Hs_0^{0 0} =  \sum_m  \II{0}_{022}^{0 -m m}  & \bigg[  \mathcal{E}^{\rm +}_{ 2 -m } \mathcal{E}^{\rm +}_{2 m} \frac{r^2}{ 2 r_s (r-r_s) } \bigg(- \frac{r^3}{r_s^3} + \frac{r^2}{r_s^2} - 2 \frac{r}{r_s} - 3  \bigg) 
    \\
   & +  \mathcal{E}^{\rm -}_{2 - m} \mathcal{E}^{\rm -}_{2 m} \frac{3 r_s}{4 r}  \bigg( 6 \frac{r^5}{r_s^5} + 2 \frac{r^4}{r_s^4} - 6 \frac{r^3}{r_s^3} - \frac{r}{r_s} + 1 \bigg) 
 \bigg]  \;.
\end{split}
\label{eq:H00m}
\ee
Moreover, the second-order constraint equations that relate $H_2$ and $K$ to $H_0$ can also be extended for $\lambda=0$, yielding
\begin{align}
\Hs_2^{00} = \ &   \sum_m  \II{0}_{022}^{0 - m m}  \frac{1}{f(r)} \frac12 \bigg[  \mathcal{E}^{\rm +}_{2 - m} \mathcal{E}^{\rm +}_{2 m}  \frac{r^2}{ r_s
   (r-r_s)}    \bigg( 7 \frac{r^3}{r_s^3}-23 \frac{r^2}{ r_s^2} +26  \frac{r}{r_s} -11  \bigg)
   \nonumber \\
   & \qquad \qquad -\mathcal{E}^{\rm -}_{2 - m} \mathcal{E}^{\rm -}_{2 m}  \frac{3 r_s}{2    r }  \bigg(98 \frac{r^5}{r_s^5} -138 \frac{r^4}{ r_s^4} +42 \frac{r^3}{r_s^3} + \frac{r}{ r_s} - 1 \bigg)
    \bigg] \;, \label{2H200} \\
    \Ks^{0 0} = \ &  \sum_m \II{0}_{022}^{0 - m m} \frac12  \bigg[  \mathcal{E}^{\rm +}_{2 - m} \mathcal{E}^{\rm +}_{2 m}  \bigg(2 \frac{r^4}{r_s^4} -4 \frac{r^3}{ r_s^3} +2 \frac{r^2}{r-s^2} -1 \bigg) 
\nonumber   \\
 & \qquad \qquad +\mathcal{E}^{\rm -}_{2 - m} \mathcal{E}^{\rm -}_{2 m}  \frac{3 r_s}{2 r} \bigg(- 20 \frac{r^5}{r_s^5} +32 \frac{r^4}{ r_s^4}-12 \frac{r^3}{r_s^3}- \frac{r}{ r_s}+ 2  \bigg)   \bigg] \; , \label{2K00}
\end{align}
{while $\Hs_1^{00} = 0$.}

{It is important to note that this procedure has implicitly fixed a gauge. Notice that when $\ell=0$ the metric components corresponding to $\mathcal{H}_0$, $\mathcal{H}_1$ and  $G$ vanish trivially, see eq.~\eqref{hPMeven4D}. Therefore, \eqref{RWgauge} does not fix any gauge: one can still perform a gauge transformation with the freedom of   choosing two arbitrary functions of $r$. In the way we took the limit, such functions have been implicitly chosen to yield precisely the solutions  written above. To understand this better, let us follow a different logic. Instead of taking the limit $\ell\rightarrow0$ in the general-$\ell$ results, let us set $\ell=0$ from the very beginning. In particular, let us  choose for instance the gauge such that $\Ks^{00} = 0$ and $\Hs_1^{00} = 0$. One shall then follow the standard procedure, and find the solutions for $\Hs_0^{00}$ and  $\Hs_2^{00}$. One can verify explicitly that these solutions are indeed related to \eqref{eq:H00m}--\eqref{2K00} via a second-order gauge transformation, up to a homogeneous solution which can be reabsorbed into a redefinition of the Schwarzschild radius in the background metric.}

\section{Worldline effective field theory}
\label{sec:EFT}
In this section we leave the framework of full GR to enter that of the EFT. 
First introduced in \cite{Goldberger:2004jt,Goldberger:2005cd}, the worldline EFT allows for  a clear and unambiguous definition of the Love numbers as Wilson coefficients of higher-dimensional operators. 
We will first briefly review its main features (see refs.~\cite{Goldberger:2006bd,Rothstein:2014sra,Porto:2016pyg,Levi:2018nxp,Goldberger:2022ebt,Goldberger:2022rqf} for more details), which we will  extend to account  for nonlinear response. We will then introduce the {background field method} \cite{DeWitt:1967ub,tHooft:1974toh,Abbott:1980hw} and explain its advantages for our calculations. In this context, the metric perturbation is computed using Feynman diagrams, which we will provide at the end of this section, and solving the resulting (loop) integrals. 

Only in the next section we will perform a similar calculation to the one  discussed in the previous section, but within the EFT context. By matching these  results  with those obtained in full GR, we will be able to uniquely determine the values of the Love number couplings induced by the quadrupole-quadrupole interaction. The extension to other multipoles will be considered in sec.~\ref{sec:HM}.

\subsection{Point-particle action and nonlinear Love numbers}
\label{ppaction}

From far away, any object appears in first approximation as a point source. For  non-rotating bodies, the action describing the dynamics of a point particle  in GR is
\begin{equation}
	S = S_{\textrm{EH}} + S_{\textrm{pp}} \, ,
	\label{eq:Spponly}
\end{equation}
where $S_{\textrm{EH}}$ is the usual Einstein--Hilbert term, while $S_{\textrm{pp}}$ is the point-particle action,
\begin{equation}
	S_{\textrm{pp}} = - M\int \D \tau = -M \int \sqrt{- g_{\mu\nu} \D x^\mu \D x^\nu} \, ,
	\label{eq:Spp}
\end{equation}
with $M$   the mass of the source, and $\tau $ the point particle's proper time.
In the second equality we used that  $\D \tau = \sqrt{- g_{\mu\nu} \D x^\mu \D x^\nu}$. In terms of the proper time, we shall define the four-velocity ${u^\mu = \D x^\mu/\D \tau}$.

Clearly, for a generic object eq.~\eqref{eq:Spponly} cannot be exact. Finite-size effects and the tidal deformation of the object can be described by adding to \eqref{eq:Spponly} a series of higher-dimensional operators, organized in a derivative expansion and localized on the particle's worldline. At quadratic order, they take the form~\cite{Goldberger:2005cd,Goldberger:2009qd}  
\begin{equation}
S_{\mathrm{T}} = \sum_{\ell=2}^\infty  \int \D \tau \left( 
Q_E^{\mu_L} E_{\mu_L} + Q_B^{\mu_L} B_{\mu_L} \right) + \dots ,
\label{Sint}
\end{equation}
where ellipses stand for nonlinear couplings between $Q_{E,B}$ and powers of $E$ and $B$, and  
where we have introduced the multi-index notation $\mu_L\equiv \mu_1\cdots \mu_\ell$ and the operators \cite{Bern:2020uwk,Riva:2023rcm}
\begin{align}
E_{\mu_1\cdots\mu_\ell}&  \equiv  \nabla^\perp_{\langle \mu_1} \cdots \nabla^\perp_{\mu_{\ell-2} } E_{  \mu_{\ell-1}\mu_\ell \rangle} \; , \label{multiEdef}
\\
B_{\mu_1\cdots\mu_\ell} & \equiv   \nabla^\perp_{\langle \mu_1} \cdots \nabla^\perp_{\mu_{\ell-2} } B_{  \mu_{\ell-1}\mu_\ell \rangle} \; , \label{multiBdef}
\end{align}
with
\begin{equation}
E_{\mu\nu} \equiv C_{\mu\rho\nu\sigma}u^\rho u^\sigma,
\qquad
B_{\mu\nu} \equiv \frac{1}{2} \epsilon_{\gamma\alpha\beta \langle \mu} {C_{\nu\rangle \delta}}^{\alpha\beta} u^\delta u^\gamma  \;,
\label{EmunuBmunu} 
\end{equation}
corresponding to the  electric and magnetic parts of the Weyl tensor  $C_{\mu\rho\nu\sigma}$, respectively. The symbol $\langle \, \cdots \,  \rangle \, $  denotes the traceless symmetrization over the enclosed indices, while 
\begin{equation}
\nabla_\mu^\perp \equiv (\delta^\nu_\mu + u^\nu u_\mu) \nabla_\nu
\end{equation}
is the covariant derivative projected on the hypersurface orthogonal to $u^\mu$.
In the following, we will mostly  work in  the rest frame of the point particle, such that the only nontrivial components of $E_{\mu\nu}$ and $B_{\mu\nu}$ become the spatial ones, i.e.~$E_{ij} = C_{titj}$ and  
 $B_{ij} \equiv \frac{1}{2} \epsilon_{tkl \langle i} {C_{j\rangle t}}^{kl}$.  In practice, we will replace $\mu_L$ in \eqref{Sint} with the spatial indices $i_L\equiv i_1 \cdots i_\ell$. 

In the action \eqref{Sint}, the quantities $Q_{E,B}^{\mu_L}$ carry all the information about the response of the body, as a result of the interaction between the worldline and the external fields $E_{\mu\nu}$ and $B_{\mu\nu}$. Such effective couplings, which have been kept  fully general so far, do not capture only the conservative tidal response of the body, but  can be used also to describe {\em dissipative} effects. In the latter case, the quantities $Q_{E,B}^{\mu_L}$ are interpreted as composite operators that depend on  additional gapless degrees of freedom, which live on the worldline and parametrize the unknown microscopic physics that is responsible for dissipation, such as the internal dynamics of the object, or absorption across the horizon in the case of black holes~\cite{Goldberger:2005cd,Goldberger:2020fot,Ivanov:2022hlo}.  In the spirit of the EFT, we will remain completely agnostic about the microscopic details of the system, which are integrated out,  and we will parametrize $Q_{E,B}^{\mu_L}$ in (nonlinear) response theory. 

Concretely, one can express the one-point function of $Q_{E,B}^{\mu_L}$ in the most general way as
\be
\begin{split}
\langle Q_{E,B}^{i_L}(\tau) \rangle 
= \sum_{n=1}^\infty \sum_{k=0}^n \int \D \tau_1 \cdots \D\tau_n  & \, {}^{(n)} \! {\cal R}_{E,B}^{i_L\vert j_{L_1} \cdots   \,  j_{L_k} \vert j_{L_{k+1}} \cdots   \,  j_{L_n}  }(\tau-\tau_1,\dots,\tau-\tau_n) 
\\
& \times E_{j_{L_1}}(\tau_1) \cdots E_{j_{L_k}}(\tau_k)   
B_{j_{L_{k+1}}}(\tau_{k+1}) \cdots B_{j_{L_n}}(\tau_n)   ,
\label{QREn}
\end{split}
\ee
where ${}^{(n)} \! {\cal R}_{E,B}$ represents the $n^{\text{th}}$-order response function. 
As an example, when $n=1$, eq.~\eqref{QREn} gives the linear response of the system and ${}^{(1)}\!{\cal R}_{E,B}$ reduces to the standard retarded Green's function~\cite{Goldberger:2020fot,Ivanov:2022hlo}. For higher values of $n$, eq.~\eqref{QREn} captures instead the nonlinear response.\footnote{The case $n=0$ captures the intrinsic multipoles of the object, which are irrelevant for our discussion below.} This is similar to nonlinear polarization theory in electromagnetism, where the nonlinear polarization of an optical medium is parametrized as an expansion in powers of the external field (see, e.g., \cite{10.1093/acprof:oso/9780198702764.001.0001}).
Note also that, as opposed to the linear case where the even and odd sectors are decoupled, at higher nonlinear orders an electric response (analogously, a magnetic response) can be induced by combinations of both external electric and magnetic fields~\cite{Bern:2020uwk,Riva:2023rcm,Hadad:2024lsf}. 
Besides obvious symmetry properties under the exchange of its multi-indices, there are some nontrivial constraints on the form of the tensor ${}^{(n)}\!{\cal R}$. One comes from causality,  which forces ${}^{(n)}\!{\cal R}$ to vanish  whenever any of its arguments $\tau-\tau_j$, for $j=1,\dots,n$, turns negative.
Moreover, the tensorial structure of ${}^{(n)}\!{\cal R}$ is further dictated by the building blocks in  the theory: these  are  the metric, the Levi-Civita tensor and possibly the spin vector, in the case of rotating objects.

 Since we are interested here in performing the matching with Schwarzschild black holes at leading order in the static limit, a certain number of simplifications occur. In the absence of rotation, the tensorial form of ${}^{(n)}\!{\cal R}$ boils down to simply a product of Kronecker deltas, involving only spatial indices in the rest frame of the point particle~\cite{Bern:2020uwk,Riva:2023rcm}. In addition, in the limit of static tides, the system is non-dissipative and  the time dependence in ${}^{(n)}\!{\cal R}$ reduces to a product of Dirac delta-functions~\cite{Goldberger:2005cd,Goldberger:2020fot,Ivanov:2022hlo,Saketh:2023bul}.
For instance, to leading order in the number of gradients,  one has, schematically,  
\begin{equation}
{}^{(n)}\!{\cal R}^{ij\vert i_1 j_1 \cdots i_n j_n}
=  \sum_k \lambda_k \, {\rm perm}_k \left[ \delta^{\langle i}_{\vert j_n \rangle} \delta^{j \rangle}_{\langle i_1 } \delta^{\langle i_2}_{j_1 \rangle} \delta^{ j_2\rangle}_{\langle i_3}   \cdots \delta^{j_{n-1} \rangle }_{\langle i_n\vert} \right]  \delta (\tau-\tau_1)\cdots  \delta (\tau-\tau_n) ,
\end{equation}
where the sum runs over non-redundant permutation $k$ of the indices (see, e.g., discussion in ref.~\cite{Bern:2020uwk}; see also sec.~\ref{sec:matchingL} for higher orders in derivatives). Plugging the nonlinear response theory solution \eqref{QREn} into the action \eqref{Sint} including nonlinear terms, one then finds a series of effective interactions which, under the previous assumptions, take the form of local operators involving contractions of $E_{ij}$ and $B_{ij}$, and derivatives thereof. 

For concreteness,  up to cubic order in the number of fields and  to leading order in derivatives, the finite-size action $S_{\rm T}$ reads
\be
\begin{split}
	S_{\rm T} =  & \int \D \tau \bigg[  
	\lambda_2^{(E)} E_{ij} E^{ij} + 
	\lambda_2^{(B)} B_{ij} B^{ij}  \\
	& +
	\lambda_{2 2 2}^{(E^3)} E^{i}{}_{j} E^{j}{}_{k} E^{k}{}_{i} 
	+ \lambda_{2 2 2}^{(B^3)} B^{i}{}_{j} B^{j}{}_{k} B^{k}{}_{i}	 +
	\lambda_{2 2 2}^{(E^2B)} E^{i}{}_{j} E^{j}{}_{k} B^{k}{}_{i} +
	\lambda_{2 2 2}^{(E B^2)} E^{i}{}_{j} B^{j}{}_{k} B^{k}{}_{i} 
	\bigg] .
	\label{eq:ST}
\end{split}
\ee
The coefficients $\lambda^{(E)}_2$ and $\lambda^{(B)}_2$ are the linear Love number couplings, which are famously zero for black holes in GR \cite{Damour:2009vw,Binnington:2009bb,Fang:2005qq,Kol:2011vg,Chakrabarti:2013lua,Gurlebeck:2015xpa, Porto:2016zng,LeTiec:2020spy,LeTiec:2020bos,Hui:2020xxx,Chia:2020yla,Charalambous:2021mea,Rai:2024lho}. In ref.~\cite{Riva:2023rcm}, we verified that this result is robust against nonlinear corrections, and we further showed that  $\lambda_{2 2 2}^{(E^3)} = 0$.  On the other hand, for parity reasons, one immediately  concludes that    $\lambda_{2 2 2}^{(B^3)} = \lambda_{2 2 2}^{(E^2B)} =0$ in GR~\cite{Hadad:2024lsf}. In sec.~\ref{sec:match}, we will extend our previous result \cite{Riva:2023rcm} to the odd sector: we will explicitly perform  the matching at the next-to-leading order in powers of $r_s$ including the operator $E^{i}{}_{j} B^{j}{}_{k} B^{k}{}_{i}$, and show that  $\lambda_{2 2 2}^{(E B^2)} = 0$ for Schwarzschild black holes. We will finally generalize the matching to all orders in the number of spatial derivatives in sec.~\ref{sec:HM}.

\subsection{Background field method}
\label{sec:BGF}

Given the EFT presented in the previous section, the starting point of our computation is the following action 
\begin{equation}
	S[g, x(\tau)] = S_{\textrm{EH}}[g]+ S_{\textrm{pp}}[g, x(\tau)] + S_{\textrm{T}}[g, x(\tau)] \, ,
	\label{eq:effS}
\end{equation}
where the terms on the right-hand side are given in eqs.~\eqref{EHaction}, \eqref{eq:Spp} and \eqref{eq:ST}, respectively. The goal is to compute the response induced by an external field. More precisely, we separate the full metric as 
\begin{align}
	g_{\mu\nu} & = \bar{g}_{\mu\nu} + \kappa h_{\mu \nu} \, ,
	\label{eq:exp_BG}
\end{align}
where $\kappa \equiv \sqrt{32 \pi G}$. Here $\bar{g}_{\mu\nu}$ is an external  ``background'' metric which includes the tidal field and, therefore,  has to satisfy the full nonlinear  Einstein equations in vacuum, i.e.,
\begin{equation}
	\bar{R}_{\mu\nu}-\frac{1}{2}\bar{g}_{\mu\nu} \bar{R} = 0 \, ,
	\label{eq:EEV}
\end{equation}
with $\bar{R}_{\mu\nu}$ and $\bar{R}$ representing the Ricci tensor and Ricci scalar of $\bar{g}_{\mu\nu}$, respectively. The second term, $\kappa h_{\mu \nu}$, denotes the response.

Given this split, it is  convenient to use the background  field method \cite{DeWitt:1967ub,tHooft:1974toh,Abbott:1980hw}. Specifically, we insert $g_{\mu \nu}$ from eq.~\eqref{eq:exp_BG} into eq.~\eqref{eq:effS} and rewrite the action into two pieces, 
$S[g, x(\tau)] = S_{0}[\bar{g}, x(\tau)] + S_{h}[\bar{g}, h, x(\tau)]$, where the first term, $S_{0}$, is independent of the response field $h_{\mu \nu}$.
Then, we treat the massive object as an external source of the gravitational field. This allows us to compute the effective action of the system, $W[\mathcal{J}]$, by formally performing the following path integral,
\begin{equation}
	\e^{i W[\mathcal{J}]} = \e^{i S_{0}[\bar{g}, x(\tau)]} \int \mathcal{D}[h] \exp\left\{ i S_{h}[\bar{g}, h, x(\tau)] + i S_{\textrm{GF}}[\bar{g}, h] + i \int \!\!\D^4 y \sqrt{-\bar{g}} \mathcal{J}^{\mu\nu}(y) h_{\mu\nu}(y)\right\} \, .
	\label{eq:PIexp}
\end{equation}
The additional action, $S_{\textrm{GF}}$, is the usual gauge-fixing term arising from a Faddeev--Popov procedure. We will work in the background de Donder gauge, i.e.~we define 
\begin{equation}
	S_{\textrm{GF}} 
	\equiv -\int \D^4 x \sqrt{-\bar{g}} \bar{g}^{\mu\nu}
		\bar{\Gamma}_\mu \bar{\Gamma}_\nu \, , \qquad 
		\bar{\Gamma}_\mu \equiv \bar{g}^{\rho\sigma}\bar{\nabla}_{\rho} h_{\sigma\nu} - 
		\frac{\bar{g}^{\rho\sigma}}{2}\bar{\nabla}_\nu h_{\rho\sigma} \, ,
		\label{eq:BGGF}
\end{equation}
where $\bar{\nabla}_\rho \bar{g}_{\mu\nu} = 0$. This choice  ensures that the final result is invariant under diffeomorphisms of the external metric $\bar{g}_{\mu\nu}$. 

We are interested in computing the one-point expectation value of the field $h_{\mu\nu}$, which in terms of the effective action $W[\mathcal{J}]$ above is given by\footnote{Since  $S_0$ is a constant term, we shall drop it in \eqref{eq:PIexp}.}
\begin{equation}
	\langle h_{\mu\nu}(x)\rangle = \frac{1}{\sqrt{-\bar{g}}}  \frac{\delta  W[\mathcal{J}]}{i\delta \mathcal{J}^{\mu\nu}(x)}  \bigg|_{\mathcal{J}=0} \, .
\end{equation}
In practice, we will compute the above equation by considering all Feynman diagrams with one external $h_{\mu\nu}(x)$ field. 
Although we are performing a path integral, we are ultimately interested  in the classical limit.
Following~\cite{Goldberger:2004jt}, we enforce this limit from the outset by employing a saddle-point approximation. Diagrammatically, this means neglecting all diagrams that contain close loops of the field $h_{\mu\nu}$. This is also why we do not need to add any ghost fields in eq.~\eqref{eq:PIexp}.

One last step before proceeding is to write
\begin{equation}
	\bar{g}_{\mu\nu} = \eta_{\mu\nu} + H_{\mu\nu} \, ,
\label{defH}
\end{equation}
where $\eta_{\mu \nu}$ is the flat Minkowksi metric and $H_{\mu\nu}$ is the external (static) tidal field, which is determined  by solving the vacuum Einstein equations and requiring regularity at the origin. We will discuss the form of $H_{\mu\nu}$ in more in details in sec.~\ref{sec:TidalH}. Equation \eqref{defH} allows to perform a perturbative expansion of the action in both the tidal field amplitude and in the coupling constant $\kappa$.
By performing this expansion, one schematically obtains, for the bulk action,
\begin{equation}
\label{eq:BA}
\begin{split}
	S_{\textrm{EH}}[g] + S_{\textrm{GF}}[\bar{g}, h] & =  -\frac{1}{2}\int \D^4 x \, 
		\left[ 
		\partial_{\mu} h_{\alpha \beta} \partial^\mu h^{\alpha \beta} 
			-\frac{1}{2}\partial_{\mu} h \partial^\mu h
		\right] \\
		& \qquad + 
		 \int \D^4 x \left[ H (\partial h)^2 + h (\partial H) \partial h + \kappa h ( \partial h)^2 \right]   \\
		& \qquad +  \int \D^4 x \left[ H^2 (\partial h)^2 + H h (\partial H) \partial h + \kappa H h ( \partial h)^2 + \kappa^2 h^2 ( \partial h)^2 \right] + \dots \, ,
\end{split}
\end{equation}
where here and below indices are raised and lowered  with  $\eta_{\mu \nu}$.  
From the first term on the right-hand side we obtain the usual graviton propagator in de Donder gauge, while the higher-order terms in the second and third lines give the self-interaction vertices (which we have written here schematically without making index contractions  explicit). Notice that, on the right-hand side, we have not explicitly written any terms containing only the external field $H_{\mu\nu}$, as they belong to $S_{0}[\bar g,x(\tau)]$ and are immaterial for the following calculations. Furthermore, since $H_{\mu\nu}$ satisfies the vacuum nonlinear Einstein equations, we do not have operators with a  single $h_{\mu\nu}$.   Additionally, for the choice of the gauge-fixing term \eqref{eq:BGGF}, all interaction vertices containing at least one tidal field $H_{\mu\nu}$ receive a contribution from $S_{\textrm{GF}}$, while the self-interactions of $h_{\mu\nu}$ remain unaffected by $S_{\textrm{GF}}$.

Finally, we have to expand the matter part of the action, i.e.~$S_{\textrm{pp}}$ and $S_{\textrm{T}}$. Rather than using the proper time, it is convenient to choose a frame in which the massive object is at rest at the origin,  i.e.~we choose the time $t$ such that
\begin{equation}
	x^{\mu}(t) = (t, 0, 0, 0) \, , \qquad v^\mu(t) \equiv \frac{\D x^\mu}{\D t} = (1, 0, 0, 0) \, .
\end{equation}
The point-particle action then takes the form
\begin{equation}
	S_{\textrm{pp}} = -M \int \D t\, \frac{\D \tau}{\D t} = -M \int \D t \sqrt{- g_{\mu\nu}  v^\mu  v^\nu} \, ,
	\label{eq:Sppt}
\end{equation}
and similarly for $S_{\textrm{T}}$. 

In order to write the Feynmann rules, let  us introduce the following conventions:
\begin{align}
	\raisebox{3pt}{\gravh} & =  h_{\mu\nu} \, , &  
	\raisebox{3pt}{\Csoure} & = \textrm{point-particle source}  \, , &  \notag \\ 
	 \raisebox{3pt}{\gravH} & =  H_{\mu\nu} \, , & 
	\raisebox{3pt}{\Tsoure} & = \textrm{tidal-induced source} \, .& 
	 \notag
\end{align}
Then,  introducing the tensor
\begin{equation}
	P_{\mu\nu\rho\sigma} = -\frac{1}{2}\big( \eta_{\mu \rho}\eta_{\nu\sigma} 
			+\eta_{\mu \sigma}\eta_{\nu\rho}
			-\eta_{\mu \nu}\eta_{\rho\sigma}\big) 
\end{equation}
to define the graviton propagator in de Donder gauge,
 the  Feynman rules obtained from the bulk action \eqref{eq:BA} are
\allowdisplaybreaks
\begin{align}
\proph & = \frac{i}{k^2 - i0^+} P_{\mu\nu\rho\sigma} \, ,  \label{eq:prop} \\
\cubic & = i \kappa {\ddl}^{(4)}(k_1 + k_2 + k_3) \mathcal{V}_{\alpha_1 \beta_1 \alpha_2 \beta_2 \alpha_3 \beta_3}  \, ,\\
\cubicback & = i  H_{\alpha_1 \beta_1}(-k_2 -k_3) {V_3^{\alpha_1 \beta_1}}_{\alpha_2 \beta_2 \alpha_3 \beta_3} \, , 
\label{eq:FRSbulk3}\\
\quarticback & = i \int_q H_{\alpha_1 \beta_1}(q)H_{\alpha_2 \beta_2}(-k_2 -k_3-q) {V_4^{\alpha_1 \beta_1\alpha_2 \beta_2}}_{ \alpha_3 \beta_3\alpha_4 \beta_4} \, ,
\label{eq:FRSbulk4}
\end{align}
where ${\ddl}^{(4)}(k)\equiv (2\pi)^4 {\delta}^{(4)}(k)$. $\mathcal{V}$ is the standard trilinear graviton vertex in de Donder gauge, while  $V_3$ and $V_4$ are cubic and quartic vertices  in the gauge  \eqref{eq:BGGF}.
Note that we can ignore the $i0^+$ prescription in the propagator \eqref{eq:prop} because it is irrelevant for the following static computations. 
On the other hand, we will only need the following rule for the point-particle part of the action:\footnote{There are no direct couplings with $H_{\mu \nu}$ coming from eq.~\eqref{eq:Sppt}  because $H_{\mu \nu}$ vanishes on the worldline.}
\allowdisplaybreaks
\begin{align}
\Source & = i \kappa\frac{M}{2}\int \D t \, \e^{- i k\cdot x(t)} v^\mu v^\nu \, . 
\label{eq:FRSmatter}
\end{align}

\section{Matching with the point particle}
\label{sec:match}
In this section we will compute the linear and quadratic response to an external tidal field in the EFT approach outlined above and match it to the full GR calculation of sec.~\ref{sec:QuadraticSolutions}. First, we will derive the tidal field $H_{\mu\nu}$ at both first and second order, expressing it in a form that is suitable for the calculation of Feynman diagrams in the EFT approach.

%%%%%%%%%%%%%%%%%%%%%%%%%%%%%%%%
\subsection{The external field}
\label{sec:TidalH}
%%%%%%%%%%%%%%%%%%%%%%%%%%%%%%%%

As mentioned above, by construction $H_{\mu\nu}$ satisfies the vacuum Einstein  equations and is regular at the location of the point particle, which we place for convenience at the origin of the coordinate system. We shall parameterize it as described in sec.~\ref{sec:RWpert}, and split it into even and odd components,
$ H_{\mu\nu}(x) = H^+_{\mu\nu}(x) + H^-_{\mu\nu}(x) $.
We can then apply the decomposition from eqs.~\eqref{RWgeven} and \eqref{RWgodd}. Specifically, we will denote the components of $H^\pm_{\mu\nu}$ with a bar and  write
\begin{align}
H_{\mu\nu}^{\rm +} &=
\begin{pmatrix}
f(r) \bar H_0 & \bar  H_1 & \nabla_A \bar  {\mathcal H}_0 \\
* & f^{-1}(r) \bar H_2 & \nabla_A \bar {\mathcal H}_1 \\
* \,\,\,\, & * & \,\,\,\, r^2 { \bar  K}
                                                   \gamma_{AB} + 
                                                   r^2(\nabla_A \nabla_B - \frac{1}{2} \gamma_{AB}) \bar  G
\end{pmatrix} \;, \label{RWgevenEFT}\\
H_{\mu\nu}^{\rm -} &=
\begin{pmatrix}
0 & 0 & - \epsilon_A{}^C \nabla_C \bar   h_0\\
* & 0 & - \epsilon_A{}^C \nabla_C \bar  h_1\\
* \,\,\,\, &  * & \,\,\,\, -{1\over 2} (\epsilon_A {}^C
                                           \nabla_C \nabla_B +
                                        \epsilon_B {}^C \nabla_C
                                           \nabla_A ) \bar  h_2
\end{pmatrix} \;. \label{RWgoddEFT}
\end{align}

Moreover, since the gauge fixing in eq.~\eqref{eq:BGGF} ensures that the final result is invariant under diffeomorphisms of $\bar{g}_{\mu\nu} = \eta_{\mu\nu} + H_{\mu\nu}$, we can choose $H_{\mu\nu}$ in any gauge we prefer. To be closer to the GR calculation, we will work in the RW gauge (see eq.~\eqref{RWgauge}). However, since all calculations using Feynman diagrams are performed in Cartesian coordinates, it is useful to first rewrite \eqref{RWgevenEFT} and \eqref{RWgoddEFT} in a Lorentz-covariant form, which will facilitate their conversion to Cartesian coordinates when required.
To  this end, it is convenient to define the  spatial radial vector  
\be
r^\mu = \mathbb{P}^{\mu}{}_{ \nu} \, x^\nu  \equiv (\delta^{\mu}{}_{\nu} + v^\mu v_{\nu}) \, x^\nu \;,
\ee
and the unit vector $\hat r^\mu = r^\mu / r$, $\hat r^\mu \hat r_\mu = 1$, where we remind the reader that in this section indices are raised and lowered using the flat metric.
As derived in app.~\ref{App:SphToCart},    eqs.~\eqref{RWgevenEFT} and \eqref{RWgoddEFT} can be written in Lorentz-covariant form as
\begin{align}
	H^+_{\mu\nu} & = \big( \bar{H}_0 + \bar{K} \big) v_\mu v_\nu + 2\bar{H}_1 v_{(\mu}\hat{r}_{\nu)}
		+\big( \bar{H}_2 - \bar{K} \big) \hat{r}_\mu\hat{r}_\nu 
		+\bar{K} \eta_{\mu\nu} \, ,
		\label{eq:Hegenc} \\
	H^-_{\mu \nu} & = - 2  v^\alpha\varepsilon_{ \alpha  \rho\sigma(\mu} v_{\nu)} \hat r^\rho \partial^\sigma \bar h_0  + 2  v^\alpha\varepsilon_{ \alpha  \rho\sigma(\mu}  \hat{r}_{\nu)} \hat r^\rho\partial^\sigma \bar h_1   \, ,
	\label{eq:Hogenc}
\end{align}
where $\varepsilon_{ \alpha  \rho\sigma\mu}$ is the flat-space Levi-Civita symbol ($\varepsilon_{0123} = +1$).

Next, we insert these expressions into the vacuum Einstein equations, expand them perturbatively in the amplitude of the tidal field, and solve for $H_{\mu \nu}$ order by order, ensuring regularity at the origin.  
To take advantage of the spherical symmetry of the background when solving the Einstein equations, it is convenient to decompose into spherical harmonics, as in sec.~\ref{sec:RWpert}. In order to work in Cartesian coordinates, we shall then express the spherical harmonics in terms of constant symmetric trace-free (STF)  tensors $\cY^{ m}_{\alpha_1 \dots \alpha_\ell}$~(see e.g., \cite{Thorne:1980ru,Maggiore:2007ulw}), as
\begin{equation}
 Y_\ell^m(\theta, \phi) =   \cY^{ m}_{\alpha_1 \dots \alpha_\ell}  \hat r^{\alpha_1} \cdots  \hat  r^{\alpha_\ell} \, .
	\label{eq:STFSphH}
\end{equation}
Here the STF tensors $\cY^{ m}_{\alpha_1 \dots \alpha_\ell}$ have covariant indices, but they are purely spatial, i.e., $\mathbb{P}_\beta{}^{\alpha_i} \cY^m_{\alpha_1 \ldots \alpha_i \ldots \alpha_\ell} = \cY^m_{\alpha_1 \ldots \beta \ldots \alpha_\ell}$, $\forall i \in \{1, \dots, \ell\}$. 
This allows us to rewrite $H^\pm_{\mu \nu}$ in Cartesian coordinates as 
\begin{equation}
	H^\pm_{\mu\nu} = \sum_{\ell} \mathcal{H}^\pm_{\mu\nu |   \alpha_1 \dots \alpha_\ell} x^{\alpha_1} \cdots x^{\alpha_\ell}	\, ,
	\label{eq:TidalGeneric}
\end{equation}
where we introduced the \textit{tidal tensors} $\mathcal{H}^\pm_{\mu\nu | \alpha_1 \dots \alpha_\ell}$, which are purely spatial constant tensors. 

To illustrate the procedure concretely, let us  expand the tidal tensor perturbatively, as $ \mathcal{H}^\pm_{\mu\nu |    \alpha_1 \dots \alpha_\ell} =  {}\ord{1}\!\mathcal{H}^\pm_{\mu\nu |  \alpha_1 \dots \alpha_k}  + {}\ord{2}\! \mathcal{H}^\pm_{\mu\nu |    \alpha_1 \dots \alpha_\ell}  + \ldots $, and  derive the  quadratic tidal tensor sourced by two linear  quadrupoles. This is what is required  to match the GR calculation performed in sec.~\ref{sec:sol}.  Solving the linear vacuum Einstein equations for $\ell = 2$ and requiring regularity at the origin, we find the even and odd linear tidal fields in RW gauge,
\begin{equation}
	{}\ord{1}\!H^\pm_{\mu\nu}(x) = {{}\ord{1}\!\mathcal{H}^\pm_{\mu\nu | \alpha \beta}}   x^\alpha x^\beta \, ,
\end{equation}
with
\begin{subequations}
\begin{align}
	{{}\ord{1}\!\mathcal{H}^+_{\mu\nu | \alpha \beta}}  &= \sum_m \frac{\mathcal{E}_{2m}^+}{r_s^2} (\eta_{\mu\nu} + 2v_\mu v_\nu)\cY^m_{\alpha\beta} \, ,  \\ 
	{{}\ord{1}\!\mathcal{H}^-_{\mu\nu | \alpha \beta}}   &= - 4 \sum_m \frac{\mathcal{E}_{2m}^-}{r_s^2} v^\tau v_{(\mu}\varepsilon_{ \nu) \alpha \tau  }{}^\sigma \cY^m_{\sigma\beta} \, ,
\end{align}	
\label{eq:TLO}%
\end{subequations}
where $\mathcal{E}^\pm_{\ell m}$ are the amplitudes of the even/odd tidal fields, chosen so to match the   $ r_s/r \ll 1$ limit of the full GR tidal field solution, see eqs.~\eqref{evenLinearQuadrupole} and \eqref{oddLinearQuadrupole}.

We now use these linear solutions to solve the vacuum Einstein equations at second order. 
As discussed in sec.~\ref{sec:solutionsQuadrpoleTimesQuadrupole}, given a $\ell = 2$ mode at linear order, the quadratic  solution is given by a superposition of $\ell=0, 2, 4$ multipoles.
Solving the vacuum Einstein equations at second order and imposing regularity, we find
\begin{equation}
	{}\ord{2}\!H^\pm_{\mu\nu} = {}\ord{2}\!\mathcal{H}^{\pm}_{\mu\nu| \alpha_1 \alpha_2 \alpha_3  \alpha_4} x^{\alpha_1} x^{\alpha_2}  x^{\alpha_3}  x^{\alpha_4}   \;,
	\label{eq:TNLOgeneric}
\end{equation}
where the tidal tensor can be divided into monopole, quadrupole and hexadecapole parts, i.e.,
\be
\ord{2}\!{\mathcal{H}^{\pm}_{\mu\nu| \alpha_1 \alpha_2 \alpha_3  \alpha_4}} = \ord{2}\!{\mathcal{H}^{\pm,\, \ell=0}_{\mu\nu| \alpha_1 \alpha_2 \alpha_3  \alpha_4}} + \ord{2}\!{\mathcal{H}^{\pm,\, \ell=2}_{\mu\nu| \alpha_1 \alpha_2 \alpha_3  \alpha_4}} + \ord{2}\!{\mathcal{H}^{\pm, \, \ell=4}_{\mu\nu| \alpha_1 \alpha_2 \alpha_3  \alpha_4}} \;.
\label{eqs:TTensorNLOpre}
\ee
Their explicit forms are given by  
\begin{subequations}
\begin{align}
	\ord{2}\!{\mathcal{H}^{+, \, \ell=0}_{\mu\nu| \alpha_1 \alpha_2 \alpha_3  \alpha_4}} & = \frac{1}{2\sqrt{\pi}}\sum_{m_1, m_2} {}^0\!{I}_{0 2 2}^{0 m_1 m_2}\sum_{p = \pm} \frac{\mathcal{E}_{2m_1}^{p}  \mathcal{E}_{2 m_2}^{p}}{r_s^4} \Big[ \Big( c_\eta\ord{0, p} \eta_{\mu\nu} + c_{v}\ord{0, p} v_\mu v_\nu \Big) \mathbb{P}_{\alpha_1 \alpha_2} + c_r\ord{0, p}  \mathbb{P}_{\alpha_1 \mu} \mathbb{P}_{\alpha_2 \nu} \Big] \mathbb{P}_{\alpha_3 \alpha_4}  \, ,\\
	\ord{2}\!{\mathcal{H}^{+, \,\ell=2}_{\mu\nu| \alpha_1 \alpha_2 \alpha_3  \alpha_4}} & =\sum_{m_1, m_2} {}^0\!{I}_{2 2 2}^{(m_1+m_2) m_1 m_2} \sum_{p = \pm}  \frac{\mathcal{E}_{2m_1}^{p}\mathcal{E}_{2m_2}^{p}}{r_s^4}  \Big[ \Big( c_\eta\ord{2,p} \eta_{\mu\nu} + c_{v}\ord{2,p} v_\mu v_\nu \Big) \mathbb{P}_{\alpha_1 \alpha_2} + c_r\ord{2,p}  \mathbb{P}_{\alpha_1 \mu} \mathbb{P}_{\alpha_2 \nu} \Big] \cY_{\alpha_3 \alpha_4}^{m_1+m_2} \, ,\\
	\ord{2}\!{\mathcal{H}^{-, \, \ell=2}_{\mu\nu| \alpha_1 \alpha_2 \alpha_3  \alpha_4}} & =  -4 \sum_{ m_1, m_2}{}^0\!{I}_{2 2 2}^{(m_1+m_2) m_1 m_2} \frac{\mathcal{E}_{2m_1}^+ \mathcal{E}_{2m_2}^-}{r_s^4}   v^\alpha v_{(\mu}\varepsilon_{\nu) \alpha \sigma \alpha_3} d\ord{2}  \mathbb{P}_{\alpha_1\alpha_2}\cY_{\sigma \alpha_4}^{m_1+m_2}  \, ,\\	
	\ord{2}\!{\mathcal{H}^{+, \, \ell=4}_{\mu\nu| \alpha_1 \alpha_2 \alpha_3  \alpha_4}} & =  \sum_{m_1, m_2} {}^0\!{I}_{4 2 2}^{(m_1+m_2) m_1 m_2}\sum_{p = \pm}  \frac{\mathcal{E}_{2m_1}^{p}\mathcal{E}_{2m_2}^{p}}{r_s^4}   \Big[ c_\eta\ord{4,p} \eta_{\mu\nu} + c_{v}\ord{4,p} v_\mu v_\nu \Big] \cY_{\alpha_1 \alpha_2 \alpha_3 \alpha_4}^{m_1+m_2} \, ,\\
	\ord{2}\!{\mathcal{H}^{-, \, \ell=4}_{\mu\nu| \alpha_1 \alpha_2 \alpha_3  \alpha_4}} & = -  8 \sum_{m_1, m_2}{}^0\!{I}_{4 2 2}^{(m_1+m_2) m_1 m_2}   \frac{\mathcal{E}_{2m_1}^+ \mathcal{E}_{2m_2}^-}{r_s^4}  v^\alpha v_{(\mu}\varepsilon_{\nu) \alpha \sigma \alpha_1} d\ord{4}  \cY_{\sigma \alpha_2 \alpha_3 \alpha_4}^{m_1+m_2} \, .
\end{align}
\label{eqs:TTensorNLO}%
\end{subequations}
The  numerical coefficients $c\ord{\ell, p}$ and $d\ord{\ell}$ appearing in these expressions are given in app.~\ref{app:source}. 

In order to perform the computation in the next sections, we need the tidal field in Fourier space. 
From eq.~\eqref{eq:TidalGeneric}, the coordinate dependence of the tidal field is given by the products $ x^{\alpha_1} \cdots x^{\alpha_\ell}$, which are contracted with spacelike tensors. Therefore, all we need to know is the Fourier transform of $ r^{\alpha_1} \cdots r^{\alpha_\ell}$, which reads\footnote{To derive this relation, note that, in the coordinate system where $r^\alpha = (0, x^i)$, we have
\begin{equation}
	 \int \! \D^4 x \,  \e^{-i q \cdot x} x^{j_1} \dots x^{j_\ell} = \ddl(q^0)  i^\ell \frac{\partial^\ell}{\partial q_{j_1} \dots \partial q_{j_\ell}} \ddl{}\ord{3}(\bm{q}) \, .
	 \label{footnote}
\end{equation}
When the right-hand side acts on a function of $q^\mu$, one gets
\begin{equation}
	\int_q \ddl(q^0) i^\ell  \frac{\partial^\ell}{\partial q_{i_1} \dots \partial q_{i_\ell}} \ddl{}\ord{3}(\bm{q}) f(q) =  (- i)^\ell \frac{\partial^\ell f(q)}{\partial q_{i_1} \dots \partial q_{i_\ell}} \bigg|_{q^\mu=0} \, .
\end{equation}
Therefore, using this relation and upon covariantization of eq.~\eqref{footnote}  one finds eq.~\eqref{eq:Fspacegen}.}
\begin{equation}
 \int \! \D^4 x \, \e^{-i q \cdot x} r^{\mu_1} \cdots r^{\mu_\ell} =  (- i) ^\ell \ddl{}\ord{4}(q)
	\mathbb{P}^{\mu_1 \nu_1} \cdots \mathbb{P}^{\mu_\ell \nu_\ell}
	\frac{\partial^\ell }{\partial q^{\nu_1} \dots \partial q^{\nu_\ell}} \, .
	\label{eq:Fspacegen}
\end{equation}
Introducing the shorthand notation
\begin{equation}
	\partial\ord{q}_{\mu_1 {\dots} \mu_\ell} \equiv ( - i )^\ell \frac{\partial^\ell }{\partial q^{\nu_1} \dots \partial q^{\nu_\ell}} \, ,
\end{equation}
the tidal field in Fourier space at linear and quadratic order  is  given by
\begin{align}
	{}\ord{1}\!H^{\pm}_{\mu\nu}(q) & = \ddl{}\ord{4}(q) \, {}\ord{1}\!\mathcal{H}^\pm_{\mu\nu|}{}^{ \alpha_1 \alpha_2} \,   \partial\ord{q}_{\alpha_1 \alpha_2} \, ,
	\label{eq:HLOf} \\
	{}\ord{2}\!H^{\pm}_{\mu\nu}(q) & = \ddl{}\ord{4}(q) \,  {}\ord{2}\!{\mathcal{H}^{\pm}_{\mu\nu|}{}^{ \alpha_1 \alpha_2 \alpha_3 \alpha_4}}    \,  \partial\ord{q}_{\alpha_1 \alpha_2 \alpha_3 \alpha_4 } 
	\label{eq:HNLOf} \;,
\end{align}
where ${}\ord{1}\!\mathcal{H}^\pm_{\mu\nu|}{}^{ \alpha_1 \alpha_2} \equiv \eta^{\alpha_ 1 \beta_1} \eta^{\alpha_ 1 \beta_1} {}\ord{1}\!\mathcal{H}^\pm_{\mu\nu| \beta_1 \beta_2}$ (see eq.~\eqref{eq:TLO}) and ${}\ord{2}\!\mathcal{H}^\pm_{\mu\nu|}{}^{ \alpha_1 \ldots \alpha_4} \equiv \eta^{\alpha_ 1 \beta_1}  \cdots \eta^{\alpha_ 2 \beta_2} {}\ord{2}\!\mathcal{H}^\pm_{\mu\nu|  \beta_1 \ldots \beta_4}$ (see eqs.~\eqref{eqs:TTensorNLOpre} and \eqref{eqs:TTensorNLO}).

\subsection{Quadratic response}
\label{sec:hEFT}

We are now ready to compute the  one-point function due to the interaction between the external tidal field and the point particle, $h_{\mu \nu} (x) \equiv \langle h_{\mu\nu} (x) \rangle$. 
This can be expanded as 
\be
h_{\mu \nu}  = \sum_{n=0}^\infty  {}^{(n)}\! h_{\mu\nu} \;, 
\ee
where $ {}^{(n)}\! h_{\mu\nu} $ is the response at $n$th-order  in the tidal field amplitude $\mathcal{E}$. Diagrammatically, $ {}^{(n)}\! h_{\mu\nu} $ can be represented as
\begin{align}
	 {}^{(n)} \! h_{\mu\nu} (x) = \raisebox{15pt}{\genhUN} + \raisebox{15pt}{\genhT}   \, ,
	\label{eq:hrespGen}
\end{align}
where the blob stands for all possible interaction vertices between $H_{\mu \nu}$ and $h_{\mu \nu}$, at a specific order in perturbation theory.
Thus, this diagram distinguishes bewteen two types of   nonlinearities. Those that arise from higher-order vertices in the bulk action (see eqs.~\eqref{eq:FRSbulk3} and \eqref{eq:FRSbulk4}), corresponding to ${\cal O}(r_s)$ corrections to the external tidal field, which are represented by the first diagram on the right-hand side. Additionally, there are nonlinearities that arise as a genuine vertex response of the worldline, receiving $r_s$-corrections from bulk interactions. They are represented by the second diagram on the right-hand side.

In the following, we will show that the first diagram on the right-hand side alone reproduces the full GR calculation at leading order in $ r_s/r $, implying that the response nonlinearities  must be absent, i.e., $\lambda=0$. The matching can be extended to higher orders in $ r_s $ by considering diagrams with multiple mass insertions on the worldline and, eventually, reconstructing the full tidal field profile computed in GR  in Sec.~\ref{sec:sol}. However, these corrections are not needed to derive our conclusions; see e.g.~\cite{Ivanov:2022hlo} for an extensive discussion  in the linear case.  Nonetheless, as a sanity check of the validity of our method at higher-order in $r_s$, we have verified that in the simplest case of $n=0 $, i.e., in the absence of the tidal field, multiple mass insertions  reconstruct the Schwarzschild metric perturbatively up to $ \mathcal{O}(r_s^2) $. The calculation is reported in details in app.~\ref{app:SchwM}. 
% 

%%%%%%%%%%%%%%%%%%%%%%%%%%
\begin{figure}[t]
\centering
\includegraphics[scale=1]{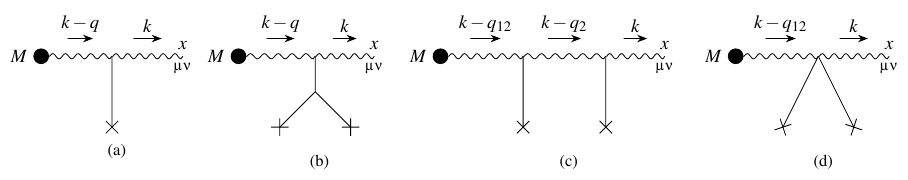}
\caption{Feynman diagrams needed for the computation of $h_{\mu\nu}$. Diagram (a) yields the order-$r_s$ correction to the linear tidal field solution. Diagrams (b), (c) and (d) represent instead order-$r_s$ corrections to the tidal source at second order in the external field amplitude.
}
\label{fig:diagsrsE}
\end{figure}
%%%%%%%%%%%%%%%%%%%%%%%%%%

We now perform the computation of $h_{\mu\nu} (x)$ up to linear order in $r_s$ and up to quadratic order in the amplitude of the tidal field $\mathcal{E}$, i.e., 
\be
h_{\mu\nu} (x) = {}^{(0)} \! h_{\mu\nu} (x) + {}^{(1)} \! h_{\mu\nu} (x) + {}^{(2)} \! h_{\mu\nu} (x)  + {\cal O} (\mathcal{E}^3) \;, 
\ee
where ${}^{(0)} \! h_{\mu\nu}(x)$ is the Schwarzschild metric expanded at  ${\cal O} (r_s)$; see in App.~\ref{app:SchwM}.  
The diagrams to compute are  represented in fig.~\ref{fig:diagsrsE}. The only diagram for the computation of ${}^{(1)} \! h_{\mu\nu}(x)$ is fig.~\ref{fig:diagsrsE}(a), while to compute ${}^{(2)} \! h_{\mu\nu}(x) $ we need \ref{fig:diagsrsE}(b,c,d).

Taking the linear tidal field to be an $\ell=2$ mode, we have
\begin{align}
	{}^{(1)} \! h_{\mu\nu}^\pm (x) & = \frac{\kappa M}{2} \int_k\frac{\ddl(k^0)\e^{i k \cdot x}}{k^2} \int_q \ddl{}\ord{4}(q) {}\ord{1}\!\mathcal{H}^{\pm
	}_{\rho\sigma,}{}^{\alpha_1 \alpha_2} \, \partial\ord{q}_{\alpha_1 \alpha_2} 
	\bigg(
	\frac{\mathcal{A}_{\mu\nu}\ord{1a}{}^{\rho\sigma}(k, q) }{(k-q)^2}
	 \bigg) \, ,
	 \label{eq:hb_int}
\end{align}
where 
\begin{equation}
	\mathcal{A}_{\mu\nu}\ord{1a}{}^{\rho\sigma}(k, q) \equiv 
		v^{\beta_1}v^{\delta_1}
		P_{\beta_1\delta_1 \beta_2 \delta_2} V^{\rho \sigma \beta_2\delta_2 \beta_3 \delta_3}P_{\beta_3 \delta_3 \mu\nu} \, .
\end{equation}
Performing the contraction and integrations using the  explicit form of the tidal tensors  in  eq.~\eqref{eq:TLO}, we eventually obtain  the even and odd part of the linear one-point function, respectively,
\begin{subequations}
\begin{align}
	\kappa {}^{(1)} \! h_{\mu\nu}^+ (x) &  = \frac{r}{r_s} 
	\sum_m \mathcal{E}_{2m}^+  \left(2 \cY^m_{\mu\nu} + \eta_{\mu\nu} \cY^m_{\rho\sigma}\hat{r}^\rho \hat{r}^\sigma \right) \, ,
	\label{eq:resp0even} 
	\\
	\kappa {}^{(1)} \! h_{\mu\nu}^- (x) & =  - 
	2 \frac{r}{r_s} \sum_m \mathcal{E}_{2m}^-   v^\alpha
	v_{(\mu }\varepsilon_{\nu) \rho \alpha}{}^{  \sigma}\cY^m_{\sigma \tau} \hat{r}^\rho\hat{r}^\tau \, .
	\label{eq:resp0odd} 
\end{align}
	\label{eq:resp0}%
\end{subequations}

We can now   compute ${}^{(2)} \! h_{\mu\nu} (x)$ from the last three diagrams  in fig.~\ref{fig:diagsrsE}. They can be recasted as 
\begin{align}
	{}^{(2)} \! h_{\mu\nu} (x)_{1b} & = \frac{\kappa M}{2} \int_k \frac{\ddl(k^0)\e^{i k \cdot x}}{k^2} \int_{q} \ddl{}\ord{4}(q) \sum_{p = \pm} {}\ord{2}\!\mathcal{H}^{p }_{ \rho_1 \sigma_1|}{}^{\alpha_1\alpha_2\alpha_3\alpha_4} \partial\ord{q}_{\alpha_1\alpha_2\alpha_3\alpha_4} \bigg( 
	\frac{\mathcal{A}_{\mu\nu}\ord{1a}{}^{ \rho_1 \sigma_1}(k, q)}{(k-q)^2}
	 \bigg) \, , \label{eq:hNLOb}\\
	 {}^{(2)} \! h_{\mu\nu} (x)_{1 {\rm x}} & = \frac{\kappa M}{2} \int_k \frac{\ddl(k^0)\e^{i k \cdot x}}{k^2} \int_{q_1, q_2} \hspace{-0.4cm} \ddl{}\ord{4}(q_1)\ddl{}\ord{4}(q_2) \sum_{p_1 = \pm} \sum_{p_2 = \pm}
	 {}\ord{1}\!\mathcal{H}^{p_1 }_{ \rho_1 \sigma_1|}{}^{\alpha_1\alpha_2}\,  {}\ord{1}\!\mathcal{H}^{p_2 }_{ \rho_2 \sigma_2|}{}^{\alpha_3\alpha_4} \partial\ord{q_1}_{\alpha_1\alpha_2} \partial\ord{q_2}_{\alpha_3\alpha_4} \notag \\
	 & \qquad \qquad  \times \bigg( 
	\frac{\mathcal{A}_{\mu\nu}\ord{2 {\rm x}}{}^{ \rho_1 \sigma_1\rho_2 \sigma_2}(k, q_1, q_2) }{(k-q_{12})^2}
	 \bigg)   \;, \qquad  {\rm x} \in \{c, d \} \;,
	 \label{eq:hNLOcd}
\end{align}
where  $q_{12} \equiv q_1 + q_2$. Clearly,  fig.~\ref{fig:diagsrsE}(b) has the same form  of fig.~\ref{fig:diagsrsE}(a) with the linear tidal field replaced by the quadratic solution. On the other hand, figs.~\ref{fig:diagsrsE}(c) and (d) are new topologies arising at quadratic order, with
\begin{align}
	\mathcal{A}_{\mu\nu}\ord{1c}{}^{ \rho_1 \sigma_1\rho_2 \sigma_2}(k, q_1, q_2) & \equiv -v^{\beta_1}v^{\delta_1} P_{\beta_1\delta_1 \beta_2\delta_2}
		V_3^{\rho_1\sigma_1\beta_2\delta_2\beta_3\delta_3} 
		\frac{P_{\beta_3\delta_3\beta_4\delta_4}}{(k-q_2)^2}
		V_3^{\rho_2\sigma_2 \beta_4\delta_4 \beta_5\delta_5} P_{\beta_5\delta_5\mu\nu} \, ,\\
	\mathcal{A}_{\mu\nu}\ord{1d}{}^{ \rho_1 \sigma_1\rho_2 \sigma_2}(k, q_1, q_2) & \equiv \frac{v^{\beta_1}v^{\delta_1}}{2} P_{\beta_1\delta_1 \beta_2\delta_2}V_4^{\rho_1\sigma_1\rho_2\sigma_2\beta_2\delta_2\beta_3\delta_3} P_{\beta_3\delta_3\mu\nu} \, .
\end{align}
Notice that there is an extra factor of 1/2 in the last amplitude from the symmetry factor of the diagram.

We  split once again the quadratic one-point function into even and odd contributions, ${}^{(2)}\! h_{\mu \nu} (x) = {}^{(2)}\! h^{+}_{\mu \nu} (x) + {}^{(2)}\! h^{-}_{\mu \nu} (x)$. The even contribution of eq.~\eqref{eq:hNLOb} comes from inserting the quadratic even tidal tensor, eqs.~\eqref{eq:TNLOgeneric} and \eqref{eqs:TTensorNLO}, while
the even contribution
 of eq.~\eqref{eq:hNLOcd} comes from replacing the tidal tensor by either two even linear modes ${}\ord{1}\!\mathcal{H}^{+}_{ \rho \sigma|}{}^{\alpha\beta}$ or two linear odd modes ${}\ord{1}\!\mathcal{H}^{-}_{ \rho \sigma|}{}^{\alpha\beta}$,  see eq~\eqref{eq:TLO}. Therefore, for each diagram,  the even one-point function is the sum of a part proportional to $(\mathcal{E}^+)^2$ (the {\em even} $\times$ {\em even} ``response''), denoted by ${}^{(2)}\!  h_{\mu\nu}^{(++)}(x)_{1 {\rm x}}$,  and a part proportional to $(\mathcal{E}^-)^2$ (the {\em odd} $\times$ {\em odd} ``response''), denoted by ${}^{(2)}\!  h_{\mu\nu}^{(--)}(x)_{1 {\rm x}}$.
 Here we report explicitly only the $tt$ component, as this is the one relevant for the matching. After using the identities involving the STF tensors $\cY_{ij}^m$  in app.~\ref{App:STFrel} and after summing over the three diagrams, one finds
\be
\begin{split}
	\kappa {}^{(2)}\! &h^{+}_{tt} (x)    =   \bigg(\frac{r}{r_s}\bigg)^3  \times \\
	   \sum_{m_1, m_2} \bigg[  \ & \mathcal{E}_{2m_1}^{+ }\mathcal{E}_{2m_2}^{+ }   
	    \bigg(   
	\frac{7}{4 \sqrt{\pi}} {}^0\!{I}_{022}^{0m_1m_2}     +
	\frac{7}{4}{}^0\!{I}_{222}^{(m_1+m_2) m_1 m_2} \cY^{m_1+m_2}_{ij} \hat{r}^i\hat{r}^j 
	-  \frac{35}{18} {}^0\!{I}_{422}^{(m_1+m_2) m_1 m_2} \cY^{m_1+m_2}_{ijkl}\hat{r}^i\hat{r}^j \hat{r}^k\hat{r}^l \bigg) \\
	 + \ & \mathcal{E}_{2m_1}^{- }\mathcal{E}_{2m_2}^{- }    \bigg(   
	\frac{3}{\sqrt{\pi}} {}^0\!{I}_{022}^{0m_1m_2}     
	- \frac34 {}^0\!{I}_{222}^{(m_1+m_2) m_1 m_2} \cY^{m_1+m_2}_{ij} \hat{r}^i\hat{r}^j 
	- \frac59 {}^0\!{I}_{422}^{(m_1+m_2) m_1 m_2} \cY^{m_1+m_2}_{ijkl}\hat{r}^i\hat{r}^j \hat{r}^k\hat{r}^l \bigg)  \bigg]\, .
	\label{eq:hNLObee} \\
\end{split}
\ee
 
 The computation of the odd modes of $h_{\mu\nu}$ is analogous. We use the odd quadratic tidal tensor for the diagram in fig.~\ref{fig:diagsrsE}(b) and the product of one even and one odd linear tidal tensors for the diagrams in fig.~\ref{fig:diagsrsE}(c) and (d). Therefore, the odd contribution to the one-point function is always proportional to $\mathcal{E}^+ \mathcal{E}^-$.
We report  the  components $ti$, that are the ones relevant for the matching. They read
\be
\begin{split}
	 \kappa {}^{(2)}\! h^{-}_{ti}(x)  =  -   \bigg(\frac{r}{r_s}\bigg)^3 \sum_{m_1, m_2} 
	\mathcal{E}_{2m_1}^+\mathcal{E}_{2m_2}^-  \varepsilon_{ij}{}^{k} & \bigg(
	 5{}^{0}\! {I}^{(m_1+m_2) m_1 m_2}_{2 2 2} \cY^{m_1+m_2}_{k l}  -\frac{13}{5}  {}^{0}\! {I}^{(m_1+m_2) m_1 m_2}_{4 2 2} \cY^{m_1+m_2}_{k l p q} \hat{r}^p \hat{r}^q
	\bigg) \hat{r}^j \hat{r}^l\, .
	\label{eq:hNLObodd}
\end{split}
\ee

Before proceeding to do the matching, as a consistency check we have  verify that the obtained $h_{\mu\nu}(x)$ satisfies the gauge  condition 
\begin{equation}
	\bar{g}^{\alpha \beta}\left( \bar{\nabla}_\alpha h_{\beta\mu} 
		-\frac{1}{2} \bar{\nabla}_\mu h_{\alpha \beta} \right) = 0 \, ,
		\label{eq:GFId}
\end{equation}
where $\bar{\nabla}_\alpha \bar{g}_{\mu\nu} = 0$, imposed by the gauge-fixing term \eqref{eq:BGGF} in the action. 
Inserting $\bar{g}_{\alpha \beta} = \eta_{\alpha\beta} + H_{\alpha \beta}$ in this equation and working at linear order in the tidal field amplitude, 
this equation becomes
\begin{equation}
	\bigg( \eta^{\alpha \beta} \partial_\alpha {}\ord{1}\!h_{\beta \mu} - \frac{1}{2}\partial_\mu {}\ord{1}\!h \bigg)
	-  {}\ord{1}\!H^{\alpha\beta} \bigg( \partial_\alpha  {}\ord{0}\!h_{\beta\mu} - \frac{1}{2}\partial_\mu  {}\ord{0}\!h_{\alpha \beta} \bigg) 
	+  {}\ord{0}\!h_{\mu\alpha} \bigg( \partial_\beta {}\ord{1}\!H^{\beta\alpha} - \frac{1}{2}\partial^\alpha {}\ord{1}\!H  \bigg)  = {\cal O}(\mathcal{E}^2) \, .
	\label{eq:GFIdLO}
\end{equation}
Since it is linear, this identity must be satisfied independently for the even and odd  perturbations. Indeed, replacing  in this equation the linear even solution, eq.~\eqref{eq:resp0even},   
the first term becomes 
\be
\eta^{\alpha \beta} \partial_\alpha {}\ord{1}\!h^{+}_{\beta \mu} - \frac{1}{2}\partial_\mu {}\ord{1}\!h^+  = \frac{\kappa}{16\pi r_s^2} 
		\sum_m \mathcal{E}_{2m}^+ \cY^m_{\alpha\beta}\hat{r}^\alpha \hat{r}^\beta   \hat{r}_\mu \;,
\ee
which cancels with the second term, while the last term vanishes. For the odd solution, eq.~\eqref{eq:resp0odd}, each of the terms in this equation vanishes individually at ${\cal O}(\mathcal{E})$, so that  the above identity is trivially satisfied.
Proceeding  analogously at  second-order, we have verified that also at that order the one-point function ${}\ord{2}\!h_{\mu\nu}(x)$ satisfies eq.~\eqref{eq:GFId}, which is a non-trivial check for both even and odd solutions.

\subsection{Matching}
\label{sec:gaugetrans}

Here we want to compare the one-point function computed in the previous subsection with the solution derived in sec.~\eqref{sec:solutionsQuadrpoleTimesQuadrupole}. 
However, the former has been derived  in the gauge that satisfies eq.~\eqref{eq:GFId} while the latter has been derived in  RW gauge.
To compare the two, we find it convenient to rewrite the EFT result in RW gauge. 
In the context of the EFT calculation, the field $ h_{\mu\nu} $ is treated as a spin-2 field propagating on the curved background $ \bar{g}_{\mu\nu} = \eta_{\mu\nu} + H_{\mu\nu} $. The linear gauge transformation $ x^\mu \to x^\mu + \xi^\mu $ that converts $ h_{\mu\nu} $ to the RW gauge is given by (see, e.g., \cite{Gleiser:1998rw})
\be
\kappa h^{\rm RW}_{\mu\nu} = \kappa h_{\mu\nu} - \xi^\rho \partial_\rho \bar{g}_{\mu\nu} - 2 \bar{g}_{\rho (\mu} \partial_{\nu)} \xi^\rho + \mathcal{O}(\xi^2),
\label{eq:GTexp}
\ee
where we have included only terms up to linear order in $ \xi $, as higher-order terms contribute only at higher order the $r_s$ expansion.%
\footnote{We need to perform a second-order gauge transformation  in $\mathcal{E}$ but at first order in $r_s$. Since the gauge parameter $\xi^\mu$ is at least of order $r_s$, any second-order term in eq.~\eqref{eq:GTexp} contributes only to higher $r_s$-corrections that we do not compute here. Note that ${}^{(1)}\!\xi$ is one  $r_s$-order  higher than the tidal field, as seen from eqs.~\eqref{eq:TLO} and \eqref{eq:xi1p}.}

For the even sector, we match the $tt$ component of the metric, since the other components are determined by it through Einstein's equations. Additionally, at this order in perturbation theory, the $ti$ components are parity-odd, allowing us to use them to match the odd sector. In the static limit, the transformations for the $tt$ and $ti$ components of the metric simplify as 
\begin{align}
	 \kappa h^{{\rm RW}}_{tt}{}^+ & = \kappa h_{tt}^+ - \xi^k \partial_k H_{tt}^+\, , 
	\label{eq:G00tr} \\
	 \kappa h^{{\rm RW}}_{ti}{}^- & = \kappa h_{ti}^- - \xi^k \partial_k H_{ti}^- - H_{k t}^- \partial_i \xi^k \, ,
	\label{eq:G0itr} 
\end{align}
where to write the right-hand sides we have used that  $H^+_{ti} = 0 = H^-_{tt}$, and that $\xi_t$  is purely even. Moreover, note that $\xi^k$ appearing in these two equations must be even because $\xi^k \partial_k H_{tt}^+$  and $\xi^k \partial_k H_{ti}^-$ must be respectively even and odd. For the gauge transformation at linear order in $\cal E$ we need $\xi^\mu = {}^{(0)}\! \xi^\mu $, where ${}^{(0)}\!\xi^\mu = \frac{r_s}{2} \hat{r}^\mu$ is the coordinate transformation of the Schwarzschild metric from de Donder to RW gauge. For the quadratic order,  we insert $\xi^\mu = {}^{(0)}\! \xi^\mu + {}^{(1)}\! \xi^\mu$, where
\be
{}^{(1)}\! \xi^\mu = \sum_m\mathcal{E}_{2m}^+\frac{r^2}{2 r_s}  
		\cY^m_{\alpha \beta} \Big(  
		2  \eta^{\beta \mu}  - \hat{r}^\mu   \hat{r}^\beta
		\Big)  \hat{r}^\alpha \, 
		\label{eq:xi1p} 
\ee
is found by imposing that ${}^{(2)}\!h^{{\rm RW}}_{\mu \nu}{}^+$ is diagonal as in the RW gauge.

Finally, since $H_{\mu \nu}$ is already in RW gauge, we can compare   the metric perturbation obtained with the EFT calculation after transforming to RW gauge, i.e., $\delta g^{\rm EFT}_{\mu \nu} =  H_{\mu \nu}  + \kappa h^{\rm RW}_{\mu \nu}$,\footnote{We match only perturbations linear or quadratic in $\cal E$, since the lowest order matches trivially with the Schwarzschild metric.}   with the exact metric perturbation obtained in the full GR calculation, converted in Cartesian coordinates. The latter reads  
\begin{align}
	 \delta g_{tt}^+ ({\bm x})& =   \sum_m f(r) \left[   \Hs_0^{0 m} (r) + \left( \Hf_0^{2 m} (r) + \Hs_0^{2 m} (r) \right) {\cal Y}^m_{\mu \nu} \hat r^\mu  \hat r^\nu + \Hs_0^{4 m} (r) {\cal Y}^m_{\mu \nu \alpha \beta} \hat r^\mu  \hat r^\nu \hat r^\alpha  \hat r^\beta \right]  \;, \\
	 \delta g_{ti}^- ({\bm x})& = -\sum_m \varepsilon_{ij}{}^{k} \hat r^j  \partial_k \left[ \left( \hf_0^{2 m} (r) + \hs_0^{2 m} (r) \right) {\cal Y}^m_{\mu \nu} \hat r^\mu  \hat r^\nu  + \hs_0^{4 m} (r) {\cal Y}^m_{\mu \nu \alpha \beta} \hat r^\mu  \hat r^\nu \hat r^\alpha  \hat r^\beta \right] \, ,
	\label{eq:G0itr-2} 
\end{align}
respectively for the $tt$ and $ti$ components.
We remind that the GR solutions are given in sec.~\ref{sec:solutionsQuadrpoleTimesQuadrupole}, see eqs.~\eqref{evenLinearQuadrupole} and \eqref{oddLinearQuadrupole} for the linear solutions and  eqs.~\eqref{eq:H02m}, \eqref{eq:h02m}, \eqref{eq:H04m}, \eqref{eq:h04m} and \eqref{eq:H00m}   for the quadratic solutions. 

%%%%%%%%%%%%%%%%%%%%%%%%%%
\begin{figure}[t]
\centering
\includegraphics[scale=1]{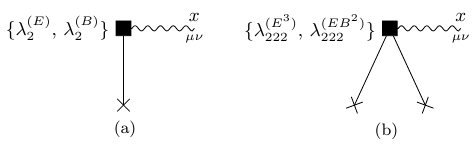}
\caption{Feynman diagrams responsible for the (a) linear ad (b) nonlinear tidal deformation. In principle, one could have terms proportional to $\lambda_2^{(E,B)}$ in the topology represented in figure (b). These are excluded because the matching at leading order in the tidal field implies $\lambda_2^{(E,B)} = 0$. See also sec.~\ref{sec:matchingL}.}
\label{fig:diagTidal}
\end{figure}
%%%%%%%%%%%%%%%%%%%%%%%%%%
As anticipated, we find that $\delta g^{\rm EFT}_{\mu \nu}$ matches the full GR solution above at leading and next-to-leading order in $r_s/r$, without the inclusion of any of the tidal operators in the action \eqref{eq:ST}. This implies that the contributions coming from the diagrams depicted in fig.~\ref{fig:diagTidal} must  vanish. As a result,
\begin{align}
	\lambda_2\ord{E} = 	\lambda_2\ord{B} = 0 \, ,
	\qquad  \lambda_{2 2 2}^{(E^3)} = \lambda_{2 2 2}^{(E B^2)} = 0 \, .
	\label{zeroqLNsrs}
\end{align}
We will extend this result to all orders in derivatives in the next section.

\section{Higher multipoles}
\label{sec:HM}

In sec.~\ref{subsec:linearsolutions}, we computed the static  solutions for the metric perturbations induced at second order  by two quadrupolar static fields. In particular, we have checked explicitly that the solutions for $H_0$ and $h_0$ (see eqs.~\eqref{eq:H02m} and \eqref{eq:h02m} for the induced quadrupole, and eqs.~\eqref{eq:H04m}--\eqref{eq:h04m} for the induced hexadecapole) take the form of simple finite polynomials with only positive powers of $r/r_s$. 
This was crucial in the matching performed in sec.~\ref{sec:match} to conclude that the quadratic Love number couplings vanish at lowest order in the number of spatial derivatives in the worldline EFT (i.e., eq.~\eqref{zeroqLNsrs}).

Our goal in this section is to show that this conclusion remains true to all orders in the derivative expansion. First of all, we will show in full generality that, for generic multipoles, the static solutions induced at second order in full GR do not contain any decaying falloff of the form $1/r^n$, with $n>1$, when expanded at large distances from the source. Then, in sec.~\ref{sec:matchingL}, will  use this result to perform the matching at generic order in the EFT and show that all quadratic Love number couplings vanish for Schwarzschild black holes in general relativity. 

\subsection{General polynomial  form of second-order static solutions}
\label{sec:1order}

We start by discussing the full GR solutions for generic multipoles. For the sake of the presentation, we will consider the even and odd induced solutions separately.

\subsubsection{Even sector} 
\label{sec:evenmatchall}

Let us start by considering eq.~\eqref{eqH0_sec} for the second-order even mode $\Hs_0^{\ell m}(r)$. The source is given in eq.~\eqref{sourceH0}. For  ease of notation, we will conveniently redefine the right-hand side of \eqref{eqH0_sec} simply as $\sum_{\ell_1, m_1, \ell_2, m_2} {\cal S}_{H_0 \ \ell \ell_1  \ell_2}^{m  m_1 m_2} (r) 
\equiv  \mathcal{S}^{\ell m}_{\mathit{H}_0}$, in such a way that \eqref{eqH0_sec} now reads
\begin{equation}
\partial^2_r \, \Hs_0^{ \ell m }  (r) +\frac{2 r-r_s}{r^2 f(r)} \partial_r \Hs_0^{\ell m} (r)-  \frac{ \lambda r^2 f(r)+r_s^2}{r^4 f(r)^2} \; \Hs_0^{\ell m} (r) =  \mathcal{S}^{\ell m}_{\mathit{H}_0}  (r) \;.
\label{eqH0_secmm}
\end{equation}
The solution is given by \eqref{H0_Greensreg},  with $H_{0,1}^\ell$ and  $H_{0,2}^\ell$ provided  in eqs.~\eqref{H01Pell} and \eqref{H02Qell}, i.e.,
\be
\begin{split}
\Hs_0^{ \ell m }  (r)  =  \frac{r_s^2}{ W^\ell_{H_0} }   \bigg[ Q^2_\ell (2r/r_s-1) & \int_{1}^{r/r_s}  {\rm d}y \, y(y-1)   P^2_\ell (2y-1)
\mathcal{S}^{\ell m}_{\mathit{H}_0} (y) 
\\
   -
P^2_\ell (2r/r_s-1) &\int^{r/r_s}  {\rm d}y \,  y(y-1)   Q^2_\ell (2y-1)
 \mathcal{S}^{\ell m}_{\mathit{H}_0} (y) 
\bigg] \;,
\label{H0PQellgeneric}
\end{split}
\ee
where we have redefined the integration variable as $y\equiv r'/r_s$.
The goal is to show that \eqref{H0PQellgeneric} does not contain any $1/r^n$  falloff, with $n>1$,  for generic $\ell$. 

To this end, let us relate $P_\ell^2 (2y -1)$ and $Q_\ell^2 (2y -1)$  in the above expression  to the Legendre functions of the first kind $P_\ell (2y -1)$. Using well known identities,\footnote{We recall the identities
\begin{equation}
P_\ell^2 (x)= (1-x^2 )  \frac{{\mathrm{d}^2} P_\ell(x)}{\mathrm{d} x^2} \, , 
\qquad
Q_\ell^2 (x)= (1-x^2 ) \frac{\mathrm{d}^2 Q_\ell(x)}{\mathrm{d} x^2} \, ,
\label{LegendreQ2}
\end{equation}
\begin{equation}
\label{legendreQ}
Q_{\ell}(x)=\frac{1}{2} P_{\ell}(x) \log \left(\frac{1+x}{x-1}\right)- R_\ell(x) \, ,
\qquad
R_\ell(x) \equiv \sum_{k=1}^{\ell} \frac{1}{k} P_{k-1}(x) P_{\ell-k}(x) \, ,
\end{equation}
\begin{equation}
\label{expression_Q2}
Q_{\ell}^2(x)=\frac{1}{2} P_{\ell}^2(x) \log \left(\frac{1+x}{x-1}\right)+\frac{2x}{1-x^2}P_{\ell}(x) + 2  P'_{\ell}(x)-(1-x^2)  R''_\ell(x) \, .
\end{equation}} 
we can write
\begin{equation}
\label{expression_P2}
 P_\ell^2 (2y-1)=- y(y-1) \partial_y^2  P_\ell (2y-1) \; ,
\end{equation}
\begin{equation}
\label{LegendreQ2_proper_variable}
Q_{\ell}^2(2y-1)=\frac{1}{2} P_{\ell}^2(2y-1) \log \left(\frac{y}{y-1}\right)+ \frac{T_{\ell}(y)}{y(y-1)} \, ,
\end{equation}
where $T_{\ell}(y)$ is a superposition of Legendre polynomials of the first kind and derivatives thereof, 
\be
\begin{split}
T_{\ell}(y) \equiv &- \frac{1}{2} (2 y -1) P_{\ell}(2 y -1) + y (y-1)  \partial_y P_{\ell}(2 y -1) \\
&- (y-1)^2 y^2 \partial_y^2 \left( \sum_{k=1}^{\ell} \frac{1}{k} P_{k-1}(2 y -1) P_{\ell-k}(2 y -1 ) \right) \;.
\end{split}
\label{Tynonpol}
\ee
The complete expression of $T_{\ell}(y)$ is not particularly important in what follows. For the moment, it will be  enough  to observe that $T_{\ell}(y)$ is a finite polynomial with only positive powers  of $y$. In particular, $T_{\ell}(y)$ is generally nonzero at $y=0$ and $y=1$. 
A second important ingredient for our argument is the form of the source $\mathcal{S}^{\ell m}_{\mathit{H}_0}$. Plugging the  linear-order physical solutions for $H_0$ (eq.~\eqref{H0linsol}) and $h_0$ (eq.~\eqref{h0linsol})
in the source~\eqref{sourceH0}, it is easy to see that $\mathcal{S}^{\ell m}_{\mathit{H}_0}$ is a rational function of the form 
\be
\mathcal{S}^{\ell m}_{\mathit{H}_0} (r) = \frac{{\rm polynomial}(r)}{(r-r_s)^2}\;.
\label{eq:sourcepoly}
\ee

Then, by using eq.~\eqref{LegendreQ2_proper_variable} in \eqref{H0PQellgeneric} we can write the second-order solution of eq.~\eqref{eqH0_secmm} as 
\begin{align}
\Hs_0^{ \ell m }  (r)  =   \ & \frac{r_s^2}{ W^\ell_{H_0} } \bigg\{ \frac12 P^2_\ell (2r/r_s-1)  \bigg[ \log \left(\frac{r}{r-r_s}\right) \int_{1}^{r/r_s}  {\rm d}y \, \mathcal{P}^{\ell m}(y)
   -
 \int^{r/r_s}  {\rm d}y \, \log \left(\frac{y}{y-1}\right)  \mathcal{P}^{\ell m}(y)
\bigg] \nonumber
\\
  & \ +
  \frac{T_{\ell}(r)}{r(r-r_s)} \int_{1}^{r/r_s} {\rm d}y \,  \mathcal{P}^{\ell m}(y)
-
  P^2_\ell (2r/r_s-1) \int^{r/r_s}  {\rm d}y \,  T_{\ell}(y)  \mathcal{S}^{\ell m}_{\mathit{H}_0} (y)  \bigg\}
\;,
\label{H04terms}
\end{align}
where eqs.~\eqref{expression_P2} and \eqref{eq:sourcepoly} imply that the quantity
\be
\mathcal{P}^{\ell m}(y) \equiv y(y-1)   P^2_\ell (2y-1) \mathcal{S}^{\ell m}_{\mathit{H}_0} (y) 
\ee
 appearing in the integrands is a finite polynomial.
Let us first focus on the square bracket on the first line of eq.~\eqref{H04terms}. By writing $\mathcal{P}^{\ell m}(y)$ in the second integral as $\mathcal{P}^{\ell m}(y)= \partial_y \int_1^y\D y' \, \mathcal{P}^{\ell m}(y')$ and integrating by parts, one can remove  the term containing $\log (\frac{r}{r-r_s})$ and write the square bracket as $  \int^{r/r_s} \frac{{\rm d} y}{y (y-1)}   \int_1^{y} {\rm d}y' \, \mathcal{P}^{\ell m}(y') = -   \log(r/r_s) \int _0^{1} {\rm d}y \, \mathcal{P}^{\ell m}(y) + \dots$,   where the ellipses denote contributions that are in the form of a finite polynomial in ${r}/{r_s}$. 
Moving to the second line, since $\mathcal{P}^{\ell m}(y)$ is a finite polynomial in $r/r_s$, the integral $ \int_{1}^{r/r_s} {\rm d}y \, \mathcal{P}^{\ell m}(y)$ is a polynomial that vanishes at $r=r_s$, cancelling the prefactor $1/(r-r_s)$.
 As a result, the first term in the second line of \eqref{H04terms}  has the structure  $\frac{1}{r} \times {\rm polynomial}(\frac{r}{r_s})$. Note that, besides positive powers of $r/r_s$, this potentially contains a $ {1}/{r}$ term, given by $ \frac{T_{\ell}(0)}{r r_s } \int _0^{1} {\rm d}y \mathcal{P}^{\ell m}(y)$. 

Finally, let us consider the last integral in the curly brackets of eq.~\eqref{H04terms}. We are going to show that it is a finite polynomial in $r/r_s$. Given the $\frac{1}{(r-r_s)^2}$ factor in the source, see eq.~\eqref{eq:sourcepoly}, it is not immediately obvious that this term does not have any decaying tail at far distances. To show this, we observe that 
it is only the part of the source involving the odd modes that generates the $\frac{1}{(r-r_s)^2}$ factor, while the part of the source coming from the even modes is always a polynomial. In particular, inserting the odd linear solutions  into \eqref{sourceH0}, 
one can recast the part of the source  from the odd modes in the form
\begin{equation}
\mathcal{S}^{\ell m}_{\mathit{H}_0} \supset \frac{1}{(r-r_s)^2} \!\! \sum_{\ell_1, m_1, \ell_2, m_2} \!\!\!\!\!\!
\mathcal{E}_{\ell_1 m_1}^-\mathcal{E}_{\ell_2 m_2}^-
\sum_{i = 0}^{\ell_1-2}\sum_{j= 0}^{\ell_2-2} \sum_{n=0}^{4} C^{n}_{i j, \ell m} \bigg( \frac{r}{r_s} \bigg)^{i + j + n}  \, ,
\label{sourceH0series}
\end{equation}
where $i$ and $j$ run over the series representation of the linear odd solution (see eq.~\eqref{h05ell-2}),  while the sum over $n$ stems from the additional $r/r_s$ terms in the source \eqref{sourceH0}. The coefficients $C^{n}_{i j, \ell m}$
can be extracted from the general expression of the source \eqref{sourceH0}.
Moreover, the terms in $T_{\ell}(y)$ that might give rise to a tail in $\Hs_0^{ \ell m }$ after integration are those in the first line on the right-hand side of eq.~\eqref{Tynonpol}---the second line  is proportional to $(y-1)^2$, which  cancels the singular prefactor in the source.
Using the  series representation of the Legendre polynomial,
\begin{equation}
P_\ell (2y-1) =  (-1)^\ell\sum_{k=0}^\ell  \begin{pmatrix}
\ell \\ k
\end{pmatrix} \begin{pmatrix}
\ell+k \\ k 
\end{pmatrix} (-y)^k \, ,
\label{Pellsrepre}
\end{equation}
and $\eqref{sourceH0series}$, the last {integral} in eq.~\eqref{H04terms} can be recast as\footnote{To obtain \eqref{intTSmm}, one can  plug \eqref{Pellsrepre} into \eqref{Tynonpol} and \eqref{sourceH0series}, and  use the identity 
\begin{equation}
y^m - y_0^m = (y-y_0)(y^{m-1} + y^{m-2}y_0 + \dots + y y_0^{m-2}+ y_0^{m-1})  \, ,
\end{equation}
with $y_0=1$, to first rewrite 
\begin{equation}
\frac{y^s}{y-1} = \frac{1}{y-1} + \dots \, ,
\qquad
\frac{(2y-1)y^s}{(y-1)^2} = \frac{1}{(y-1)^2}  + \frac{s+2}{y-1} + \ldots \, ,
\end{equation}
up to simple polynomials in $y$.
Then, one can perform the integral in $y$: the integral  $\int^{r/r_s}\frac{{\rm d}y}{(y-1)^2}$ generates terms with an inverse power of $r-r_s$, but this gets canceled when multiplied by the associated Legendre polynomial $P^2_\ell (2r/r_s-1)$ (see eq.~\eqref{expression_P2}); the integral $\int^{r/r_s}\frac{{\rm d}y}{y-1}$ produces instead the log in \eqref{intTSmm}. Finally, resumming in $k$  yields the combination $\frac{1}{2}C^{n}_{i j} \left[\ell(\ell+1)-(i + j + n)-2\right]$.}
\be
\begin{split}
  \int^{r/r_s}  \!\!\! {\rm d}y \,  T_{\ell}(y)
 \mathcal{S}^{\ell m}_{\mathit{H}_0} (y) 
= \ & \frac{1}{2}    \log\left( \frac{r}{r_s}-1 \right) \!\! \sum_{\ell_1, m_1, \ell_2, m_2}  \!\! \!\!  \mathcal{E}_{\ell_1 m_1}^-\mathcal{E}_{\ell_2 m_2}^- \\
& \times
 \sum_{i = 0}^{\ell_1-2}\sum_{j= 0}^{\ell_2-2} \sum_{n=0}^{4} C^{n}_{i j, \ell m} \left[\ell(\ell+1)-(i + j + n)-2\right]+\ldots, 
\;
\label{intTSmm}
\end{split}
\ee
where, as usual, we have dropped  polynomials  in $r/r_s$. For fixed values of $\ell$ and $m$, using the explicit expressions for the coefficients $C^{n}_{i j,\ell m}$ and summing over $n$, we get that the second line of  \eqref{intTSmm} is proportional to 
\be
\begin{split}
& \left\{  3 \ell_1(\ell_1+1) \ell_2(\ell_2+1) {}^{0} \!   I_{\ell \ell_1  \ell_2}^{m  m_1 m_2}+\left[2 + \ell(\ell+1)
   -2 \ell_1(\ell_1+1) -2 \ell_2(\ell_2+1)\right]
   {}^{1} \!   I_{\ell \ell_1  \ell_2}^{m  m_1 m_2} +2 \, {}^{2} \!   I_{\ell \ell_1  \ell_2}^{m  m_1 m_2} \right\} 
\\
& \qquad  \times \sum_{i = 0}^{\ell_1-2}\sum_{j= 0}^{\ell_2-2} (-1)^{i+j} \frac{ (\ell_1 + i +2)! }{(\ell_1 - i - 2)!}  \frac{(\ell_2 + j +2)!}{(\ell_2 - j - 2)!} \frac{   2 (i + j + 6) - \ell_1(\ell_1+1) - \ell_2(\ell_2+1)   }{ i! (i+3)!  j! (j+3)!}.
\;
\label{doubleSum}
\end{split}
\ee
One can check that the sums on the second line of \eqref{doubleSum} admit an analytic result, which is identically zero for generic, unfixed, values of $\ell_1$ and $\ell_2$.
We stress  that the vanishing of this term is not a consequence of the selection rules enforced by the first line in \eqref{doubleSum} but is a consequence of the specific radial dependence of the source.

All in all, we conclude that the second-order solution in the even sector has the structure
\begin{equation}
\Hs_0^{ \ell m }  (r)  =  - \frac{1}{W^\ell_{H_0}} \left(  r_s^2  P^2_\ell (2r/r_s-1) \log(r/r_s)   - \frac{r_s}{r} {T_{\ell}(0)}  \right) \int _0^{1} {\rm d}y \,  \mathcal{P}^{\ell m}(y) + {\rm polynomial}(r/r_s) \, .
\label{genericH0sol}
\end{equation}
This shows that the second-order even modes do not contain $1/r^n$ falloffs with $n >1$ when expanded at large $r$. Although we cannot provide a general proof here, we find that the  integral above vanishes exactly,
\begin{equation}
\int^{1}_0 \mathrm{d}y\,  \mathcal{P}^{\ell m}(y) = \int^{1}_0 \mathrm{d}y\, y(y-1)  P^2_\ell (2y-1) \mathcal{S}^{\ell m}_{\mathit{H}_0} (y)  =0 ,
\label{integral01text}
\end{equation}
for fixed  values of $\ell$, $\ell_1$ and $\ell_2$, after using the explicit expression for $\mathcal{S}^{\ell m}_{\mathit{H}_0}$. This implies that neither the $1/r$ nor the $\log(r/r_s)$ terms are present in \eqref{genericH0sol}, and that the second-order even solution $\Hs_0^{ \ell m } $ is indeed just a finite polynomial.

\subsubsection{Odd sector} 
\label{sec:oddmatchall}
Let us now turn to eq.~\eqref{eqh0oddodd} for the second-order odd field $\hs^{\ell m}_0 (r)$. The quadratic source on the right-hand side, that here we denote for simplicity as $  \sum_{\ell_1, m_1, \ell_2, m_2} {\cal S}_{h_0 \ \ell \ell_1 \ell_2}^{ m m_1 m_2}  (r) \equiv {\cal S}^{\ell m}_{h_0}  (r)$, is of mixed type, given by the product of a linear even mode and a linear odd one (see eq.~\eqref{sourceh0}).  The solution to \eqref{eqh0oddodd} is given by eq.~\eqref{h0_Greensreg}:
\begin{equation}
\hs^{\ell m}_0 (r)   = \frac{r_s^2}{ W^\ell_{h_0} }  
  \left[ h_{0,4}^\ell (r)
  \int_{1}^{r/r_s}  {\rm d}y \,  h_{0,5}^\ell (y)
   {\cal S}^{\ell m}_{h_0}  (y) 
- h_{0,5}^\ell (r) \int^{r/r_s}  {\rm d}y \,  h_{0,4}^\ell (y)
   {\cal S}^{\ell m}_{h_0}  (y) 
\right] \;,
\label{h0_Greensreg-2}
\end{equation}
where $h_{0,4}^\ell (r)$ and $h_{0,5}^\ell (r)$ can be read off from eqs.~\eqref{h04ell} and \eqref{h05ell}. 
It is worth noticing that  the odd source ${\cal S}^{\ell m}_{h_0}(r)$ computed on the linear field solutions $\hf^{ \ell m}_{0} (r)$ and $\Hf^{ \ell m}_{0} (r)$ is a finite polynomial in $r$ and vanishes in $r=0$: $\mathcal{S}^{\ell m}_{\mathit{h}_0} (r) = r \cdot {\rm polynomial}(r)$.

One important ingredient in the argument for the even case was the possibility of recasting the non-analytic terms in the quadratic solution  in the form of  the first line in eq.~\eqref{H04terms}. This was a direct consequence of eq.~\eqref{LegendreQ2_proper_variable}, i.e.~of the fact that all the $x$-dependence in the coefficient of the log in $Q_\ell^2$  is fully determined by the first Legendre solution $P_\ell^2$. Note that this is not  special of the Legendre polynomials, but it holds more in general for  solutions of generic Fuchsian-type second-order ordinary differential equations~\cite{kristensson2010second}.\footnote{This happens whenever the roots of the indicial equation for a given regular singular point ($r=0$ and $r= r_s$ in the case of \eqref{eqh0oddodd}) differ by an integer number~\cite{kristensson2010second}.}
In the case of \eqref{eqh0oddodd}, one can indeed show that the independent solution $h_{0,4}^\ell (r)$ in eq.~\eqref{h04ell}  has the structure~\cite{kristensson2010second}
\begin{equation}
h_{0,4}^\ell (r) = \alpha_\ell \log\left( \frac{r}{r-r_s} \right) h_{0,5}^\ell (r) + \frac{\zeta_\ell(r)}{r} \; , 
\end{equation}
where $h_{0,5}^\ell (r)$ is the polynomial in eq.~\eqref{h05ell-2}, while $\alpha_\ell$ are constant, $\ell$-dependent coefficients, and $\zeta_\ell(r)$ is an analytic function of $r$ in the neighbourhoods of $r=0$ and $r= r_s$. In particular, it follows from the properties of the hypergeometric function that $\zeta_\ell(r)$ is a finite polynomial in $r$ \cite{Bateman:100233}. 

Thus, we can follow the same logic  as in the case above and write
\begin{equation}
\hs^{\ell m}_0 (r)   = \frac{\alpha_\ell  r_s^2}{ W^\ell_{h_0} }    h_{0,5}^\ell (r)
  \left[ \log\left( \frac{r}{r-r_s} \right)
  \int_{1}^{r/r_s}  {\rm d}y \,  h_{0,5}^\ell (y)
   {\cal S}^{\ell m}_{h_0}  (y) 
- \int^{r/r_s}  {\rm d}y \,  \log\left( \frac{y}{y-1} \right) h_{0,5}^\ell (y) 
   {\cal S}^{\ell m}_{h_0}  (y) 
\right]+\dots
\label{h0_Greensreg-3}
\end{equation}
where we kept only terms that could potentially give rise to inverse powers of $r$ at large $r$.
From the same algebraic manipulations that we used for the first line in \eqref{H04terms}, we can then conclude that $\hs^{\ell m}_0 (r)$ has no induced decaying $1/r^n$ falloff (with $n>1$) at $r=\infty$.
As for the second-order $\Hs^{\ell m}_0$, $\hs^{\ell m}_0$ takes the form of a polynomial in $r/r_s$ with only positive integer powers, except for a $\frac{1}{r}$ and  a $\log(r)$ term, which will not affect our argument in section~\ref{sec:matchingL}. Nevertheless, the coefficients of  these terms are proportional to the integral
\begin{equation}
\int_{0}^{1}  {\rm d}y \,  h_{0,5}^\ell (y)
   {\cal S}^{\ell m}_{h_0}  (y)  = 0 ,
 \label{toprove3}
\end{equation}
which we find to be vanishing after fixing specific values of $\ell_1$, $\ell_2$ and $\ell$, although we cannot provide a general proof here.

\subsection{Matching to all orders in derivatives}
\label{sec:matchingL}

In this section, we will perform the matching between the worldline EFT and the full solutions for $\Hs^{\ell m}_0 (r) $ and $\hs^{\ell m}_0 (r) $ discussed above, for all values of $\ell$. As opposed to sec.~\ref{sec:match}, where we studied the matching for $\ell=2$ tidal fields and showed how to reconstruct the nonlinear tidal solution at the next-to-leading order in the large-$r$ expansion,  
here we will keep $\ell$ generic, while working at the leading order in $r_s$, i.e., neglecting the contribution from the point-particle action $S_{\rm pp}$ in eq.~\eqref{eq:Spp}.
This will be enough to conclude that all quadratic Love number couplings vanish in the EFT in the case of Schwarzschild black holes.

After imposing parity invariance, the generalization of the cubic Love number action \eqref{eq:ST} to higher orders in derivatives is  
\begin{equation}
\begin{split}
	S_{\rm T} = \int \D \tau \bigg\{  &  \sum_{\ell=2}^\infty  \Big[  
	 \lambda^{(E)}_\ell E_{  i_1\cdots i_\ell  } E^{  i_1\cdots i_\ell  } + 
	\lambda^{(B)}_\ell B_{  i_1\cdots i_\ell  } B^{  i_1\cdots i_\ell  }  \Big]
	  \\
	 +	& \sum_{n,m,l}  \lambda^{(E^3)}_{nml}  E_{  i_1 \cdots i_l j_1 \cdots j_n } E^{  i_1 \cdots i_l }{}_{k_1 \cdots k_m } E^{  j_1 \cdots j_n k_1 \cdots k_m }  
	\\
		+ &  \sum_{n,m,l}     \lambda^{(EB^2)}_{nml}  E_{  i_1 \cdots i_l j_1 \cdots j_n } B^{  i_1 \cdots i_l}{}_{ k_1 \cdots k_m } B^{  j_1 \cdots j_n k_1 \cdots k_m } 
	 + \mathcal{O} \left(E^4,E^2B^2,B^4 \right)
	\bigg\} \, ,
\end{split}
	\label{eq:STallL}
\end{equation}
where $E_{ i_1 \cdots i_n } $ and $B_{ i_1 \cdots i_n } $ are respectively defined in eqs.~\eqref{multiEdef} and \eqref{multiBdef}, while $n$, $m$ and $l$ are non-negative integers. The cubic terms in the second and third lines of \eqref{eq:STallL} correspond to couplings of $\ell_1 = l+n$, $\ell_2 = l+m$ and $\ell_3 = n+m$ multipoles. Since $E_{ i_1 \cdots i_n } $ and $B_{ i_1 \cdots i_n } $ are STF  tensors, in order to avoid overcounting, the sum in the second line of \eqref{eq:STallL} is constrained to run  over $l \geq n \geq m \geq 0$, which translates into $\ell_1 \geq \ell_2 \geq \ell_3$ (with $\ell_1,\ell_2,\ell_3 \geq 2$). 
The sum in the third line of \eqref{eq:STallL} runs instead over non-negative integers $n$, $m$ and $l$, such that  $l \geq n$, with again $\ell_1,\ell_2,\ell_3 \geq 2$.

 To perform the matching, we will first solve the quadratic equations obtained from the EFT action at leading order in $r_s$ and compute the induced response. We already know that the linear Love numbers vanish~\cite{Damour:2009vw,Binnington:2009bb,Fang:2005qq,Kol:2011vg,Gurlebeck:2015xpa,Hui:2020xxx,Rai:2024lho}, so we can set $\lambda^{(E)}_\ell=0=\lambda^{(B)}_\ell$ in \eqref{eq:STallL}. This will considerably simplify the analysis below. In particular, as we will see,  it will be enough to work with the linearized  components of the Weyl tensor.  In the static case, they read  
\begin{equation}
E_{i j} = -\frac{1}{2} \partial_i \partial_j \delta g_{tt} + \mathcal{O}(\delta g^2),
\qquad
B_{ij} 
= - \frac{1}{2} {\varepsilon_{ti}}^{kl}  \partial_j\partial_k \delta g_{it} + \mathcal{O}(\delta g^2) \, ,
\label{expansionEBlinear}
\end{equation}
where here $\delta g_{\mu\nu}\equiv g_{\mu\nu}- \eta_{\mu\nu} $.  
Then, the variation of the  action $S = S_{\rm EH}+ S_{\rm{T}}$ with respect to $g_{\mu\nu}$, including only quadratic corrections, yields schematically  
\be
\begin{split}
\Mpl^2 G_{\mu\nu}  \sim  
 \bigg[ &  \sum_{\ell_1,\ell_2,\ell_3} \lambda^{(E^3)}_{\ell_1\ell_2\ell_3} (-1)^{\ell_1}  \partial_{i_1} \cdots \partial_{i_{\ell_1}}\left( \delta\ord{3}(\bm{x})   E_{   L_2  } E_{  L_3  }
\right) 
 	\\
	  	+  & \sum_{\ell_1,\ell_2,\ell_3}  \lambda^{(EB^2)}_{\ell_1\ell_2\ell_3} (-1)^{\ell_1}  \partial_{i_1} \cdots \partial_{i_{\ell_1}}\left( \delta\ord{3}(\bm{x})   B_{  L_2 } B_{  L_3 }
\right) + \text{perms} \bigg] \delta_\mu^t\delta_\nu^t
  \\
	  	+ \bigg[  &  \sum_{\ell_1,\ell_2,\ell_3}  \lambda^{(EB^2)}_{\ell_1\ell_2\ell_3} (-1)^{\ell_2} 
	\varepsilon_{t j}{}^{ i_{1} i_2 } \partial_{i_1} \cdots \partial_{i_{\ell_2}}\left( \delta\ord{3}(\bm{x})   E_{  L_1 } B_{  L_3 } 
\right) + \text{perms} \bigg]  \delta_{(\mu}^j\delta_{\nu)}^t 
  \\
	& \equiv J^+(\bm{x})  \delta_\mu^t\delta_\nu^t+ 
	 J^-_j (\bm{x}) \delta_{(\mu}^j\delta_{\nu)}^t \, ,
\label{EeqsS}
\end{split}
\ee
where the precise form of the contractions among the tensors on the right-hand side will not be important for our argument. We are interested in solving the previous perturbation  equations on flat space at second order in the fields. Since the right-hand side is already quadratic in the Weyl tensor, as anticipated, the linear expansions \eqref{expansionEBlinear} will suffice. 

Equations \eqref{EeqsS} can be also written as 
\begin{equation}
\Mpl^2  R_{\mu\nu} \sim J^+(\bm{x}) \left(  \delta_\mu^t\delta_\nu^t  + \frac{1}{2}\eta_{\mu\nu} \right)
+ 
 J^-_j(\bm{x})   \delta_{(\mu}^j\delta_{\nu)}^t \, .
\label{EeqsS2}
\end{equation}
Let us  decompose the metric fluctuation as 
\begin{equation}
\delta g_{\mu\nu} = H_{\mu\nu} + h_{\mu\nu} \, ,
\end{equation}
where $H_{\mu\nu}$ represents the external tidal field, while $h_{\mu\nu}$ is the response. Each field is in turn expanded in perturbation theory as $H_{\mu\nu}= {}^{(1)}\! H_{\mu\nu}+{}^{(2)}\! H_{\mu\nu}+\dots $, and $h_{\mu\nu}= {}^{(1)}\! h_{\mu\nu}+{}^{(2)}\! h_{\mu\nu}+\dots $, to be plugged into \eqref{EeqsS2}. By definition, $H_{\mu\nu}$ solves the homogeneous vacuum Einstein equations i.e., $R_{\mu\nu}[h=0]=0$. Since the linear response vanishes, ${}^{(1)}\! h_{\mu\nu}=0$, the left-hand side of  \eqref{EeqsS2} is  simply $\Mpl^2 \, {}^{(1)} \!\delta R_{\mu\nu} [{}^{(2)}\! h]$, where ${}^{(1)} \! \delta R_{\mu\nu}$ denotes the first-order variation of the Ricci tensor, evaluated on the second-order perturbation ${}^{(2)}\! h_{\mu\nu}$. Instead, the right-hand side of \eqref{EeqsS2} depends quadratically  on ${}^{(1)}\! H_{\mu\nu}$, after the Weyl tensors are evaluated on the linear tidal field solution. All in all, \eqref{EeqsS2} becomes at second order an inhomogeneous equation for ${}^{(2)}\! h_{\mu\nu}$, with a fully known source, which can be solved through standard Green's function methods. 

To this end, it is convenient to choose the  de Donder gauge, defined by
\begin{equation}
\partial^\mu \delta g_{\mu\nu} - \frac{1}{2} \partial_\nu \delta g =0 \, ,
\label{dedonderflat}
\end{equation}
with $\delta g \equiv \eta^{\mu\nu} \delta g_{\mu\nu}$.\footnote{Here the gauge fixing for $H_{\mu \nu}$ is slightly different from the one that we chose in sec.~\ref{sec:EFT}, where $H_{\mu\nu}$ was in RW gauge.}  
In such a gauge,  a certain number of simplifications occur on flat space,    in particular $\delta^{(1)} \! R_{\mu\nu} = -\frac{1}{2}\square \delta g_{\mu\nu}$.
For the static linear tidal field---to plug in the expressions \eqref{expansionEBlinear}, which we will need to evaluate the sources $J^\pm$ in \eqref{EeqsS2}---we shall thus take~\cite{Hui:2020xxx,Rai:2024lho}
\begin{equation}
{}^{(1)}\! H_{tt}(\vec x) = {\cal E}^{+}_{i_1\cdots i_{\ell}}x^{i_1}\cdots x^{i_\ell}\, ,
\qquad
{}^{(1)}\! H_{tj}(\vec x) = {\cal E}^{-}_{j | i_1\cdots i_\ell}x^{i_1}\cdots x^{i_\ell}\, .
\label{tidalh}
\end{equation}
The relevant equations in \eqref{EeqsS2} boil down to
\begin{equation}
  \vec{\nabla}^2 \,  {}^{(2)}\! h_{tt} = - \Mpl^{-2} J^+(\bm{x})  \, ,
  \qquad
\vec{\nabla}^2 \,  {}^{(2)}\! h_{tj} = - \Mpl^{-2} J^-_j(\bm{x})  \, ,
\label{EeqsS2-1}
\end{equation}
where the sources $J^\pm_j(\bm{x})$, evaluated on the tidal fields \eqref{tidalh}, are  proportional to derivatives of the delta-function~\cite{Hui:2020xxx}.
The solutions can be straightforwardly obtained by going to momentum  following~\cite{Hui:2020xxx,Rai:2024lho}, and are found to scale at large $r\equiv \vert \bm{x} \vert$ as ${}^{(2)}\! h_{tt}\sim 1/r^{\ell+1}$, and similarly for ${}^{(2)}\! h_{tj}$, with $\ell$ corresponding to the multipole number of the induced second-order field and such that $\ell\geq2$. (This scaling can be also understood by simple power counting, by looking at the $\bm{x}$-dependence in the sources in \eqref{EeqsS}.)
These solutions should then be compared with the expressions for $\Hs_0^{ \ell m }$  and $\hs^{\ell m}_0$ in eqs.~\eqref{H0PQellgeneric} and \eqref{h0_Greensreg-2}, obtained in full general relativity. Note that in general such a comparison requires some care, as the two calculations are  done in two different gauges, as we extensively discussed in sec.~\ref{sec:gaugetrans}. However, as long as one is interested in the matching at leading order in the flat-space limit, one can easily check that, under the assumptions of vanishing linear response (i.e., ${}^{(1)}\! h_{\mu\nu}=0$) and static limit, the metric perturbations ${}^{(2)}\! \delta g_{tt}$ and ${}^{(2)}\! \delta g_{tj}$ in the RW and de Donder (dD) gauge coincide, i.e.,\footnote{This can be seen also by using eqs.~\eqref{eq:G00tr}--\eqref{eq:G0itr} and carefully taking the $r_s\rightarrow0$ limit in eq.~\eqref{eq:xi1p}. 
Indeed, by dimensional analysis the scale of the tidal field is set by  $\mathcal{E}_{\ell m}/r_s^\ell$, which should be kept  finite when taking this limit. In this case ${}\ord{1}\!\xi^\mu=0$, barring residual large gauge transformations satisfying $\square {}\ord{1}\!\xi^\mu=0$, which are allowed in de Donder gauge.}
\begin{equation}
{}^{(2)}\! \delta g_{tt}^{\rm RW}={}^{(2)}\! \delta g_{tt}^{\rm dD}\, ,
\qquad {}^{(2)}\! \delta g_{tj}^{\rm RW}={}^{(2)}\! \delta g_{tj}^{\rm dD} \, .
\end{equation}
This makes the matching of ${}^{(2)}\! h_{tt}$ and ${}^{(2)}\! h_{tj}$ from \eqref{EeqsS2-1} with their counterpart solutions in the full theory straightforward. In particular, since we showed in sec.~\ref{sec:1order} that $\Hs_0^{ \ell m }$  and $\hs^{\ell m}_0$ take a simple form with no $1/r^n$ ($n>1$) falloff at large $r$, one then concludes that all the Love number coefficients $\lambda^{(E^3)}$ and $\lambda^{(EB^2)}$ in \eqref{eq:STallL} vanish to all orders in derivatives. Subleading corrections in $r_s$ to the nonlinear tidal field solution are expected to be reconstructed by including contributions from the point-particle action $S_{\rm pp}$, as we did in sec.~\ref{sec:match} for an external quadrupolar tidal field.

\section{Conclusions}

In this work, we provided the first comprehensive analysis of static quadratic  perturbations and quadratic Love numbers of Schwarzschild black holes in general relativity, extending the previous results of \cite{Riva:2023rcm}. We studied both even and odd sectors, with no restriction on the angular dependence of the modes. We derived the second-order equations in  Regge--Wheeler gauge and obtained an analytic expression for the sources (see sec.~\ref{sec:quad}). We showed that, when evaluated on the linear solutions, the sources take   a simple form, which is in turn  inherited by the induced second-order metric perturbations. In full generality, we proved  that the latter take the form of simple polynomials with only positive powers of $r$, except  for potentially a $1/r$ and a $\log(r)$ term; although we could not provide a general proof of the absence thereof, explicit checks for fixed values of the angular momentum quantum numbers of the modes show that their coefficients generically vanish (see end of secs.~\ref{sec:evenmatchall} and \ref{sec:oddmatchall}). 
We then used the worldline effective field theory \cite{Goldberger:2004jt,Goldberger:2005cd,Goldberger:2006bd,Rothstein:2014sra,Porto:2016pyg,Levi:2018nxp,Goldberger:2022ebt,Goldberger:2022rqf} to defined the nonlinear tidal deformability and quadratic Love numbers of compact objects.  By performing an explicit matching with the full solutions derived in sec.~\ref{sec:sol}, we concluded that all quadratic (even and odd) Love numbers vanish for Schwarzschild black holes.

Our results raise a series of interesting questions on the fundamental nature of black holes in general relativity. First of all, the vanishing of the quadratic Love numbers appears as a fine tuning in the worldline EFT, mirroring the case of the linear Love numbers~\cite{Rothstein:2014sra,Porto:2016zng}. In the latter case,  symmetries acting on the static sector of the linearized perturbations have been found~\cite{Hui:2021vcv,Hui:2022vbh,Rai:2024lho} and proposed as a resolution to the puzzle. Our findings here are  suggestive that the ladder symmetries of~\cite{Hui:2021vcv,Hui:2022vbh,Rai:2024lho} admit an extension to nonlinear order, which could be used to explain  the vanishing of the nonlinear tidal response of black holes in general relativity~\cite{Combaluzier-Szteinsznaider:2024sgb,Kehagias:2024rtz}. In addition, the simplicity of the obtained results, in particular the form of the sources and of the second-order solutions (see sec.~\ref{sec:1order}) suggests that one might be able to reorganize perturbation theory in a way that takes full advantage of the symmetries of the system and simplifies the intermediate steps. One notable example of this is given by axisymmetric (even) perturbations: for this subset of modes, 
the full metric, including perturbations, can be recast in the Weyl form, which has been used to show that there is no static, axisymmetric induced response to all order in perturbations~\cite{Gurlebeck:2015xpa,Poisson:2021yau, Combaluzier-Szteinsznaider:2024sgb,Kehagias:2024rtz}. This applies to the even sector and for modes with zero magnetic quantum number. On the other hand, we showed here that also the odd sector displays a very simple structure. It will be interesting to understand this further, and find a more systematic way to perform the matching, and generalize our results to cubic and higher orders in perturbations beyond axisymmetry.
Finally,  it will be interesting to study subleading tidal effects, in particular nonlinearities,  for different black hole solutions, such as Kerr spacetime,  black holes with charge, and solutions in higher spacetime dimensions \cite{Hui:2020xxx, Charalambous:2021mea, Rodriguez:2023xjd,Charalambous:2023jgq,Rosen:2020crj}. We leave all  these aspects for future investigations.

\vspace{-.3cm}
%=======================================
\paragraph{Acknowledgements} 
It is a pleasure to thank Eric Poisson for interesting comments on our paper. The research of L.S.~has been funded, in part, by the French National Research Agency (ANR) under project ANR-24-CE31-1097-01, and by the Programme National GRAM of CNRS/INSU with INP and IN2P3 co-funded by CNES.
M.M.R.~is partially funded by the Deutsche Forschungsgemeinschaft (DFG, German
Research Foundation) under Germany's Excellence Strategy -- EXC 2121 ``Quantum Universe'' -- 390833306, by the ERC Consolidator Grant ``Precision Gravity: From the LHC to LISA'' provided by the European Research Council (ERC) under the European Union's H2020 research and innovation programme (grant agreement No.~817791) and by the DOE
DE-SC0011941.

\appendix

%%%%%%%%%%%%%%%%%%%%%%%%%%
\section{Integration of tensor harmonics} \label{App: harmonics}

In this appendix we provide  details on how to perform the angular integrations of tensor harmonics, eqs.~\eqref{intE0}--\eqref{intJ2}. We remind the reader that $\nabla_A$ is the covariant derivative  on the unit 2-sphere $S^2$, $\epsilon_{AB}$ is the Levi-Civita tensor (with non-vanishing components $\epsilon_{\theta \phi} = - \epsilon_{\theta \phi}  = \sin\theta$) and that we use the metric on $S^2$,  
$\gamma_{AB}$,  to raise and lower indices.   

We will use the so-called ``pure-spin'' harmonics approach, following ref.~\cite{Brizuela:2006ne}.    
We construct symmetric trace-free (STF) tensor harmonics by acting on  scalar spherical harmonics with covariant derivatives $\nabla_A$ and with the Levi-Civita symbol $\epsilon_{AB}$. In particular, 
we define even and odd STF $s$-index tensors of spin $s \geq 1$ ($s$ is integer), angular momentum $\ell$ and magnetic number $m$, respectively  as
\begin{align}
    Z_{\ell \ A_1 \ldots A_s}^{m} &\equiv (\nabla_{A_1} \ldots \nabla_{A_s} Y_{\ell}^{m})_{\rm STF} \; , \\
    X_{\ell \  A_1 \ldots A_s}^{m} &\equiv (\epsilon_{A_1}^{\  \; B} Z_{\ell \  B A_2 \ldots A_s}^{m})_{\rm STF},
\end{align}
where by the index STF we denote the  part. We discuss the case $s=0 $ separately below. 
We can rewrite the integrals in compact form as
 \begin{align}
{}^{s} \!   I_{\ell \ell_1  \ell_2}^{m  m_1 m_2} & \equiv  \int {\rm d} \Omega \; Y_{\ell}^{m}   Z_{\ell_{1} \  A_1 \ldots A_s}^{m_{1}}  Z_{\ell_{2}}^{m_{2} \  A_1 \ldots A_s} \label{int1} \;, \\
{}^{s} \!   J_{\ell \ell_1  \ell_2}^{m  m_1 m_2}  & \equiv \int {\rm d} \Omega \; Y_{\ell}^{m}   X_{\ell_{1} \  A_1 \ldots A_s}^{m_{1}}  Z_{\ell_{2}}^{m_{2} \ A_1 \ldots A_s} \label{int2} \;.
\end{align}
Then, we can express the products of two spin-$s$ spherical tensors with all the indices contracted, into spherical harmonics,    i.e.~\cite{Brizuela:2006ne},   
\begin{align}
    Z^{m_{1}}_{\ell_{1} \  A_1 \ldots A_s} Z^{m_{2}, A_1 \ldots A_s}_{\ell_{2}}  &=  (-1)^s  \sum_{\ell = |\ell_{1} - \ell_{2}|}^{\ell_{1} + \ell_{2}} \left[ 1 + (-1)^{\ell + \ell_{1} +\ell_{2}}\right] E^{s\;  \ell_{2} \; m_{2}}_{-s\;  \ell_{1} \; m_{1} \;  \ell} \;  Y_{\ell}^{m_{1}+m_{2}} \;,  \label{decompTI1}\\
    X^{m_{1}}_{\ell_{1} \ A_1 \ldots A_s} Z^{m_{2}, A_1 \ldots A_s}_{\ell_{2}}  &= - i (-1)^s   \sum_{\ell = |\ell_{1} - \ell_{2}|}^{\ell_{1} + \ell_{2}} \left[ 1 - (-1)^{\ell + \ell_{1} +\ell_{2}}\right] E^{s\;  \ell_{2} \;  m_{2}}_{-s\; \ell_{1} \;  m_{1}\; \ell} \; Y_{\ell}^{m_{1}+m_{2}} \;,\label{decompTI2}
\end{align}
where the factor $\left[ 1 \pm (-1)^{\ell + \ell_{1} +\ell_{2}}\right]$  ensures that only terms with a specific $(\ell + \ell_{1} +\ell_{2})$-parity (even for $+$ and odd for $-$) are picked up in the sum.
The coefficients  of these expansions, $E^{s \;  \ell_{2} \;  m_{2}}_{-s \;  \ell_{1}\;  m_{1} \ell}$, can be written in terms of Clebsch--Gordan coefficients $\langle  \ell_1 \; m_1 \; \ell_2 \; m_2 | \ell \; m \rangle$. Specifically, they are given by
\begin{equation}
    E^{s_2 \;  \ell_{2} \; m_{2}}_{s_1 \;  \ell_{1} \;  m_{1} \;  \ell} = 
    \frac{k(\ell_{1}, |s_1|) k(\ell_{2}, |s_2|)}{k(\ell, |s_1+s_2|)} \; C^{m_{1}  m_{2}  (m_{1}+m_{2})}_{\ell_{1}   \ell_{2}  \ell} \; C^{s_1   s_2   (s_1+s_2)}_{\ell_{1}   \ell_{2}   \ell} \;,
    \label{Edef}
\end{equation}
where $C^{m_{1}  m_{2}   m}_{\ell_{1}  \ell_{2}  \ell} \equiv \langle  \ell_1 \; m_1 \; \ell_2\;  m_2 | \ell \; m \rangle$
and
\be
k(\ell,s) \equiv \sqrt{\frac{(1+2\ell)(\ell+s)!}{2^{s+2}\pi(\ell-s)!}}.
\ee

We can now solve the integrals \eqref{int1} and \eqref{int2}. 
Using the expansions eqs.~\eqref{decompTI1}-\eqref{Edef},  the orthogonality condition, $\int {\rm d} \Omega  \; Y_{\ell}^m \; Y_{\ell '}^{m'}{}^* = \delta_{\ell \ell'} \delta^{m m'}$, 
and the relation
\be
\frac{k(\ell_{1}, |s|) k(\ell_{2}, |s|)}{k(\ell, 0)}  = 2^{- |s|} \sqrt{\frac{(1+2 \ell_1) (1+2 \ell_2) }{4 \pi (1+2 \ell)} \frac{(l_1+|s|)! (l_2+|s|)! }{(l_1-|s|)! (l_2-|s|)!}} \;,
\ee
we obtain,  
\begin{align}
{}^{s}\!   I_{\ell \ell_1 \ell_2}^{m m_1 m_2}  &=  \frac12 \left[ 1 + (-1)^{\ell + \ell_{1} +\ell_{2}}\right]  \; \EE{s}_{\ell \ell_1 \ell_2}^{m m_1 m_2}     \;, \label{IntI}\\
 {}^{s}\!   J_{\ell \ell_1 \ell_2}^{m m_1 m_2} &=  - i \frac12 \left[ 1 - (-1)^{\ell + \ell_{1} +\ell_{2}}\right]  \; \EE{s}_{\ell \ell_1 \ell_2}^{m m_1 m_2}   \;, \label{IntJ}
\end{align}
where
\begin{align}
\EE{s}_{\ell \ell_1  \ell_2}^{m  m_1 m_2}  \equiv  2^{1-|s|}   (-1)^s    \sqrt{\frac{(1+2 \ell_1) (1+2 \ell_2) }{4 \pi (1+2 \ell)} \frac{(l_1+|s|)! (l_2+|s|)! }{(l_1-|s|)! (l_2-|s|)!}}  \; C^{m_{1}   m_{2}  m }_{\ell_{1}  \ell_{2}  \ell}  \; C^{-s   s   0}_{\ell_{1}  \ell_{2}   \ell} \;.
\end{align}
The case $s=0$ is well known, i.e.,
\begin{align}
\EE{0}_{\ell \ell_1  \ell_2}^{m  m_1 m_2}  \equiv      \sqrt{\frac{(1+2 \ell_1) (1+2 \ell_2) }{4 \pi (1+2 \ell)} }  \; C^{m_{1}   m_{2}    m }_{\ell_{1}    \ell_{2}   \ell}  \; C^{0   0   0}_{\ell_{1}    \ell_{2}    \ell} \;,
\end{align}
so the following definition,
\be
\EE{s}_{\ell \ell_1  \ell_2}^{m  m_1 m_2}  \equiv  (2 - \delta_{0s}) 2^{-s}   (-1)^s    \sqrt{\frac{(1+2 \ell_1) (1+2 \ell_2) }{4 \pi (1+2 \ell)} \frac{(l_1+|s|)! (l_2+|s|)! }{(l_1-|s|)! (l_2-|s|)!}}  \; C^{m_{1}   m_{2}  m }_{\ell_{1}  \ell_{2}   \ell}  \; C^{-s   s   0}_{\ell_{1}   \ell_{2}  \ell} \;,
\ee
holds for $s\ge 0$.

We note the following symmetry properties of the Clebsch--Gordan coefficients:
\be
C^{m_{1}   m_{2}   m }_{\ell_{1}  \ell_{2}  \ell}  = (-1)^{\ell + \ell_1 + \ell_2} C^{ m_{2} m_{1}    m }_{ \ell_{2}  \ell_{1}  \ell}  = (-1)^{\ell + \ell_1 + \ell_2} C^{- m_{1}   - m_{2}   - m }_{\ell_{1}   \ell_{2}   \ell} \;. \label{CGprop}
\ee
Since $C^{ - s    s  0 }_{\ell_{1}  \ell_{2}   \ell}  =  C^{ - s  s   0 }_{ \ell_{2}  \ell_{1}   \ell} $,  $\EE{s}_{\ell \ell_1  \ell_2}^{m  m_1 m_2}$ has the same symmetries as $C^{m_{1}   m_{2}   m }_{\ell_{1}   \ell_{2}   \ell}$,
i.e., $\EE{s}_{\ell \ell_1  \ell_2}^{m  m_1 m_2}$ is symmetric under the exchange \(\ell_1 m_1 \leftrightarrow \ell_2 m_2\) if \(\ell + \ell_1 + \ell_2\) is even, and antisymmetric if $\ell + \ell_1 + \ell_2$ is odd.
Therefore,  eqs.~\eqref{IntI} and \eqref{IntJ} can be rewritten as
 \begin{align}
{}^{s}\!I_{\ell \ell_1  \ell_2}^{m  m_1 m_2}  &=  \frac12 (\EE{s}_{\ell \ell_1  \ell_2}^{m  m_1 m_2}  + \EE{s}_{\ell \ell_1  \ell_2}^{m  m_1 m_2} ) \; , \label{Ipart} \\
{}^{s}\!J_{\ell \ell_1  \ell_2}^{m  m_1 m_2} &= - \frac{i}2 ( \EE{s}_{\ell \ell_1  \ell_2}^{m  m_1 m_2} -  \EE{s}_{\ell \ell_1  \ell_2}^{m  m_1 m_2}) \; , 
\label{Jpart}
\end{align}
respectively.
Note also that, if  $\ell + \ell_1 + \ell_2$ is odd, eq.~\eqref{CGprop} implies
$\EE{s}^{ 0 \  0   0 }_{\ell_{1}  \ell_{2}   \ell}  =  0 $   and $\EE{s}^{ m_1  m_1   m }_{\ell_{1}   \ell_{1}   \ell}  =  0 $. The same properties hold for ${}^{s}\!J_{\ell \ell_1  \ell_2}^{m  m_1 m_2}$.

Finally, note that  the $s = 1$ integrals can be expressed in terms of the $s = 0$ ones using integrations by parts. For instance, 
\be
\begin{split}
{}^{1} \!   I_{\ell \ell_1  \ell_2}^{m  m_1 m_2} &=    \int {\rm d } \Omega \; Y_{\ell}^{m} \nabla_{A} Y_{\ell_{1}}^{m_{1}} \nabla^{A} Y_{\ell_{2}}^{m_{2}} =  \frac{1}{2}\int_{S^2} Y_{\ell}^m \left [ \Delta_{S^2} \Big( Y_{\ell_{1}}^{m_{1}} Y_{\ell_{2}}^{m_{2}} \Big) - Y_{\ell_{1}}^{m_{1}} \Delta_{S^2} Y_{\ell_{2}}^{m_{2}} - Y_{\ell_{2}}^{m_{2}} \Delta_{S^2} Y_{\ell_{1}}^{m_{1}}  \right ]  
    \\
  &   =\frac{1}{2}\Big[\ell_{1}(\ell_{1}+1) + \ell_{2}(\ell_{2}+1) - \ell(\ell+1)\Big] {}^{0} \!   I_{\ell \ell_1  \ell_2}^{m  m_1 m_2}  \;.
\end{split}
\ee

\section{Coefficients}
\label{app:source}

Here we provide the explicit expressions of  coefficients appearing in the text, starting from those in the sources in sec.~\ref{sec:quad}.
 The coefficients appearing in eq.~\eqref{sourceH0} are
\begin{equation}
\begin{split}
   \Apz  & =  \frac{ 1}{4    }   (\lambda -2)  (\lambda - \lambda_1 -\lambda_2 ) \lambda   \;, \\
   \Apu   &= -   \frac{\lambda^3}{4       }  
  + \frac{ \lambda^2 (4 +  \lambda_1(4 - 4 \lambda_1 + \lambda_1 \lambda_2)   -  \lambda_1 \lambda_2)  }{2    \left(\lambda _1-2\right) \left(\lambda _2-2\right) }  
  - \frac{ \lambda_1\lambda (20 - 10 \lambda_1 - 6 \lambda_2 + 3 \lambda_1 \lambda_2)  }{2      \left(\lambda _1-2\right) \left(\lambda _2-2\right) }   + \frac{ 1}{2    }  \lambda_1 (\lambda_1 - \lambda_2)    \\
    \Apd   &= -   \frac{\lambda_1(\lambda_1-\lambda_2) }{4     } +  \frac{\lambda^2  (2 \lambda_1^2  - \lambda_1 \lambda_2 - 4)}{4    \left(\lambda _1-2\right) \left(\lambda _2-2\right) } +  \frac{ \lambda(8   + \lambda_1 (4 - 6 \lambda_1 + \lambda_1 \lambda_2)) }{4    \left(\lambda _1-2\right) \left(\lambda _2-2\right) }    \\
  \Bpd   & =  -  \frac{  (\lambda_1 - \lambda_2) (2 + \lambda_1 -\lambda_2) }{4     }  + \frac{  \lambda( \lambda_1^2 + \lambda_1 + 2 \lambda_2 -6 )}{2 (\lambda_1 -2)     } - \frac{  \lambda^2 ( 3 \lambda_1 + 2 \lambda_2 - 6 )}{4 (\lambda_1 -2)    }   \;, \\
  \Amz &= \frac12  \lambda_1 \lambda_2 (\lambda -2) \lambda  \;, \\ 
\Amu &=   \frac14  \{  \lambda ^3 -4 \lambda ^2  -12 \lambda + 2 \lambda [ (\lambda _1 + \lambda _2 )   ( 5  -3 \lambda ) +  2 \lambda _1 \lambda _2  ( 6  -  \lambda ) ] + ( \lambda _1^2 +  \lambda _2^2) (- 6 + 11 \lambda)       \nonumber  \\ 
& \qquad -6 ( \lambda _1^3 + \lambda _2^3 ) +12 (\lambda _1+ \lambda _2)^2 \}   \;, \\
\Amd & = \frac{1}{8} \{ -5 \lambda ^3 +44 \lambda ^2 +92 \lambda +  ( \lambda _1 + \lambda _2)  ( -92 -120 \lambda + 21\lambda^2  ) - \lambda _1 \lambda _2 ( 132 + 56 \lambda - 4 \lambda ^2) \nonumber \\ 
 & \qquad +  (\lambda _1^2 + \lambda _2^2) (76  -35 \lambda)         +19 ( \lambda _1^3+ \lambda _2^3)     - \lambda_1 \lambda_2 (\lambda_1 + \lambda_2 ) \} \;, \\ 
\Amt  &=  \frac{1}{8} \{ 3 \lambda ^3 -58 \lambda ^2 -112 \lambda + (\lambda _1+ \lambda_2) ( 112+ 164\lambda -9 \lambda ^2) +13 (\lambda _1^2 +\lambda_2^2)\lambda    + \lambda _1 \lambda _2  (124 + 16 \lambda)  \nonumber   \\ 
& \qquad -7 (\lambda _1^3 + \lambda _2^3) -106  (\lambda _1^2 + \lambda_2^2)    +\lambda _1 \lambda _2 (\lambda_1+\lambda_2)   \} \;, \\ 
\Amq & = \frac{1}{8} \{ 19 \lambda ^2 +38 \lambda -  (\lambda _1 + \lambda_2) (38 + 58 \lambda)    +39 (\lambda _1^2+  \lambda _2^2)  -40 \lambda _1 \lambda _2 \} \;, \\
\Bmz &= \frac{1}{4} ( \lambda ^3 +4 \lambda ^2 +20 \lambda +2 \lambda _1 \lambda ^2+3 \lambda _2 \lambda ^2-5 \lambda _1^2 \lambda -9 \lambda _2^2 \lambda -22 \lambda _1 \lambda -10 \lambda _1 \lambda _2 \lambda -16 \lambda _2 \lambda  +2 \lambda _1^3+5 \lambda _2^3 \nonumber \\
&\qquad +18 \lambda _1^2-2 \lambda _1 \lambda _2^2+12 \lambda _2^2-20 \lambda _1+3 \lambda _1^2 \lambda _2-26 \lambda _1 \lambda _2-20 \lambda _2 )   \;, \\
\Bmu & = -\frac{\lambda^3}{4}+\frac{1}{4} \left(-2 \lambda_1-3 \lambda_2-12\right) \lambda ^2+\frac{\lambda}{4} \left(5 \lambda_1^2+10 \lambda_2 \lambda_1+48 \lambda_1+9 \lambda_2^2+42 \lambda_2-36\right) + \\
&+ \frac{1}{4} \left(-2
\lambda_1^3-3 \lambda_2 \lambda_1^2-36 \lambda_1^2+2 \lambda_2^2 \lambda_1+46 \lambda_2 \lambda_1+36 \lambda_1-5 \lambda_2^3-30 \lambda_2^2+36 \lambda_2\right) \;, \\
\Bmd & = \frac{11}{4}\lambda^2+\frac{\lambda}{4} \left\{22 -32 (\lambda_1+ \lambda_2)\right\} +\frac{1}{4} \left\{21
(\lambda_1^2+\lambda_2^2) -22 (\lambda_1+\lambda_2) -20 \lambda_1 \lambda_2 \right\} \;, \\
\Ez &= -\frac{\lambda^3}{4}+\frac{\lambda^2}{8} \left(2\lambda_1+2 \lambda_2+4\right) +\frac{\lambda}{8} \left(-4 \lambda_1-4 \lambda_2\right) \;, \\
\Eu &= \frac{\lambda^3}{4}+\frac{ \lambda^2}{8} \left(-2
\lambda_1-2 \lambda_2-12\right)+\frac{\lambda}{8} \left(26 \lambda_1+26
\lambda_2-16\right) +\frac{1}{8}
\left(-14 \lambda_1^2+12 \lambda_2 \lambda
_1+16 \lambda_1-14 \lambda_2^2+16 \lambda
_2\right) \\
\Ed &= \frac12 \lambda^2+\frac{ \lambda}{8} \left(-13 \lambda_1-13
\lambda_2+8\right) +\frac{1}{8}
\left(9 \lambda_1^2-10 \lambda_2 \lambda
_1-8 \lambda_1+9 \lambda_2^2-8 \lambda
_2\right) \\
\Cmz &= 6-\lambda +6 \lambda_1 \;, \qquad
\Cmu = \frac12 ( 3 \lambda -9 \lambda_1-46 ) \;, \qquad
\Cmd = \frac12 (56-\lambda +3 \lambda_1) \;, \\
\Dmz &= -10 + \lambda - \lambda_1-3 \lambda_2 \;, \qquad
\Dmu = 18 -\lambda +\lambda_1+3 \lambda_2 \;, 
\end{split}
\end{equation}

The coefficients appearing in eq.~\eqref{sourceh0} are
\be
\begin{split}
\Cz & = \lambda - \lambda_1 - \lambda_2 + \lambda_1  \lambda_2 \;, \qquad
\Cu  =  \frac{ 
 \lambda (2 -  \lambda_1 + 2 \lambda_2)}{    \lambda_1-2    } +  \lambda_1  + \lambda_2  - \lambda_1 \lambda_2    \;, \\
 \Cd & =  - \frac{ \lambda (2 - \lambda_1 + 2  \lambda_2)}{  2 ( \lambda_1-2)    }  -   \frac12 (\lambda_1 - \lambda_2 )    \;, \\
 \Dz & = - \lambda + \lambda_1 - \lambda_2  \;, \qquad
 \Du  = - \frac12  \left( \frac{\lambda (6 - 3 \lambda_1 - 2\lambda_2)}{   \lambda _1-2 } +  3 (\lambda_1 - \lambda_2) \right) \;,
\end{split}
\ee

The numerical coefficients appearing in eq.~\eqref{eqs:TTensorNLO} are given by 
\begin{subequations}
\begin{align}
	& c\ord{0,+}_\eta = 1 \, , & & c\ord{0,+}_{v} = \frac{1}{2} \, , & & c\ord{0,+}_{r} = \frac{5}{2} \, , & &  {c}\ord{0,-}_\eta = -15 \, , & &  {c}\ord{0,-}_{v} = -\frac{21}{2} \, , & &  {c}\ord{0,-}_{r} = -\frac{117}{2} \, , \\
	& c\ord{2,+}_{\eta} = -\frac{1}{8} \, , & & c\ord{2,+}_{v} = -\frac{5}{8} \, , & &  c\ord{2,+}_{r} = -\frac{7}{8} \, , & &  {c}\ord{2,-}_{\eta} = \frac{63}{8} \, , & &  {c}\ord{2,-}_{v} = \frac{99}{8} \, ,& &  {c}\ord{2,-}_{r} = \frac{105}{8} \, ,\\
	& c\ord{4,+}_{\eta} = \frac{97}{36} \, , & & c\ord{4,+}_{v} = \frac{85}{18} \, , & &   & &  {c}\ord{4,-}_{\eta} = -\frac{7}{36} \, , & &  {c}\ord{4,-}_{v} = -\frac{73}{18} \, , & &  \\
	& d\ord{2} = -\frac{1}{2} \, ,& &  d\ord{4} = \frac{8}{5} \, . & &  & &   
\end{align}
\end{subequations}

%%%%%%%%%%%%%%%%%%%%%%%%%%%%
\section{From spherical to Cartesian coordinates}
\label{App:SphToCart}
%%%%%%%%%%%%%%%%%%%%%%%%%%%%

The goal of this  appendix is to express the tidal field in the RW gauge in a Lorentz-covariant form, making it easy to express it in Cartesian coordinates, which is the most convenient form for Feynman diagrams. As discussed in sec.~\ref{sec:TidalH}, the tidal field represents a perturbation around Minkowski spacetime, i.e., $H_{\mu\nu} = \bar{g}_{\mu\nu} - \eta_{\mu\nu}$. We can decompose this perturbation into even and odd modes, denoted as $H^{+}_{\mu\nu}$ and $H^{-}_{\mu\nu}$, respectively. In the RW gauge and in spherical coordinates, these take the following form:
\begin{align}
	H^{+}_{\mu\nu} & = 
	\begin{pmatrix}
		\bar{H}_0 & \bar{H}_1 & 0 & 0 \\
		\bar{H}_1 & \bar{H}_2 & 0 & 0 \\
		0 & 0 & r^2 \bar{K} & 0 \\
		0 & 0 & 0 & r^2 \sin^2\theta \bar{K} 
	\end{pmatrix} \, , 
	\label{eq:Hegen}\\
	H^{-}_{\mu\nu} & = 
	\begin{pmatrix}
		0 & 0 & -\frac{\partial_\phi \bar{h}_0}{\sin\theta} & \sin\theta \partial_\theta \bar{h}_0 \\
		0 & 0 & -\frac{\partial_\phi \bar{h}_1}{\sin\theta} & \sin\theta \partial_\theta \bar{h}_1 \\
		-\frac{\partial_\phi \bar{h}_0}{\sin\theta} & \sin\theta \partial_\theta \bar{h}_1 & 0 & 0  \\
		-\frac{\partial_\phi \bar{h}_0}{\sin\theta} & \sin\theta \partial_\theta \bar{h}_1 & 0 & 0 
	\end{pmatrix} \, .
	\label{eq:Hogen}
\end{align}

To rewrite them in covariant form, we remind the reader that $v^\mu$ is the four-velocity of the massive object, which defines our time-like direction. In its rest frame, where we work,  $v^\mu = (1,0,0,0)$.
The radial direction is defined by the spatial vector  ${r}^\mu \equiv (\delta^\mu{}_\nu + v^\mu v_\nu) x^\nu$, or by the unit vector parallel to it, $\hat r^\mu = r^\mu /\sqrt{ r^\rho r_\rho} = (0,1,0,0)$, where the indices are raised and lowered with the Minkowski metric. Finally, projecting orthogonally to these two directions, we obtain the metric on the unit 2-sphere, i.e., 
\be
\Omega_{\mu\nu} = \eta_{\mu\nu} + v_\mu v_\nu -\hat{r}_\mu\hat{r}_\nu 	\, .  \label{sphercoor}
\ee

Therefore, the even part of the tidal field simply reads
\begin{align}
	H^{+}_{\mu\nu} & = \bar{H}_0 v_\mu v_\nu + 2\bar{H}_1 v_{(\mu} \hat{r}_{\nu)} + \bar{K} \Omega_{\mu\nu} \notag \\
	& = \big( \bar{H}_0 + \bar{K} \big) v_\mu v_\nu + 2\bar{H}_1 v_{(\mu}\hat{r}_{\nu)}
		+\big( \bar{H}_2 - \bar{K} \big) \hat{r}_\mu\hat{r}_\nu 
		+\bar{K} \eta_{\mu\nu} \, ,
\end{align}
where for the last equality we have used   eq.~\eqref{sphercoor}.

To obtain the odd part of the tidal field, it is convenient to introduce the two basis vectors on the unit 2-sphere
\be
e_A^\mu = \frac{\partial \hat r^\mu}{\partial x^A} \;, \qquad x^A \in \{ \theta, \phi \} \; ,
\ee
which are obviously orthogonal among each other and to the other two vectors, and satisfy 
\be
\eta_{\mu \nu} e^\mu_A e^\nu_B = \Omega_{AB} \;.
\ee
Since by eq.~\eqref{eq:Hogen} only  $H^-_{0A} = v^\mu e_A^\nu H^-_{\mu \nu}$ and $H^-_{rA} = \hat r^\mu e_A^\nu H^-_{\mu \nu}$ are nonzero, we can write
\be
H^-_{\mu \nu} = - 2 v_{(\mu} e^A_{\nu)} H^-_{0 A} + 2 \hat r_{(\mu} e^A_{\nu)} H^-_{r A}  
=  2 v_{(\mu} e^A_{\nu)} \epsilon_{AB} \Omega^{BC} \partial_C  \bar h_0/ - 2  \hat r_{(\mu} e^A_{\nu)} \epsilon_{AB} \Omega^{BC} \partial_C \bar h_1  \;.
\ee
Replacing $\epsilon_{AB}$ on the right-hand side using the identity $\epsilon_{AB} = v^\alpha \hat r ^\beta e^\rho_A e^\sigma_B \varepsilon_{\alpha \beta \rho \sigma} $, we finally obtain
\begin{equation}
	H^{-}_{\mu \nu} = -2  v^\alpha\varepsilon_{ \alpha  \rho\sigma(\mu}v_{\nu)} \hat{r}^\rho\partial^\sigma \bar{h}_0 
	+2  v^\alpha\varepsilon_{ \alpha  \rho\sigma(\mu}  \hat{r}_{\nu)} \hat{r}^\rho\partial^\sigma \bar{h}_1 \, .
\end{equation}
%

%%%%%%%%%%%%%%%%%%%%%%%%%%
\section{Symmetric trace-free tensor relations} 
\label{App:STFrel}
%%%%%%%%%%%%%%%%%%%%%%%%%%

Here we derive some useful relations involving the STF tensors $\cY_{i_1 \dots i_\ell}^{ m}$ introduced in sec.~\ref{sec:TidalH}. We  recall 
their definition,
\begin{equation}
 Y_\ell^m(\theta, \phi) =   \cY^{ m}_{i_1 \dots i_\ell}  \hat r^{i_1} \cdots  \hat r^{i_\ell} =  \cY^{ m}_{i_L} \, \hat r^{i_L}  \, ,
\end{equation}
where in the second equality we use the multi-index notation introduced in sec.~\ref{ppaction}.

Using this, we rewrite  the normalization relation of two spherical harmonics with the same $\ell$ in terms of the STF tensors,
\be
	\cY_{i_L}^{m *} \cY_{j_{L}}^{m'} 
	\int \D\Omega \, \hat{r}^{i_{2 L}}    
	 = \delta^{m m'}  \, .
\ee
Solving the angular integral,\footnote{We use
$	\int \D\Omega \,  \hat{r}^{i_1}\hat{r}^{i_2} \dots \hat{r}^{i_{2\ell-1}} \hat{r}^{i_{2\ell}} = \frac{4\pi}{(2\ell +1)!!}
	\big(
	\delta_{i_1 i_2} \dots \delta_{i_{2\ell -1} i_{2\ell}}
	+ \text{perms}
	\big) $.}
we obtain 
\be
\cY_{i_L}^{m *} \cY_{j_{L}}^{m'} 
	\big(	\delta_{i_1 i_2} \cdots \delta_{i_{2\ell -1} i_{2\ell}}
	+ \text{perms}	\big)    =  \frac{ (2 \ell+1) !! }{4 \pi} \delta^{m m'}  \;,
\ee
from which we can derive  
\be
 \cY^{m *}_{ijkl} \cY^{m'}_{ijkl} = \frac{315}{32\pi} \delta^{m m'} \, ,
\ee
which is needed for the computation  in sec.~\ref{sec:hEFT}.

We can follow a similar procedure for the products of three STF tensors.
In particular, rewriting  the definition of ${}^0\!\mathcal{I}_{\ell \ell_1 \ell_2}^{m m_1 m_2}$ (eq.~\eqref{intE0})  in terms of the STF tensors and solving the angular integral for $\ell + \ell_1 +\ell_2$ even, we obtain
\begin{align}
	\cY_{i_L}^{m *} \cY_{j_{L_1}}^{m_1}\cY_{k_{L_2}}^{m_1} \big(
	\delta_{i_1 i_2} \cdots \delta_{i_{2\ell -1} i_{2\ell}}
	+ \text{perms}
	\big) = \frac{(\ell+\ell_1+\ell_2 +1)!!}{4\pi}
	{}^0\! {I}_{\ell \ell_1 \ell_2}^{m m_1 m_2} \, .
\end{align}
For the computation  in sec.~\ref{sec:hEFT} we need the following contractions 
\begin{align}
	 \cY^0\cY^{m_1}_{ij} \cY^{m_2}_{ij} &= \frac{15}{8\pi} {}^0\! {I}_{0 2 2}^{0 m_1 m_2} \, , \\
	 \cY_{ij}^{m *}\cY^{m_1}_{jk} \cY^{m_2}_{ki} &= \frac{105}{32\pi} {}^0\! {I}_{2 2 2}^{m m_1 m_2} \, , \\
	 \cY_{ijkl}^{m *}\cY^{m_1}_{ij} \cY^{m_2}_{kl} &= \frac{315}{32\pi} {}^0\! {I}_{4 2 2}^{m m_1 m_2} \, .
\end{align}

Finally, in order to simplify the final expression for the one-point function computed in sec.~\ref{sec:hEFT}, one needs to decompose the following tensor $\cY^{m_1}_{ij} \cY^{m_2}_{kl}$. For symmetry reasons, one necessarily has
\begin{align}
	\cY^{m_1}_{ij} \cY^{m_2}_{kl} = 
	c_1 \delta_{ij}\delta_{kl} + 2 c_2  \delta_{i(k}\delta_{l) j}
	+ c_3 \cY_{ij}^m \delta_{kl} + c_4 \cY_{kl}^m \delta_{ij}
	+ 2 c_5 \Big(
	\cY_{i(k}^m \delta_{l) j} + \cY_{j(k}^m \delta_{l) i}
	\Big)
	+ c_6 \cY^m_{ijkl} \, .
\end{align}
The coefficients $c_n$ can now be find by contracting the above expression with all possible combinations of $\delta_{ij}$, $\cY^{m*}_{ij}$ and $\cY^{m*}_{ijkl}$. One eventually obtains
\be
	\cY^{m_1}_{ij} \cY^{m_2}_{kl}  = 
	\frac{{}^0\! {I}_{0 2 2}^{0 m_1 m_2}}{4\sqrt{\pi}} 
	\left(
	3\delta_{i(k}\delta_{l) j} - \delta_{ij}\delta_{kl} 
	\right) 
	- {}^0\! {I}_{2 2 2}^{m m_1 m_2} 
	\left(
	 \cY_{ij}^m \delta_{kl} +   \cY_{kl}^m \delta_{ij} - \frac32\cY_{i(k}^m \delta_{l) j} - \frac32 \cY_{j(k}^m \delta_{l) i}
	\right) 
	+ {}^0\! {I}_{4 2 2}^{m m_1 m_2}  \cY^m_{ijkl} \, .
\ee

\section{Recovering the Schwarzschild metric }
\label{app:SchwM}

%%%%%%%%%%%%%%%%%%%%%%%
\begin{figure}[t]
\centering
\includegraphics[scale=1]{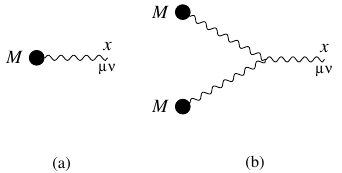}
\caption{Feynman diagrams that reconstruct the Schwarzschild metric up to order $r_s^2$.}
\label{fig:diagsrs}
\end{figure}
%%%%%%%%%%%%%%%%%%%%%%%

In this appendix we show how  one can consistently  reconstruct, from the EFT worldline approach,  the Schwarzschild metric in de Donder coordinates. We perform the explicit calculation up to ${\cal O}(r_s^2)$, in which case the metric takes the form (see e.g.~\cite{Mougiakakos:2020laz})
\begin{equation}
	\D s^2 =  
	- \left(1 - \frac{r_s}{r} + \frac12 \frac{r_s^2}{r^2  }  \right) \D t^2 
	+ \left(1+\frac{r_s}{r} + \frac54 \frac{r_s^2}{r^2}   \right)    \D {\bm x} \cdot \D {\bm x} +  \frac74 \frac{r_s^2}{r^2}    \frac{   ({\bm x} \cdot \D {\bm x})^2  }{r^2} + {\cal O}(r_s^3)\, .
	\label{dedon}
\end{equation}
See also \cite{Goldberger:2004jt,Mougiakakos:2020laz,Jakobsen:2020ksu} for previous results and  \cite{Damgaard:2024fqj,Mougiakakos:2024nku} for a recent reconstruction of the Schwarzschild metric to all orders. 
For the calculation we will assume dimensional regularisation in $d=D+1$ dimensions.

Since for this computation   $H_{\mu\nu} = 0$, we  only need the two diagrams  in fig.~\ref{fig:diagsrs}. 
At leading order we only have the point-particle sourcing $h_{\mu\nu}$, i.e. the diagram \ref{fig:diagsrs}(a). This gives
\be
\begin{split}
	\kappa {}^{(0)}\!h_{\mu\nu}(x)  & = - 8 \pi r_s \int \D t \, v^\rho v^\sigma \int_k 
			\frac{\e^{-i k \cdot ( x(t) - x )}}{k^2}P_{\rho\sigma\mu\nu} + {\cal O} \left((r_s/r)^2\right) \\
			& =8 \pi r_s 
			\left(v_\mu v_\nu +\frac{1}{2}\eta_{\mu\nu}\right)
			\int_{\bm{k}}
			\frac{\e^{i \bm{k} \cdot \bm{x} }}{\bm{k}^2}  + {\cal O} \left((r_s/r)^2\right) \\
			&= \frac{r_s}{ r}
		\left(\eta_{\mu\nu} + 2 v_\mu v_\nu\right) + {\cal O} \left((r_s/r)^2\right) \;,
			\label{eq:SCHLO}
\end{split}
\ee
where $r = \sqrt{\bm{x}^2}= \sqrt{\delta_{ij} x^i x^j }$ and we used $\int_{\bm k}
		\frac{\e^{i \bm{k} \cdot \bm{x} }}{(\bm{k}^2)^\alpha} = 
		\frac{1}{(4\pi)^{D/2}}\frac{\Gamma\left(\frac{D}{2}-\alpha\right)}{\Gamma(\alpha)}
		\left(\frac{\bm{x}^2}{4}\right)^{\alpha - \frac{D}{2} }$ for the last equality. 
One can check that $g_{\mu \nu} = \eta_{\mu\nu} + \kappa  {}^{(0)}\! h_{\mu\nu}(x) $ reproduces 
the Schwarzschild metric in de Donder gauge \eqref{dedon} expanded up to ${\cal O}(r_s)$.

At next-to-leading order order we have to compute the diagram \ref{fig:diagsrs}(b). Using the rules defined in the previous section, we explicitly get
\begin{align}
	\kappa  {}^{(0)}\!h_{\mu\nu}(x) _{1b} & = -  32 \pi^2 r_s^2 \int \D t_1  \D t_2 \, 
	v^{\rho_1}v^{\sigma_1}v^{\rho_2}v^{\sigma_2} 
	\int_{q, k} \frac{\e^{- i q \cdot ( x(t_1)-x(t_2) ) - i k \cdot ( x(t_1) - x )}}{q^2 k^2 (k - q)^2} \notag \\
	& \qquad \quad P_{\rho_1 \sigma_1 \alpha_1 \beta_1} P_{\rho_2 \sigma_2 \alpha_2 \beta_2}
	 \mathcal{V}^{\alpha_1 \beta_1 \alpha_2 \beta_2 \alpha_3 \beta_3}
	 P_{\alpha_3 \beta_3 \mu\nu} \notag \\
	 & =   2 \pi^2 r_s^2 \int_k \ddl(k^0) \frac{\e^{i k \cdot x}}{k^2} 
	\big(7 k_\mu k_\nu + 3 k^2 \eta_{\mu\nu} - k^2 v_\mu v_\nu\big)
	 \int_q  \frac{\ddl(q^0)}{q^2(k-q)^2} \, ,
\end{align}
where the second equality follows from integrating in $t_1$ and $t_2$ and from tensorial reduction of the integrands. Solving the integrals in the last line\footnote{These two integrals are useful
\begin{align}
\int_{\bm k} \frac{1}{[\bm{q}^2]^a[(\bm{k}-\bm{q})^2]^b} &=
	\frac{(\bm{q}^2)^{\frac{D}{2}-a-b}}{(4\pi)^{\frac{D}{2}}}
	\frac{\Gamma\left( a + b - \frac{D}{2} \right)}{\Gamma\left( a \right) \Gamma\left( b \right)}
	\frac{\Gamma\left( \frac{D}{2} - a \right) \Gamma\left( \frac{D}{2} - b \right) }{\Gamma\left( D - a - b \right)} \;, \\
	\int_{\bm k}
		\frac{k^a k^b k^c}{(\bm{k}^2)^\alpha} \e^{i \bm{k} \cdot \bm{x} } &= 
		\frac{i}{(4\pi)^{D/2}}\frac{\Gamma\left(\frac{D}{2} - \alpha + 2\right)}{\Gamma(\alpha)}
		\left(\frac{\bm{x}^2}{4}\right)^{\alpha - \frac{D}{2}  - 2}  \bigg\{\left( \alpha - \frac{D}{2}  -2\right)\frac{x^a x^b x^c}{2\bm{x}^2} + \frac{\delta^{ab}x^{c}}{4} + \frac{\delta^{c (a}x^{b)}}{2} \bigg\} \;.
\end{align} }
and adding this contribution to the leading order one, we finally obtain
\be
\kappa {}^{(0)}\!  h_{\mu\nu}(x)= \frac{r_s}{ r}
		\left(\eta_{\mu\nu} + 2 v_\mu v_\nu\right) + \frac{ r_s^2}{
	4 r^2}
	\left(5 \eta_{\mu\nu} + 3 v_\mu v_\nu - 7 \hat{r}_\mu \hat{r}_\nu \right) +   {\cal O} \left((r_s/r)^3\right)  \, ,
	\label{eq:Schwarz2}
\ee
where we recall that the normalised vector in the radial direction is $\hat{r}^\mu = (0, x^i/r )$.
This result reproduces 
the Schwarzschild metric in de Donder gauge \eqref{dedon} expanded up to ${\cal O}(r_s^2)$ and agrees with refs.~\cite{Jakobsen:2020ksu, Mougiakakos:2020laz}.

\newpage
\renewcommand{\em}{}
\bibliographystyle{utphys}
\addcontentsline{toc}{section}{References}
\bibliography{biblio}

\end{document}